\documentclass[]{revtex4-1}
\pdfoutput=1
\usepackage[ruled,vlined]{algorithm2e}
\usepackage{graphicx}
\usepackage{dcolumn}
\usepackage{bm}
\usepackage{float}
\usepackage{subfigure}
\usepackage{amssymb,amsmath}
\usepackage{mathtools}
\usepackage{url}
\newcommand{\tsize}{0.42}
\begin{document}

\title{Weighted Community Detection and Data Clustering Using Message Passing}

\author{Cheng Shi$^{1,3}$, Yanchen Liu$^{2,3}$ and Pan Zhang$^3$}
\email{To whom correspondence should be addressed. Email: panzhang@itp.ac.cn}

\affiliation{
$^{1}$CompleX Lab, Web Sciences Center, University of Electronic Science and Technology of China, Chengdu 611731, People’s Republic of China\\
$^{2}$Network Science Institute, Northeastern University, 177 Huntington Avenue, Boston, MA, 02115, United States of America\\
$^{3}$CAS key laboratory of theoretical physics, Institute of Theoretical Physics, Chinese Academy of Sciences, Beijing 100190, China
}

\date{\today}

\begin{abstract} 
Grouping objects into clusters based on similarities or weights between them is one of the most important problems in
science and engineering. 
In this work, by extending message passing algorithms and spectral algorithms proposed for unweighted community detection problem, we develop a non-parametric method based on statistical physics, by mapping the problem to Potts model at the critical temperature of spin glass transition and applying belief propagation to solve the marginals corresponding to the Boltzmann distribution. 
Our algorithm is robust to over-fitting and gives a principled way to determine whether there are significant clusters in the data and how many clusters there are.
We apply our method to different clustering tasks.
In the community detection problem in weighted and directed networks, we show that our algorithm significantly outperforms existing algorithms. In the clustering problem when the data was generated by mixture models in the sparse regime we show that our method works all the way down to the theoretical limit of detectability and gives accuracy very close to that of the optimal Bayesian inference. In the semi-supervised clustering problem, our method only needs several labels to work perfectly in classic datasets. 
Finally, we further develop Thouless-Anderson-Palmer equations which reduce heavily the computation complexity in dense-networks but gives almost the same performance as belief propagation.
\end{abstract}

\maketitle

\section{\label{sec:intro}Introduction}
Clustering is one of the core problems in unsupervised learning and plays an important role in science and engineering~\cite{hartigan1975clustering}. It aims to group nodes into clusters in such a way that similar items are put into the same group, while different items are put into different groups.
A common approach is to encode the similarity relationships into a similarity graph, with nodes representing items and weighted edges carrying the similarities. Thus the problem is closely related to clustering in the similarity graph, so-called community detection in weighted networks. 

With a long history, many different clustering methods have been proposed~\cite{jain1999data,Frey2007,rodriguez2014clustering}. These include algorithms that optimize an objective function (with an example of the famous K-means algorithm~\cite{jain2010data} which optimizes distances between data points to their putative centers using a scalable expectation-maximization method); generative modeling \cite{holland1983stochastic} which assumes that the data points are generated by an underlying distribution then turns the clustering problem into a statistical inference problem; and spectral clustering method \cite{Luxburg2007,Shi1997,Ng2001,zhang2016robust} which uses spectral properties of the linear operator associated with the data.

Despite a large variety of methods, there is so far no single algorithm works perfectly in all applications. K-means algorithm is fast and is guaranteed to converge, however, it does not work if the clusters are non-spherical. 
The expectation-maximization algorithm used by the K-means algorithm typically has many possible fixed points, making its performance depend on the initial condition. 
Methods based on optimizing a goodness-to-fit objective are prone to overfit, finding clusters even when there is no significant structure~\cite{Zhang2014pnas}.
Generative modeling is widely used in unsupervised learning tasks, however, it is not easy to choose a correct model having good capability while being easy to perform statistical inference. One also has to be very careful about the over-fitting problem when the model has many parameters.
Moreover, a typical drawback for most clustering algorithms is that they do not provide a principled way to choose the number of clusters, as the objective function is usually a monotone function of the number of groups. This is because for clustering algorithms which optimize an objective function, the objective function can usually be converted to a log-likelihood (or negative log-likelihood) of a generative model, thus optimizing the objective function amounts to maximum-likelihood inference of a generative model. We refer to \cite{Zhang2014pnas,newman2016community} for examples.
The likelihood function describes how well the data is fit by the model.
It is well known that without proper regularizations or model selection method such AIC or BIC, maximum likelihood value is usually an increasing function of number of parameters, i.e. more complex the model is, the better power it could fit. Increasing number of groups results to increasing number of parameters of the model, hence results to increase of the maximum (or minimum) possible value of objective function. 
Thus one needs to specify the number of clusters based on hypothesis testing or adopt an alternative model selection procedure.

Another challenge for standard methods is the data sparsity, that with $n$ items we do not have all $n^2/2$ pairwise measurements, but only a small fraction of them. That is, the similarity graph is sparse. The data sparsity is motivated when the similarity measurements are expensive or difficult to obtain (consider computing similarity of two high-resolution images), or hard to store when the number of items is huge. Another motivation is to speed up the computation using a sampling of measurements~\cite{achlioptas2007fast}, when a small fraction of pairwise measurements is enough to reveal the clustering structures.

In this work, we propose to solve the problem of data sparsity, overfitting, and determining the number of groups jointly using tools of statistical physics. We map the problem onto the Potts model, giving a Gibbs measure at a finite temperature,
then solve the marginals of the Boltzmann distribution using belief propagation (BP) algorithm, which is derived from cavity method of statistical physics. 
By analyzing the stability of the fixed points of belief propagation equations, we can separate the phase diagram into paramagnetic phase, retrieval phase and spin glass phase. This gives us a principled method of deciding whether there are statistically significant clusters in the data, detecting the clusters, and determining the number of clusters.
The linearization of the BP algorithm also gives a fast spectral clustering algorithm that works almost as good as BP, with analytically analyzable spectrum properties.
Our algorithms have computational complexity linear in a number of pairwise similarities, hence is highly scalable if the number of measurements is linear to a number of items.
When the number of measurements is large (i.e. the weighted or similarity graph is dense), we also develop the Thouless-Anderson-Palmer equation where the computational complexity linear to system size.

We notices that the Potts model have been used for a long time for the inference task, such as in \cite{Weigt2009}, and for clustering by Blatt, Wiseman and Domany in $1996$ \cite{blatt1996superparamagnetic,blatt1996clustering,wiseman1998superparamagnetic}. However their clustering methods based on Potts model is much different from this work. First, in \cite{blatt1996superparamagnetic,blatt1996clustering,wiseman1998superparamagnetic} the interactions of the Potts model are non-negative, leading to a ferromagnetic model, while our model uses introduces both positive and negative interactions, hence displays a richer phase diagram including spin glass phase. Moreover, the method in \cite{blatt1996superparamagnetic,blatt1996clustering,wiseman1998superparamagnetic} adopts Monte-Carlo methods using Swendsen-Wang algorithm. The algorithm is known to work well on plainer graphs in, for example, $2$ dimensional lattices. However for general weighted/similarity clustering problems defined in a high-dimensional space, or on real-world (usually sparse) similarity graphs, block updates such as Swendsen-Wang algorithm have a low acceptance rate and hence is not guranteed to be efficient. In contrast, our method relies on mean-field theory and message passing equations, which give linear computational complexity in sparse graphs (using belief propagation) and dense graphs (using TAP equations).

Our approach is built upon recently developed message passing algorithm~\cite{Zhang2014pnas} and spectral algorithms~\cite{Krzakala2013} for community detection in unweighted networks. 
In the unweighted networks the Modularity~\cite{Girvan2002} is a standard measure of community structure and corresponds to log-likelihood of the degree-corrected stochastic block model~\cite{Zhang2014pnas,newman2016community}.
However Modularity does not apply to our case where similarities (weights) are present. Actually we do not have a standard measure of clustering, thus sometimes we need semi-supervision to guide the algorithm, as we demonstrate in Sec.~\ref{sec:semi}. Although the spectral algorithm using non-backtracking matrix~\cite{Krzakala2013} can be trivially extended to weighted graphs by using straightforwardly the weights rather than $1$ in defining the non-backtracking operator, it is clearly sub-optimal in some well-defined clustering problems. We will show that the weighted non-backtracking operator obtained by linearizing the proposed belief propagation algorithm resolves this problem.

To evaluate the performance of our proposed algorithm, we consider the synthetic model of Gaussian mixtures in the sparse regime where optimal statistical inference algorithms exist. We show that our algorithm works almost as good as the optimal inference algorithm, all the way down to the theoretical limit of detection, while do not require parameters of the model, as opposed to the optimal Bayesian inference methods which do require parameters of the model.
Then we show applications of our method in several clustering problems, including semi-supervised clustering on standard datasets, and community detection in weighted and directed networks where we demonstrate that our method significantly outperforms existing methods such as Louvain, Infomap and Oslom. 

The following text is organized as follows.
In Sec.~\ref{sec:method} we describe in detail our methods. These include belief propagation algorithm for data clustering, its associated non-backtracking operator, analysis of phase diagram, and how to determine the number of clusters. In 
Sec.~\ref{sec:app} we apply our method to several problems, including clustering in the mixture of Gaussians in sparse regime, semi-supervised clustering in classic problems and community detection in weighted and directed networks. We conclude in Sec.~\ref{sec:con}.

\section{method}\label{sec:method}
We consider clustering of $n$ nodes into $q$ groups. An item $i$ takes a discrete group number $t_i\in \{1,...,q\}$, interacting with another item $j$ with the pairwise similarity measurement $\omega_{ij}$. 
In this work we consider that some pairs of similarities are unknown, so the similarity graph is sparse. Therefore we can treat the system as a sparse Potts model, with each group assignment $\{t\}$ being a Potts configuration.
We note that our method obviously works as well when all $n(n-1)/2$ pairs of similarities are given, i.e. the similarity graph is fully connected. In Sec.~\ref{sec:con} we discuss the possible speed-up of our method in the fully-connected case.
We set the chosen objective function $Q(\{t\})$ that we want to optimize as negative Hamiltonian $-E(\{t\})=Q(\{t\})$ of the Potts model, and assign to partitions a Boltzmann distribution 
at a finite inverse temperature ${\beta}$: 
\begin{equation}\label{eq:P}
P(\{t\})=\frac{1}{Z}e^{\beta Q(\{t\})},
\end{equation}
with partition function written as 
$Z=\sum_se^{\beta Q(\{s\})}$.
The function of the introduced temperature is tuning the energy level so that significant clustering can emerge, as we will demonstrate in Sec.~\ref{sec:significance}.
It is easy to see that with $\beta=0$, i.e. at an infinite temperature, every partition has the same probability $\frac{1}{q^n}$, whereas at zero temperature $\beta\to\infty$, only partitions that optimize the $Q(\{t\})$ have finite measure.

Here we treat similarities as random variables.
Notice that in the real-world data where we usually observe different similarities between pairs of variables without accessing the real process of generating the similarities, random variables would be the most unbiased assumption. Of course, if we have prior knowledge on the similarities, we would be able to use them.
Actually, our method do not necessarily need the similarities to be a random variable. An example is that in the Stochastic Block Model with a uniform similarities on a (structured) random graph, our algorithm reduces to known algorithms which works as good as optimal algorithms.

Once the set of similarities $\{\omega_{ij}\}$ between pairs of nodes are given as weights, we use the internal-weights as the objective function to measure the quality of the partition:
\begin{equation}
Q(\{t\})=\frac{1}{m}(\sum_{\left \langle ij \right \rangle\in\mathcal{E}}\omega_{ij} \delta_{t_{i}t_{j}}-\sum_{\left \langle ij \right \rangle}\overline{\omega}\delta_{t_{i}t_{j}})
\end{equation}
where $\mathcal{E}$ represents the set of all edges in the similarity graph, $m$ is the number of edges, and $\overline{\omega}=2\sum_{\left \langle ij \right \rangle\in\mathcal{E}}\omega_{ij}/n^2$ is the averaged mean weight. We can check that the definition of internal-weights guarantees that for a random partition $\tilde t$, the expectation of internal-weights is

\begin{align}
m\mathbb{E} \left (Q(\{\tilde t\})\right ) &=\mathbb{E} \left (\sum_{\left \langle ij \right \rangle\in\varepsilon}\omega_{ij} \delta_{\tilde t_{i}\tilde t_{j}}\right )-\mathbb{E} \left ( \sum_{\left \langle ij \right \rangle}\overline{\omega}\delta_{\tilde t_{i}\tilde t_{j}}\right) =0.\nonumber
\end{align}
Therefore, larger value of $Q(\{t\})$ represents larger deviation of $\{t\}$ from a random partition.

\subsection{\label{sec:bp}Belief propagation algorithm}
The marginal distribution of Boltzmann distribution Eq.~\eqref{eq:P} is in general difficult to solve, so we aim to approximately compute marginal distribution $\{\psi_{t_i}\}$ using approximations, then
assign to each node its most likely group according to its marginal distribution. 
As we consider the case where the similarity graph is sparse, the best approximation that we could take is the Bethe approximation which approximate the Boltzmann distribution Eq.~\eqref{eq:P} as
\begin{equation}\label{eq:Bethe}
P(\{t\})\approx\hat P(\{t\})=\frac{\prod_{\langle ij\rangle}\psi_{t_i,t_j}}{\prod_i\psi_{t_i}^{d_i}}.
\end{equation}
Here $\psi_{t_it_j}$ denotes two-point marginals of variable $i$ and $j$, and $d_i$ is the degree (number of neighbors) of node $i$ in the similarity graph.
The Bethe approximation is exact when the similarity graph is a tree and a good approximation when the similarity graph is sparse. Based on Eq.~\eqref{eq:Bethe}, one can write the Bethe free energy as a good approximation to the true free energy $-\frac{1}{\beta}\log Z$, then derive a message passing algorithm, so-called \textit{belief propagation}, that minimizes the Bethe free energy~\cite{Mezard2009, Yedidia2001}.

Using the conditional marginals (or messages) $\psi_{t_i}^{i \rightarrow j}$, BP equation is written as 
\begin{align}\label{eq:bp:iter}
\psi_{t_i}^{i\rightarrow k}
=\frac{1}{Z_{i\rightarrow k}}\prod_{j\in \partial i \backslash k}\sum _{t_j=1}^{q} e^{\beta \omega_{ij} \delta_{t_i t_j}}\psi_{t_j}^{j \rightarrow i} 
\prod_{l\neq i}\sum _{t_l=1}^{q} e^{-\beta\overline{\omega} \delta_{t_i t_l} }\psi_{t_l}^{l \rightarrow i}
\end{align}
where $\partial i \backslash k$ denotes the neighbors of node $i$ except for node $k$, and $Z_{i\rightarrow k}$ is the normalization constant.
In the language of cavity method~\cite{Mezard2009}, the message $\psi_{t_i}^{i\to j}$ means the marginal probability that node $i$ belongs to group $q_i$ in the absence of node $j$. The first term in (4) represents the interactions between node $i$ and its neighbors, while the second term represents the weak interactions between node $i$ and all others. Since there is interaction between each pair of nodes, the computational complexity of one update of all messages is $O(n^2)$. In the case of sparse similarity graphs on which we mainly focus in our work, we can greatly reduce the computational complexity by doing  mean-field approximation to the overall weak interactions between every pair of nodes (detailed derivations can be found in Appendix~\ref{sec:derive}):
\begin{equation}\label{eq:bp}
\psi_{t_i}^{i\rightarrow k}
\approx \frac{e^{h(t_i)}}{Z_{i\rightarrow k}}\prod_{j\in \partial i \backslash k} (1+\psi_{t_i}^{j \rightarrow i}(e^{\beta\omega_{ij}}-1))  
\end{equation}
where $h(t)=-\beta \overline{\omega}\sum_i \psi_{t}^{i}$ is the field representing the effect of all the other nodes to the node $i$. With this approximation the computational complexity is reduced to $O(m)$, i.e. linear in the number of edges.

After iterating the messages until they converge or exceed a specified maximum number of iterations, we can compute marginals of each node using
\begin{equation}\label{eq:bp:marginal}
\psi_{t_i}^i=\frac{ e^{h(t_i)}}{Z_{i}}\prod_{j\in \partial i } (1+\psi_{t_i}^{j \rightarrow i}(e^{\beta\omega_{ij}}-1)) ,
\end{equation}
then obtain the partition $\{\hat{t}\}$ by assigning each node to the group of which its marginal is the largest. We define the internal weight of the obtained partition $Q\{\hat t\}$ \textit{retrieval weights}.
From Eq.~\eqref{eq:bp} we observe that BP equation always has a fixed point  
\begin{equation}
\psi_{t_i}^{i \rightarrow j}=\psi_{t_i}^{i }=\frac{1}{q},
\end{equation} 
that we call the \textit{factorized fixed point} or \textit{paramagnetic fixed point}. 
This fixed point actually comes from the permutation symmetry of the system, only carrying information that each node is equally likely to be in every group. 
One can show that the paramagnetic fixed point is the only fixed point when the temperature is high, i.e. $\beta$ is close to $0$. With a large $\beta$ the paramagnetic fixed point could be unstable. Then there may exist another fixed point that dominates the Gibbs measure, or too many fixed points that BP jumps back and forth hence fails to converge. 
Depending on convergence properties of BP and the value of the retrieval weights $Q(\{t\})$, we can divide the phase diagram into three phases:
\begin{itemize}
\item{Paramagnetic phase, where BP converges to the paramagnetic solution, and every partition has the same weight hence by definition retrieval weights $Q\{\hat t\}=0$.}
\item{Non-convergence phase, where BP does not converge. If the underlying graph is a sparse graph, the non-convergence of BP reflects that the spin-glass susceptibility diverges, and replica symmetry is broken~\cite{Zdeborova2009}. In this phase $Q\{\hat t\}$ has a fluctuating value and depends on the initial conditions.}

\item{Retrieval phase, where BP converges to a non-factorized fixed point, with a non-vanishing retrieval weights $Q\{\hat t\}$. }
\end{itemize}

These phases represent different structures in the data: the paramagnetic phase reflects the permutation symmetry of the system, spin glass phase represents the random structure of noise, and the retrieval phase represents the significant clustering structure in the data.
The typical phase diagram is that at low $\beta$ regime system is in the paramagnetic phase. With $\beta$ increases, the paramagnetic fixed point becomes unstable and the system enters either the retrieval phase or the spin glass phase,
depending on whether there exists statistically significant clustering structure and how strong it is relative to the random structure. It is known that in some models with a clustering structure, e.g. the stochastic block model~\cite{holland1983stochastic}, the model is not distinguishable from a random graph from any test, in other words, contiguous to random graphs~\cite{mossel2015reconstruction}, if the planted partition exists but is not strong enough.

\subsection{Existence of the statistically significant clustering structure}\label{sec:significance}
The goal of our algorithm is to find the statistically significant clustering structure, which is represented by the retrieval phase. The first step is to determine whether the retrieval phase exists. Naively we can scan the whole range of $\beta$, trying to find the retrieval phase where BP converges and $Q\{\hat t\}>0$. This procedure is actually fast because one can use a binary search. However, we can be smarter by making use of the fact that when the paramagnetic fixed point becomes unstable, the system will jump to either the retrieval phase or the spin glass phase. This is to say that we only need to check the inverse temperature where the system is supposed to enter spin glass phase from the paramagnetic phase, the spin-glass transition $\beta^*$: if BP converges at $\beta^*$ and $Q\{\hat t \} >0$, we conclude that system has a retrieval phase, otherwise system has no retrieval phase.

The spin-glass transition can be calculated analytically by analyzing stability of the paramagnetic fixed point under random perturbations. Suppose we give a perturbation with zero mean and unit variance to the paramagnetic fixed point as
\begin{equation}\label{eq:noise_1}
\psi_{t_k}^{k\to i}=\frac{1}{q}+\epsilon_{t_k}^{k\to i}.
\end{equation}
After one step of iteration of BP equation Eq.~\eqref{eq:bp}, the perturbation is propagated to neighbors of each node with
\begin{equation}\label{eq:noise}
\epsilon_{t_i}^{i \rightarrow j}=\sum\limits_{k \in\partial i\backslash j}\sum\limits_{t_k}T^{i \rightarrow j, k \rightarrow i}_{t_i,t_k}\epsilon_{t_k}^{k \rightarrow i},
\end{equation}
where the $q\times q$ matrix $T$ encodes the derivatives of messages at the paramagnetic fixed point
\begin{align}
T^{i \rightarrow j, k \rightarrow i}_{t_i,t_k}&=\left.\frac{\partial \psi_{t_i}^{i \rightarrow j}}{\partial \psi_{t_k}^{k \rightarrow i}} \right|_{\psi = \frac{1}{q}} 
\end{align}
The derivatives can be calculated using Eq.~\eqref{eq:bp}:
\begin{align}\label{eq:T}
T=\begin{bmatrix}
    \frac{e^{\beta \omega_{ij}}-1}{e^{\beta \omega_{ij}}+q-1}\frac{q-1}{q} & \frac{1-e^{\beta \omega_{ij}}}{e^{\beta \omega_{ij}}+q-1}\frac{1}{q}  & \dots  & \frac{1-e^{\beta \omega_{ij}}}{e^{\beta \omega_{ij}}+q-1} \frac{1}{q}\\
    \frac{1-e^{\beta \omega_{ij}}}{e^{\beta \omega_{ij}}+q-1}\frac{1}{q} & \frac{e^{\beta \omega_{ij}}-1}{e^{\beta \omega_{ij}}+q-1}\frac{q-1}{q}  & \dots  & \frac{1-e^{\beta \omega_{ij}}}{e^{\beta \omega_{ij}}+q-1} \frac{1}{q}\\
    \vdots & \vdots & \ddots & \vdots \\
    \frac{1-e^{\beta \omega_{ij}}}{e^{\beta \omega_{ij}}+q-1}\frac{1}{q} & \frac{1-e^{\beta \omega_{ij}}}{e^{\beta \omega_{ij}}+q-1} \frac{1}{q}& \dots  &\frac{e^{\beta \omega_{ij}}-1}{e^{\beta \omega_{ij}}+q-1} \frac{q-1}{q}
\end{bmatrix},
\end{align}
with the largest eigenvalue being
\begin{equation}
\eta_{ij}=\frac{e^{\beta \omega_{ij}}-1}{e^{\beta \omega_{ij}}+q-1}.
\end{equation}

Since here we consider a sparse similarity graph, the graph is assumed to have a locally-tree-like property.
Then after $\tau$ steps iteration of BP equation Eq.~\eqref{eq:bp}, the perturbation is propagated to (on average) $\hat c^\tau$ neighbors through $\hat c^\tau$ distinct paths. For each path the perturbation that arrives at the end would be $\epsilon_i \eta_{1}\eta_{12}\eta_{23}...\eta_{(\tau-1)\tau}$.
Sum of all perturbations at the end of each path is a random variable with mean $0$ and variance
\begin{equation}
\mathbb{V_\tau}=\langle\eta^2_{ij}\rangle^{\tau}_{\omega_{ij}}\hat {c}^\tau.
\end{equation}
Here $\langle\cdot\rangle_{\omega_{ij}}$ denotes averaging over $\omega_{ij}$, $\hat c$ denotes average excess degree given degree distribution $p(d)$: 
\begin{equation}
\hat c=\sum_{d=1}^\infty\frac{dp(d)}{c}(d-1)=\frac{\langle d^2
\rangle}{c}-1.
\end{equation}
If $\mathbb{V_\tau}>1$, random noise will propagate in the system, and paramagnetic fixed point is not stable. Thus the spin-glass transition temperature can be obtained by solving the following equations.
\begin{align}\label{eq:beta_star}
\left \langle \left( \frac{e^{\beta^* \omega_{ij}}-1}{e^{\beta^* \omega_{ij}}+q-1}\right)^2 \right \rangle_{\omega_{ij}} \hat c =1.
\end{align}

\subsection{\label{sec:nonbacktracking} The non-backtracking operator}
As what has been investigated in~\cite{Krzakala2013,dogmat}, when a system has permutation symmetry, linearizing the BP equation at the paramagnetic fixed point results to an efficient spectral algorithm using the so-called non-backtracking operator. 
Again we analyze the stability of the factorized solution by linearizing the cavity marginals:
\begin{equation}\label{eq:Delta}
\psi_{t_i}^{i \rightarrow j}=\frac{1}{q}+\Delta_{t_i}^{i \rightarrow j}. 
\end{equation}
Notice that the difference between the last equation to Eq.~\eqref{eq:noise_1} is that $\Delta_{t_i}$ does not represent a zero-mean noise, but a perturbation in the direction of the clustering structure.
The iterating equation of linearized belief propagation can be written as
\begin{equation}
\Delta_{t_i}^{i \rightarrow j}\coloneqq\sum\limits_{k \in\partial i\backslash j}\sum\limits_{t_k}T^{i \rightarrow j, k \rightarrow l}_{t_i,t_k}\Delta_{t_k}^{k \rightarrow l},
\end{equation}
where $\coloneqq$ denotes equality up to a constant. The matrix $T$ is given by Eq.~\eqref{eq:T}, and can be re-written as
\begin{equation}
T^{i \rightarrow j, k \rightarrow l}_{t_i,t_k}=B_{i \rightarrow j, k \rightarrow l}\otimes I_{q\times q}+\frac{1}{q}B_{i \rightarrow j, k \rightarrow l}\otimes J_{q\times q},
\end{equation}
where $\otimes$ denotes the tensor product, $I_{q\times q}$ and $J_{q\times q}$ are $q\times q$ identity matrix and all-one matrix respectively.
Here we introduce the $q\times q$ non-backtracking matrix $B$
\begin{equation}\label{eq:B}
B_{i \rightarrow j, k \rightarrow l}=\delta_{il}(1-\delta_{kj})\frac{e^{\beta \omega_{ij}}-1}{e^{\beta \omega_{ij}}+q-1}.
\end{equation}
Therefore, if we represent the vector of deviations (whose length is $2m\times q$) by
$ \Delta =\{\Delta_{t}^{i \rightarrow j}|t\in \{1,2,...,q\}, (i,j)\in\mathcal E\}$,
its update equation is written as 
\begin{equation}\label{eq:update_1}
\Delta\coloneqq B \otimes I_{q\times q} \cdot\Delta +\frac{1}{q}B\otimes J_{q\times q} \cdot\Delta.
\end{equation}
By making use of the normalization condition
$\sum_q\psi_i^{i\to j}=1$
and Eq.~\eqref{eq:Delta}, we have equality
$\sum\limits_{t_i} \Delta_{t_i}^{i \rightarrow j}=0$.
Then the update equation is simplified to 
\begin{equation}\label{eq:update_2}
\Delta_t\coloneqq B \cdot\Delta_t.
\end{equation}
Here $\Delta_t = \{\Delta_t^{i \to j}|t(i,j)\in\mathcal E\}$ is the vector of deviations on group $t$, which is of size $2m$.

From the last equation we can see that linearizing BP becomes the eigenvector problem of the non-backtracking matrix $B$. 
It is easy to see that $B$ is asymmetric, so its eigenvalues and eigenvectors could be complex numbers. Analogous to the non-backtracking operator of a graph \cite{Krzakala2013}, on the complex plane, its spectrum is mainly composed of two parts: a circle that most of the eigenvalues are confined in, and possible real-eigenvalues outside the circle. The radius of the circle, so-called the \textit{edge of bulk}, together with the leading eigenvalue $\lambda_1$ describe completely the stability of the paramagnetic fixed point of BP as well as the phases diagram of the system: 
\begin{itemize}
\item{If absolute value of the leading eigenvalue $|\lambda_1|<1$, then system is in the paramagnetic state.}
\item{If $\lambda_1$ is complex with $|\lambda_1|>1$, system is in the spin-glass state. }
\item{If $\lambda_1$ is real and greater than $1$, the system is in the retrieval state. The associated real-eigenvectors can be used to detect the clustering structure.}
\end{itemize}

Using technique of ~\cite{Krzakala2013} we can compute the edge of the bulk as follows.
For any matrix $B$, its eigenvalues satisfy the following inequality
\begin{equation}\label{eq:inequal}
\sum_{i=1}^{2m}\left|{\lambda_i}\right|^{2r}\leqslant Tr B^r(B^T)^r
\end{equation}
where $\{\lambda_i\}$ are the $2m$ eigenvalues of $B$. Note  that $[B^r(B^T)^r]_{i\rightarrow j, i\rightarrow j}$ goes from node $i$ through a $r$-length path to node $j$ and then come back to $i$ through the same path
\begin{equation}
[B^r(B^T)^r]_{i\rightarrow j, i\rightarrow j}=\sum_{\mathcal{P}} \prod_{(x\rightarrow y) \in \mathcal{P}} \eta_{xy}^2
\end{equation}
where $\mathcal{P}$ denotes one of these paths, and  $(x\rightarrow y) \in \mathcal{P}$ means all the edges in the path $\mathcal{P}$. Under the assumptions of sparsity and that correlations between edges are week, we can write the right hand side of (14) as
\begin{equation}
\sum_{\mathcal{P}} \prod_{(x\rightarrow y) \in \mathcal{P}} \eta_{xy}^2 = \sum_{\mathcal{P}} \left\langle \eta_{xy}^2 \right\rangle^l_{\omega_{ij}}.
\end{equation}
Because there are $c^r$ such paths, take  expectation of Eq.~\eqref{eq:inequal} results to   
\begin{equation}
E \left|{\lambda_i}\right|^{2r}\leqslant c^r \left\langle \eta_{ij}^2 \right\rangle^l_{\omega_{ij}}.
\end{equation}
Therefore, the edge of bulk of eigenvalues of $B$ is 
\begin{equation} \label{eq:edge}
\kappa=\sqrt{\hat c  \langle\eta_{ij}^2 \rangle_{\omega_{ij}}}.
\end{equation}
Since the eigenvalues of the non-backtracking matrix is closely related to the stability of paramagnetic phase of belief propagation, it characterizes the phases of the system. 
In more detail, when the leading eigavalue is real, the point that it begins to be larger than $1$ indicates the transition from paramagnetic to retrieval phase. If the leading eigenvalue is complex, then it means there is no stable BP fixed point, thus the point where its absolute value becoming larger than $1$ indicates the paramagnetic to spin glass transition.
We can see that the transition point where edge of bulk equals to $1$ is identical to the condition of the spin-glass transition Eq.~\eqref{eq:beta_star}. This relation connects the edge of bulk in the spectrum of the non-backtracking operator to the spin glass transition of the Potts system, and gives another explanation on why it is sufficient to check only at $\beta^*$ for looking for the retrieval state: if system has a retrieval phase, it is easily identifiable at $\beta^*$ as the real-eigenvalue representing it has to be larger than $1$.

\subsection{\label{sec:choose} Determine the number of groups}
Choosing the number of groups is a classic model selection problem. A common approach is setting $q$ by optimizing the objective function. However, it is known to be prone to overfitting. For example, the objective function is always optimizable even in a random graph, and is an increasing function of $q$, whereas the correct group number is $q=1$ in the random graphs. Other approaches use penalty term such as entropy or description length~\cite{Peixoto2014} to regularize the model. However entropy penalizations, which actually choose $q$ by minimizing the free energy, usually requires that the related generative model is very close to the true model that generated the data~\cite{Decelle2011}. Penalty terms adopting the minimum description length is good at controlling the complexity of the model, but usually over constraint the model and predicts a smaller $q$~\cite{Peixoto2014}.

Here we propose to choose the correct number of groups, $q*$, at the value where the retrieval weights $Q(\{\hat t\})$ achieves its maximum. For $q\leq q*$, retrieval weights increase with $q$, while for $q>q^*$, either the retrieval phase disappears or $Q(\{\hat t\})$ stays the same indicating that increasing number of groups is no more helpful to obtain a better objective function. 

\section{Applications\label{sec:app}}
In this section, we apply our methods to several different clustering problems: the mixture model in the sparse regime where exact statistical inference results exist; clustering in two-dimensional geometric clustering datasets where the similarity function is hard to define; and community detection in weighted and directed graphs.

\subsection{Mixture model in sparse regime}
\begin{figure}[h]
\subfigure[]{
\includegraphics[width=\tsize\textwidth]{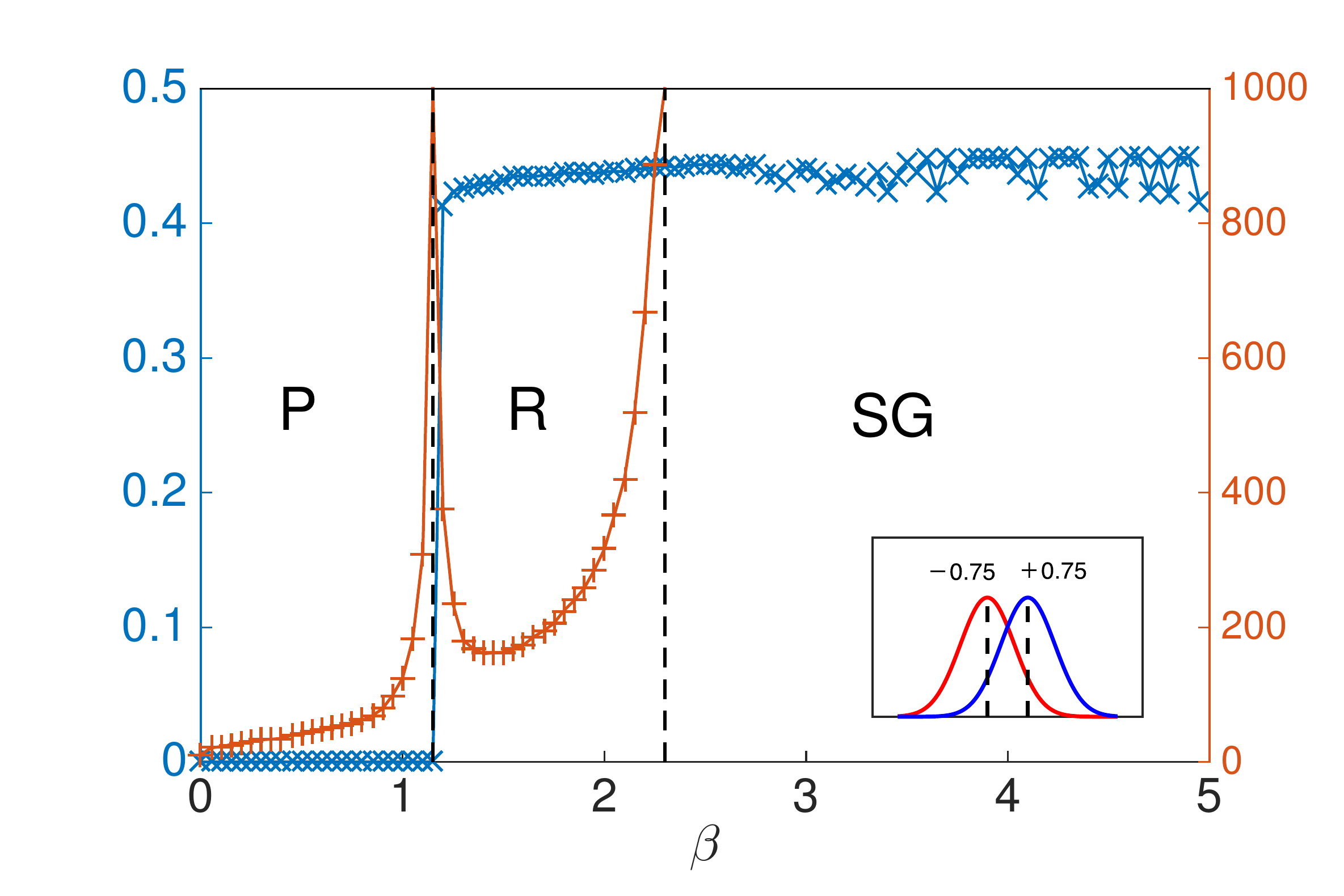}
}
\subfigure[]{
\includegraphics[width=\tsize\textwidth]{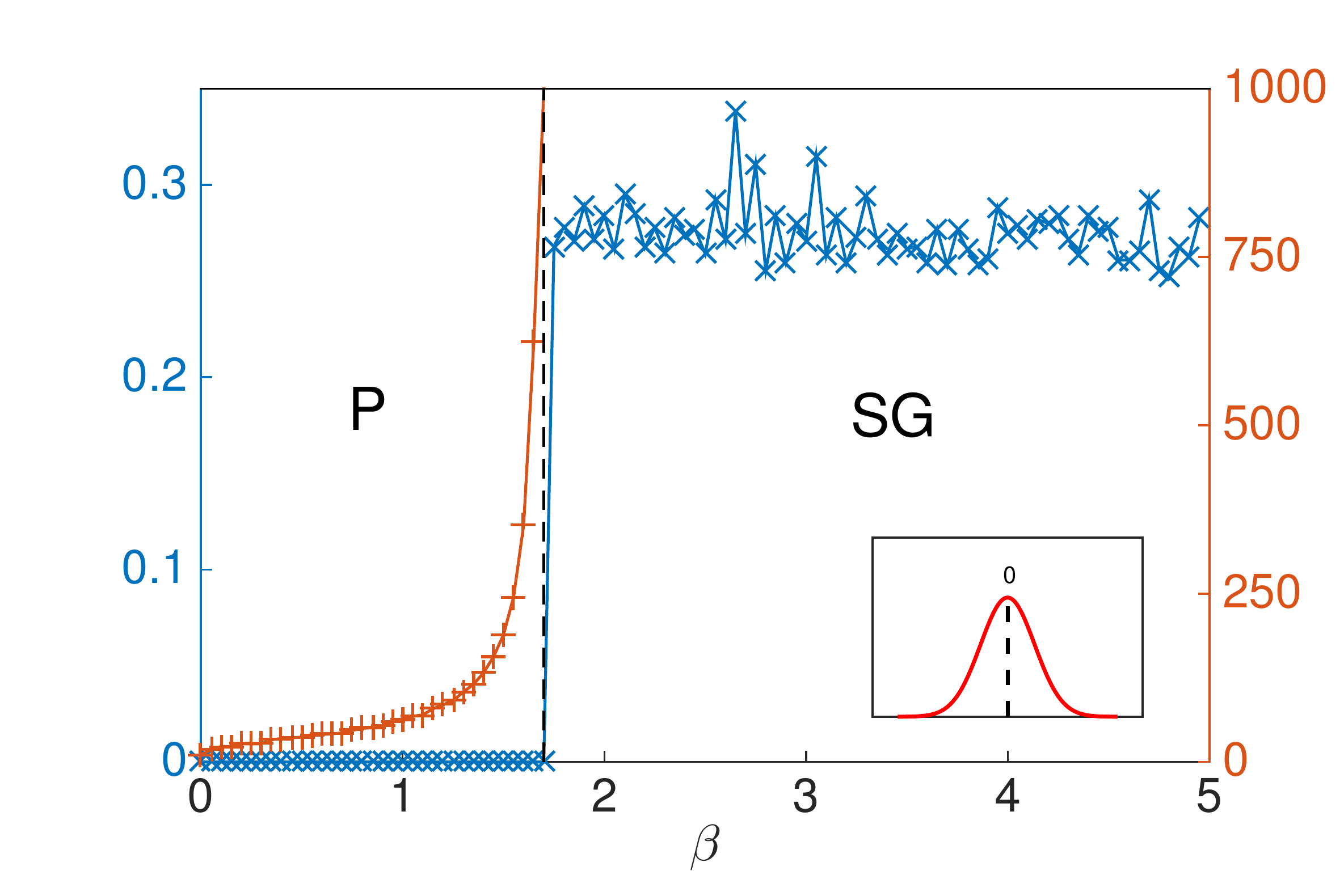}
}
\caption{Retrieval weights (blue $\times$, left y axis) and BP convergence time (red +, right y axis) on the datasets generated by Gaussian mixture model~\cite{saade2016clustering} in the undetectable regime (a) and in the detectable regime detectable regime (b). 
Both datasets have $n = 10000$ items, average number of measurements per item $c = 4$. The Gaussian weight distributions $p_{\textrm{in}}$ and $p_{\textrm{out}}$ have unit variance and zero mean for (a) (two Gaussian distribution coincide, as illustrated in the inset), and mean $0.75, -0.75$ for (b), as illustrated in the inset. In panel (a), the dashed line denotes the theoretical paramagnetic to spin glass transition happens at $\beta^*$ Eq.~\eqref{eq:tran}. In the panel (b), two dashed lines denote two experimental transitions : paramagnetic to retrieval transition at $\beta_R\approx 1.15$ and retrieval to spin  glass transition at $\beta_{SG}\approx 2.3$.
\label{fig:phases}
}
\end{figure}

We use the model proposed in ~\cite{saade2016clustering}, which can be seen as a generalization of the labeled stochastic block model~\cite{heimlicher2012community}. In the model, data is constructed with $n$ items in $q$ preassigned clusters of equal size. Each item has a group label $t_i^*$. Thus $\{t_i^*\}$ is the ground truth or planted partition, the partition that we want to recover. Pairwise measurements are chosen uniformly at random, on a Erd\"os-R\'enyi random graph with average degree $c$. The pairwise similarity $S_{ij}$, between item $i$ and $j$, is a random variable sampled from a probabilistic distribution $\mathcal P_{ij}(\omega_{ij})$ depending on $t_i^*$ and $t_j^*$. 
Following \cite{saade2016clustering}, we use the simplest setting  that 
\begin{align}
\mathcal P_{ij}(\omega_{ij})=
\begin{cases}
\mathcal P_{\textrm{in}}(\omega_{ij}),&\textrm{   if } t^*_i=t_j^*,\\
\mathcal P_{\textrm{out}}(\omega_{ij}),&\textrm{ otherwise.}\\
\end{cases}
\end{align}
And two probability distributions are chosen as Gaussian with different means, as plotted in insets of Fig.~\ref{fig:phases}.

It has been conjectured in \cite{saade2016clustering} that there is a detectability threshold on average number of measurements $c^*$ per item, given by
\begin{equation}\label{eq:tran}
c^*=q\left( \int d\omega \frac{(P_{\textrm{in}}(\omega)-P_{\textrm{out}}(\omega))^2}{P_{\textrm{in}}(\omega)+(q-1)P_{\textrm{out}}(\omega)}  \right)^{-1}.
\end{equation}
Below the threshold no algorithm can detect the planted partition with success better than a random guess. While with $c>c^*$, authors showed that the Bayes-optimal estimate approximated by belief propagation algorithm using correct parameters of the model achieves this threshold, and provide an asymptotically optimal accuracy.

On this model, we first examine whether our method can find retrieval state in both detectable and undetectable phases.
Figure~\ref{fig:phases} provides the retrieval weights $Q(\{t\})$ and convergence time of belief propagation as functions of $\beta$ on two instances (one instance in the undetectable phase while the other one in the detectable phase) generated by the model of \cite{saade2016clustering}.
In the top panel, the data is in the undetectable
phase, meaning no algorithm can find information of the planted partition better than a random guess. We can see from the figure that our algorithm finds that the phase diagram is composed of paramagnetic phase and non-convergence (NC) phase (which is actually the spin glass phase in this random sparse graph), separated by the transition $\beta^*\approx 1.7$, with a very good agreement with the predicted spin-glass transition Eq.~\eqref{eq:beta_star}. So our algorithm decides that there is no retrieval phase in the whole regime of $\beta$, hence reports that there is no significant clustering structure in the data.
In the bottom panel of Fig.~\ref{fig:phases} the data is in the detectable phase, then we see that our algorithm indeed finds the retrieval phase in between the paramagnetic phase and non-convergence phase, which gives information of the planted partition. 
Fig.~\ref{fig:phases} also shows that $\beta^*$ indeed locates at the retrieval phase, hence our algorithm running at $\beta^*$ will find the retrieval phase as well as correct information about the planted partition.
This confirms our theory in ~\ref{sec:significance}  that in practice we only need to run our algorithm \textit{once} at $\beta^*$ Eq.~\eqref{eq:beta_star} rather than checking full regime of $\beta$. 

To evaluate the performance of our algorithm in accuracy of detecting the planted partition, we define the overlap between the inferred partition $\{\hat t\}$ and the true partition $\{t^*\}$
\begin{equation}\label{eq:ovl}
\mathcal O(\hat t,t^*)=\max_{\pi}\frac{\frac{1}{n}\sum_{i=1}^n\delta(t_i^*,\pi(\hat{t}_i))-\frac{1}{q}}{1-1/q}
\end{equation}
as a measure of accuracy. Here $\pi(\hat t)$ is one permutation (among totally $q!$ permutations) of the partition $\hat t$. 
Clearly we have $0\leq\mathcal O(\hat t,t^*)\leq 1$, and $ \mathcal O(\hat t,t^*)=0$ means the inferred partition is not correlated with the planted one while $\mathcal O(\hat t,t^*)=1$ indicates that the planted partition is fully recovered. 

\begin{figure}[h]
\centering
\includegraphics[width=0.6\textwidth]{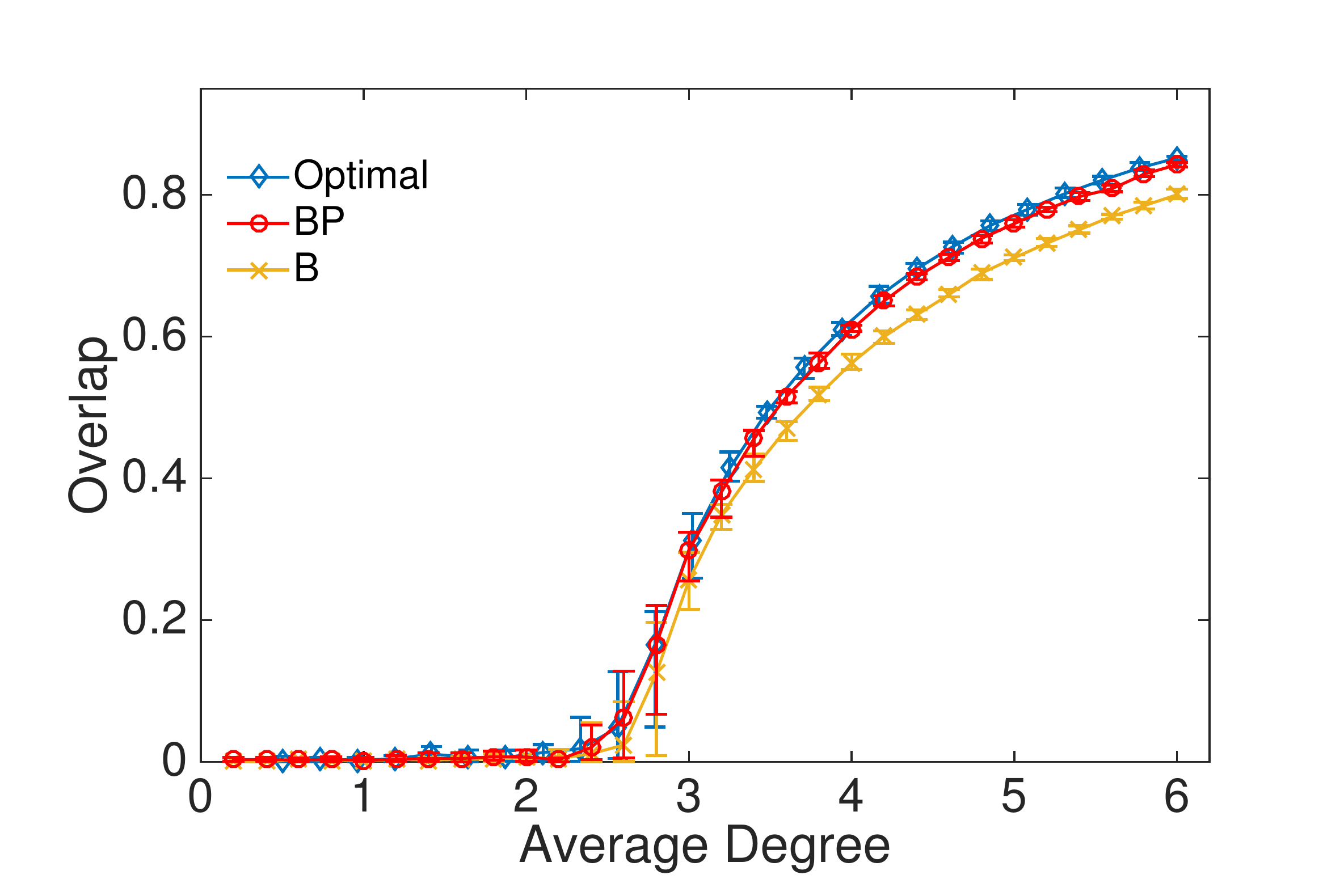}
\caption{Comparison of accuracy (evaluated using overlap Eq.~\eqref{fig:ovl}) of belief propagation (BP), spectral clustering using the non-backtracking operator (B) and Bayes-optimal results (Optimal) in~\cite{saade2016clustering}. Each point is averaged over $10$ realizations of size $n=100,000$, and Gaussian weight distributions $\omega_{in}$,$\omega_{out}$ with means $0.75$ and $-0.75$ respectively and unit variance. The theoretical transition~Eq.~\eqref{eq:tran} is at critical average degree $c^*\approx 2.63$. \label{fig:ovl}} 
\end{figure}

In Fig.~\ref{fig:ovl} we plot the overlap between the planted partition and those obtained using BP and from spectral clustering algorithm using leading eigenvector of the non-backtracking operator~Eq.\eqref{eq:B}, both at the spin-glass transition temperature $\beta^*$. The performances are compared to the optimal Bayesian inference results in~\cite{saade2016clustering}.
We can see from the figure that  BP and the non-backtracking operator both work all the way down to the detectability transition Eq.~\eqref{eq:tran} at $c^*\approx 2.63$. While the non-backtracking operator works worse in the regime away from the transition, BP works really close to the optimal accuracy, particularly in the regime close the phase transition. Remarkably, while the optimal Bayesian inference algorithm~\cite{saade2016clustering} requires the knowledge of correct parameters of the model, our methods, both BP and the non-backtracking operator, do not need to know the parameters.

\begin{figure}[h]
\centering
\subfigure{
\label{Fig.sub.1}
\includegraphics[width=\tsize\textwidth]{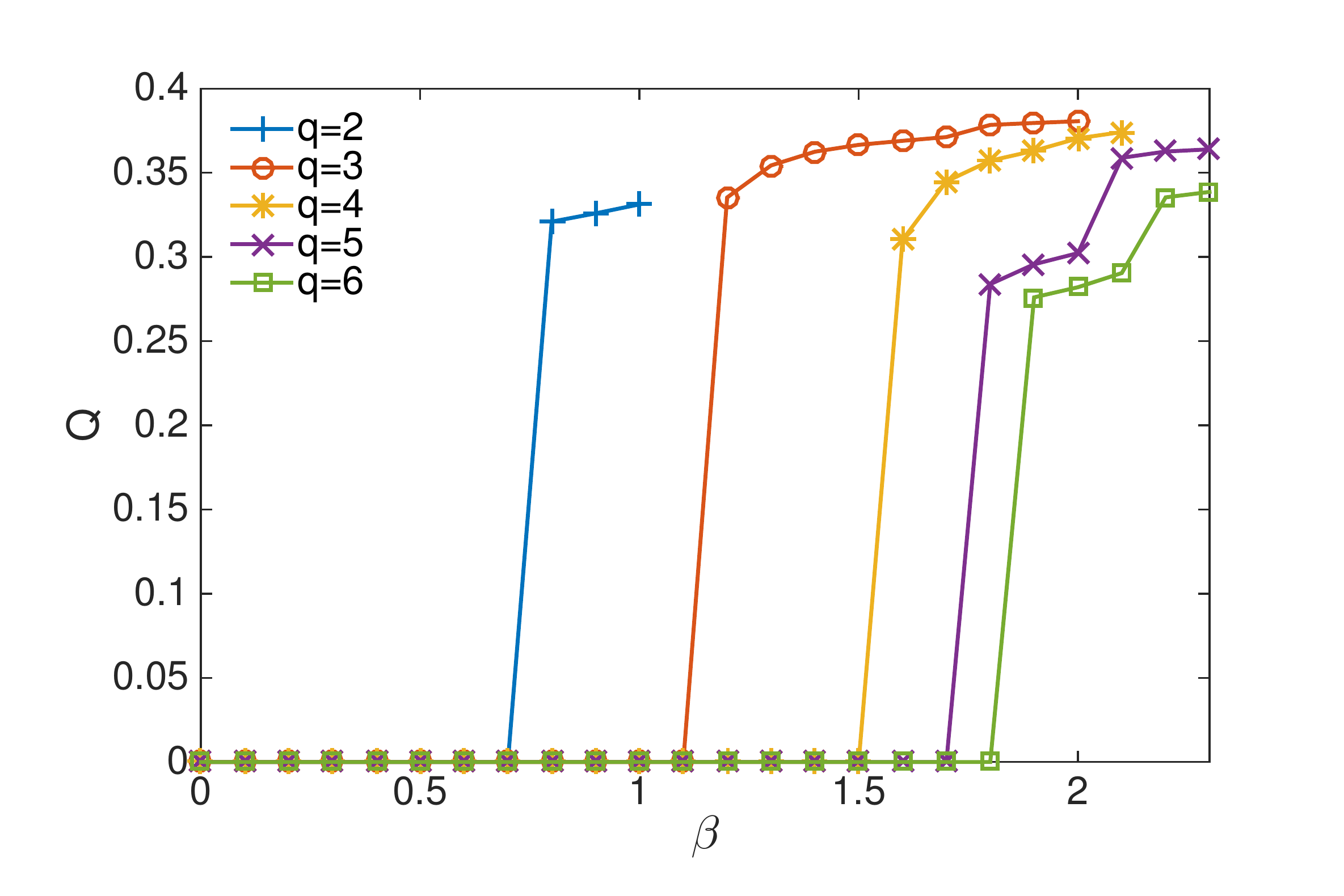}
}
\subfigure{
\label{Fig.sub.2}
\includegraphics[width=\tsize\textwidth]{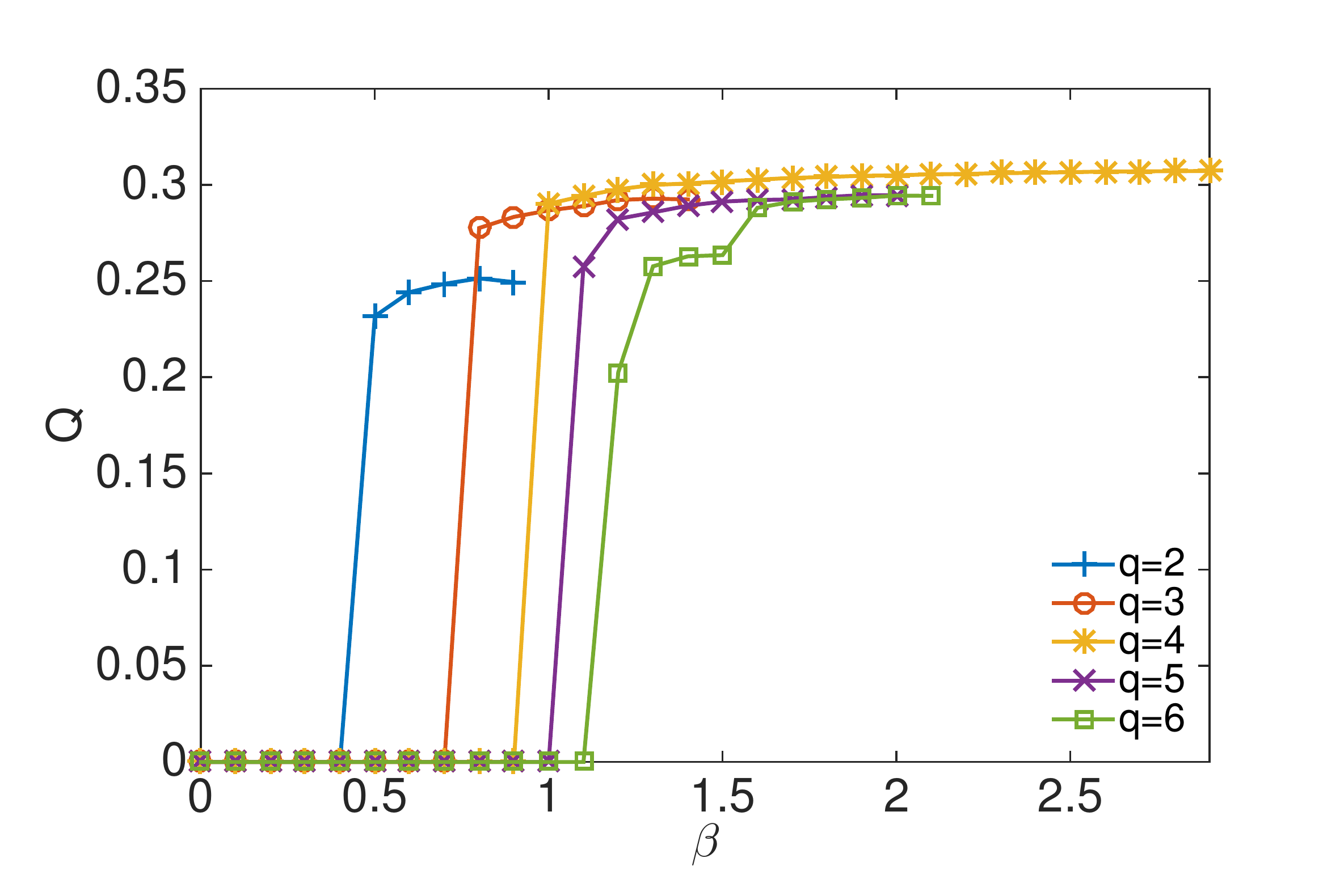}
}
\caption{Retrieval weights for different values of $q$ as a function of $\beta$ on two datasets generated by Gaussian mixtures~\cite{saade2016clustering}. Two datasets are both in the detectable regime with $n = 10000$ items, average measurements per item $c = 6$ and correct number of groups $q^*=3$ for the top panel and $c=12, q^* = 4$ for the bottom panel. \label{fig:choose_q}}
\end{figure}

In Fig.~\ref{fig:choose_q} we plot retrieval weights $Q$ for different values of group number $q$ as a function of $\beta$ for two datasets generated by the Gaussian mixture model with $3$ and $4$ groups respectively. As we can see that, in both networks, the retrieval weights at the correct number of groups $Q$ archives its maximum value, and the retrieval phase has the widest regime. 
Therefore on these two examples, our method of determining a number of groups (as in Sec.~\ref{sec:choose}) gives correct group number.

\begin{figure*}
\centering
\subfigure[]{
\label{Fig.sub.1}
\includegraphics[width=0.42\textwidth,height=0.28\textwidth]{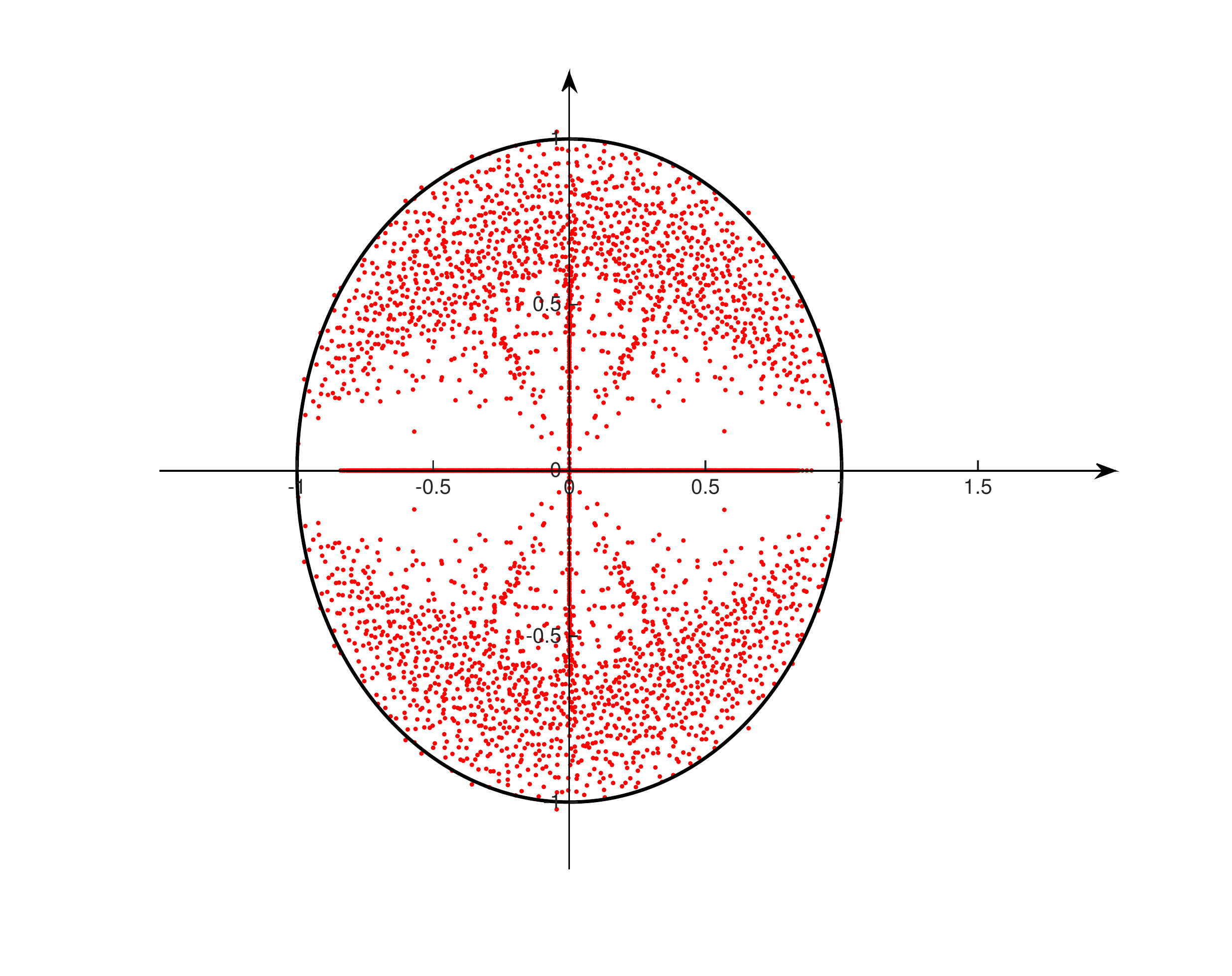}}
\subfigure[]{
\label{Fig.sub.2}
\includegraphics[width=0.42\textwidth,height=0.28\textwidth]{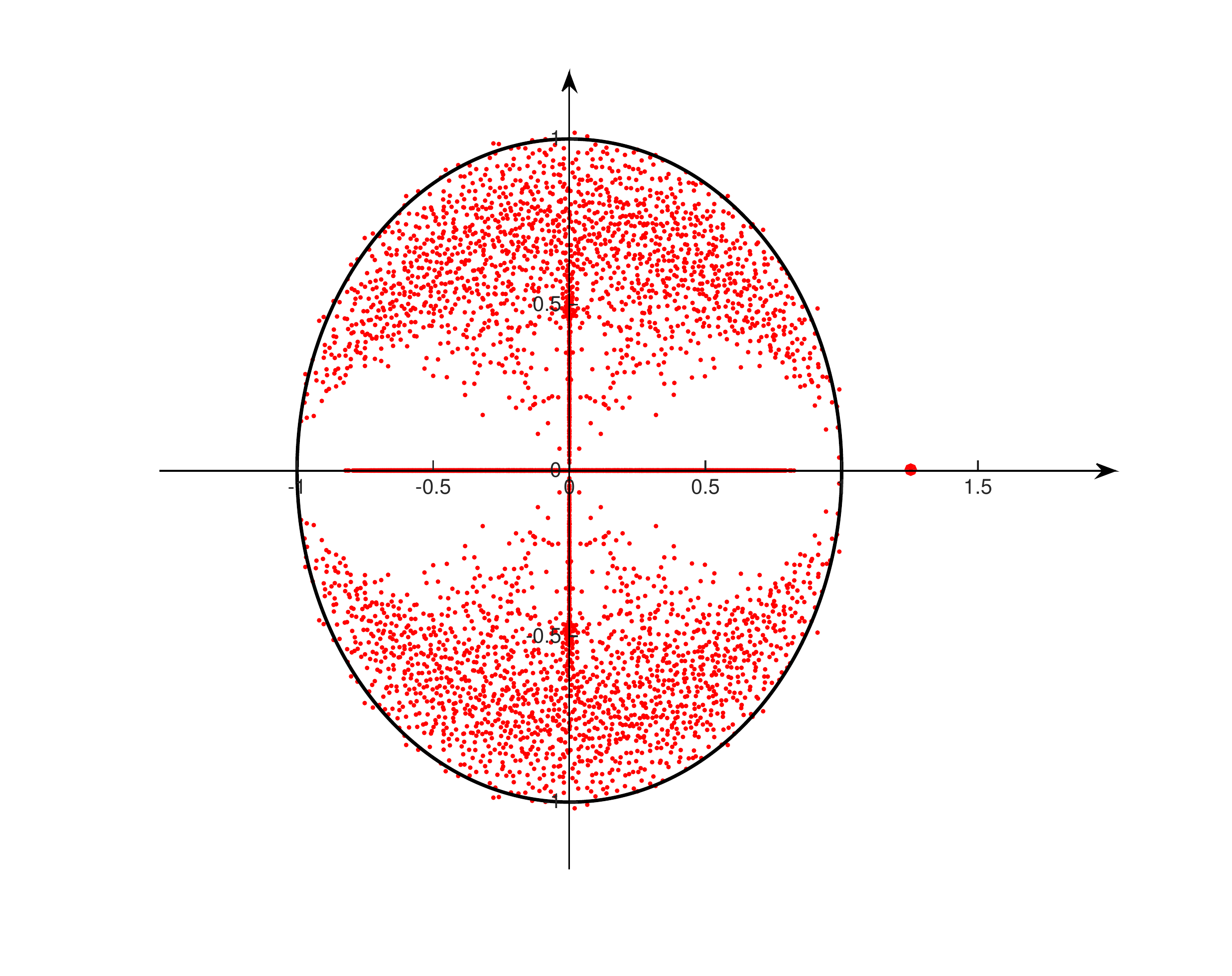}
}
\subfigure[]{
\label{Fig.sub.3}
\includegraphics[width=0.42\textwidth,height=0.28\textwidth]{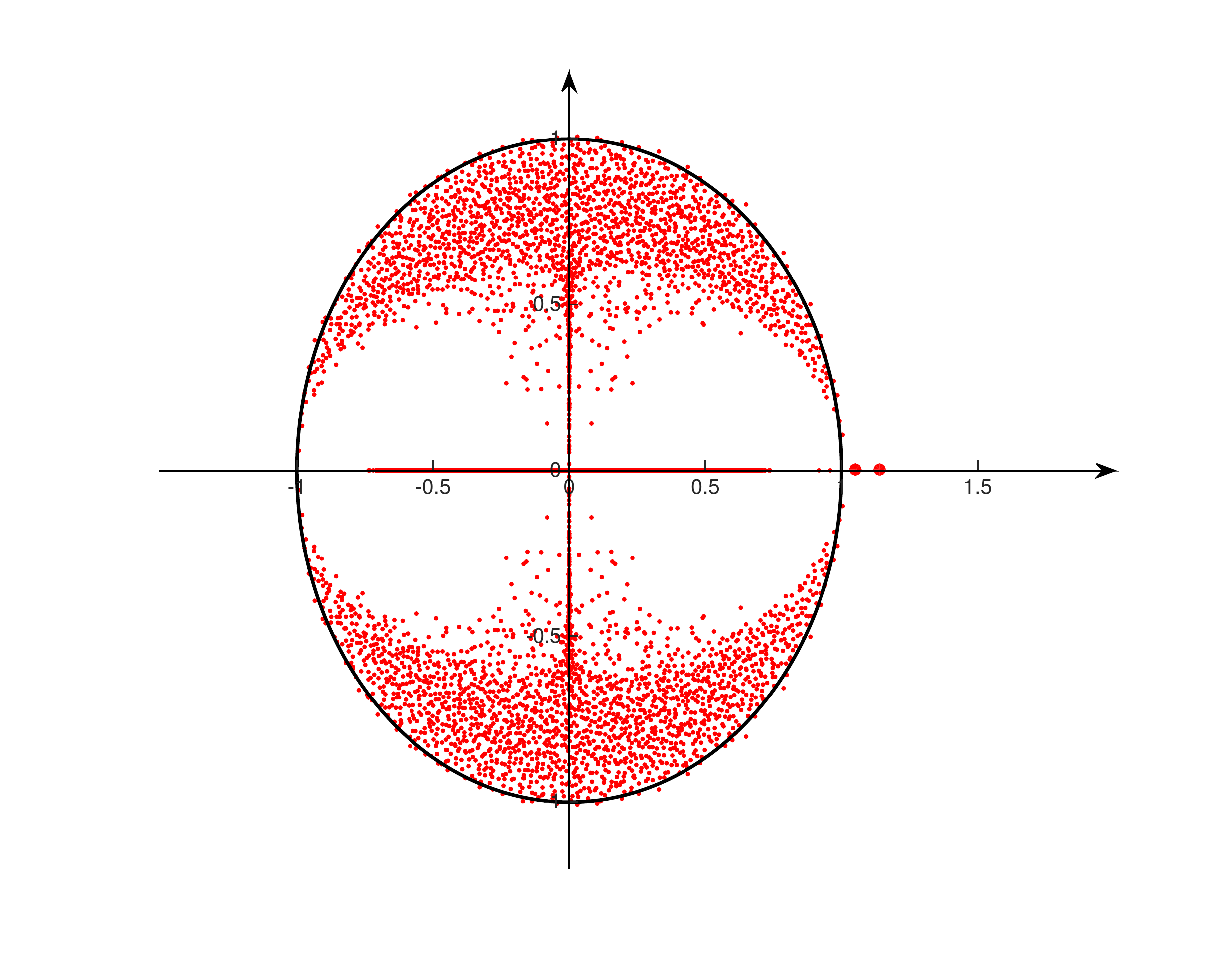}
}
\subfigure[]{
\label{Fig.sub.4}
\includegraphics[width=0.42\textwidth,height=0.28\textwidth]{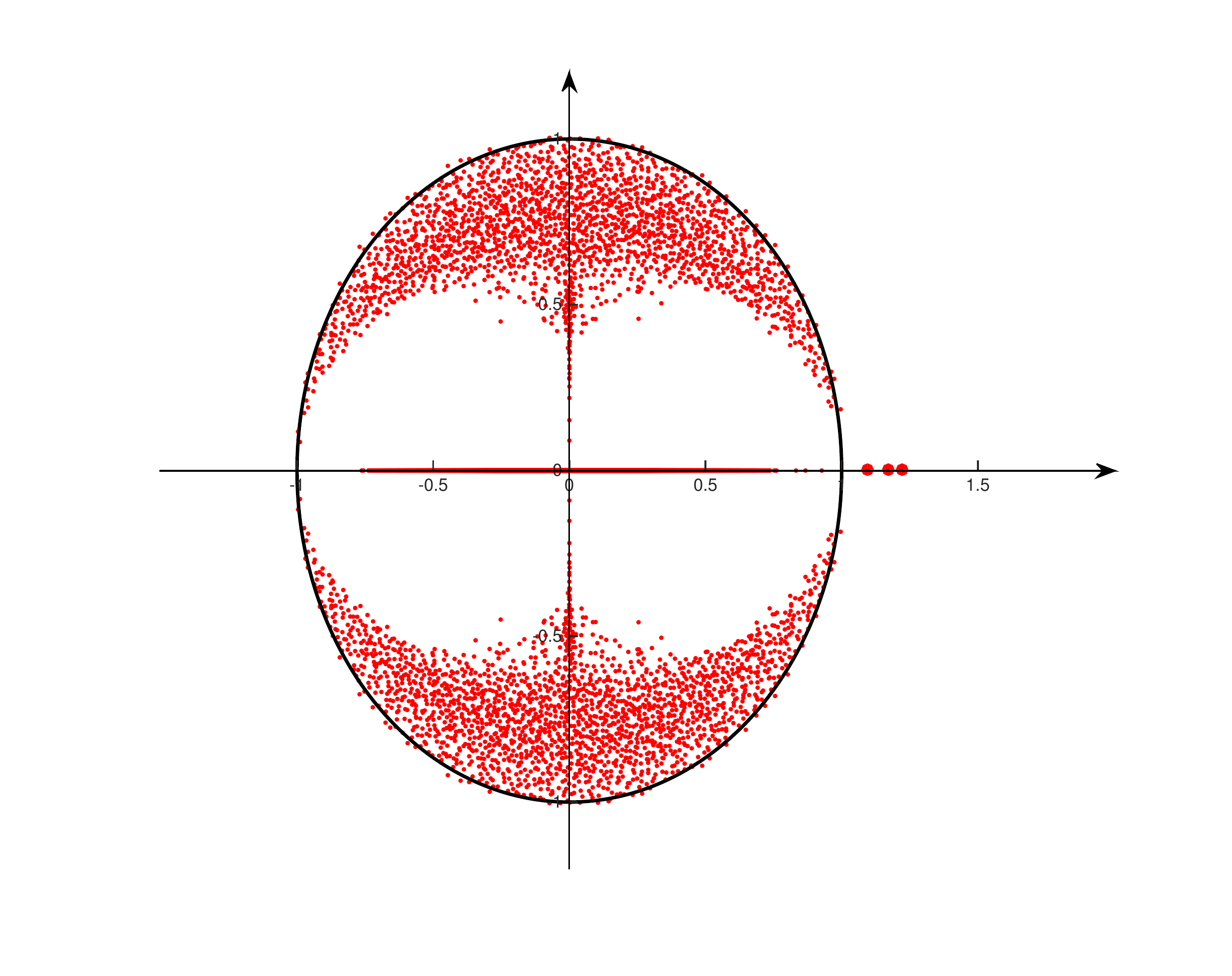}
}
\caption{Spectrum of non-backtracking matrix~Eq.\eqref{eq:B} (at the spin-glass transition $\beta^*$) on the complex plane for datasets generated by the Gaussian mixtures~\cite{saade2016clustering}. These datasets have $1000$ items. For (a),(b),(c) and, (d), the average measurements per item are $4$, $4$, $7$, the number of groups are $1$, $2$, $3$, and $4$ respectively. The black circles are theoretical predictions~Eq.~\eqref{eq:edge} of edge of the bulk.\label{fig:spectrum}}
\end{figure*}

In Fig.~\ref{fig:spectrum} we draw spectrum of the non-backtracking operator $B$ (Eq.~\eqref{eq:B}) at the spin-glass transition $\beta^*$ of four datasets generated using Gaussian mixtures with $q=1, 2, 3, 4$ groups respectively. Since matrix $B$ is asymmetric, the spectrum is defined on the complex plane. 
The black circle in each figure is the theoretical prediction of edge of bulk given by
Eq.~\eqref{eq:edge}, and we can see that it fits very well with the numerical results in all $4$ figures.
The figures also illustrate that the number of real eigenvalues outside the bulk gives the number of groups in the datasets. We have checked that this is nicely consistent with results given by maximizing the retrieval weights as proposed in Sec.~\ref{sec:choose}.

\begin{figure}[h]
    \centering
    \includegraphics[width=0.6\textwidth]{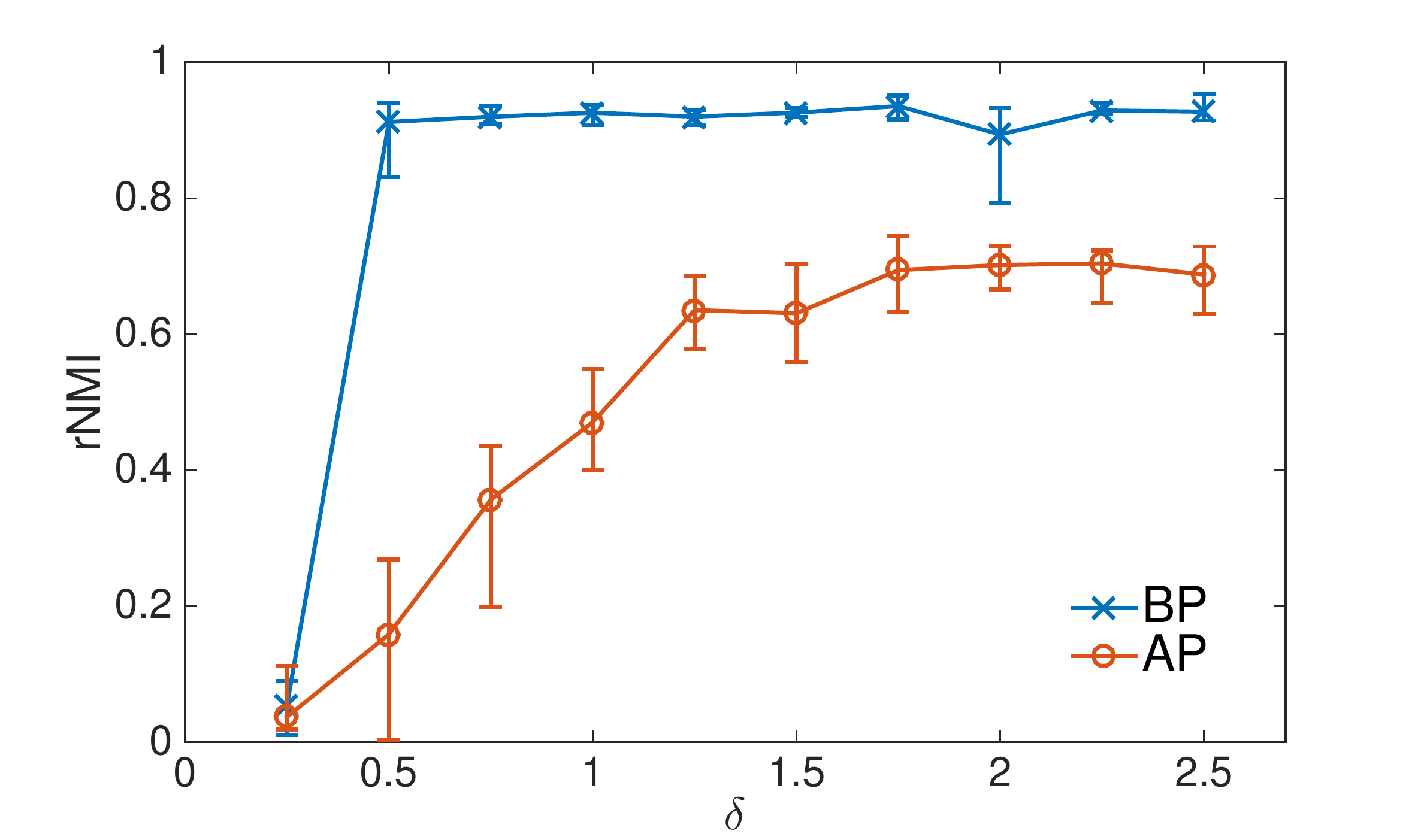}
    \caption{Performance comparison between our belief propagation algorithm (BP) and affinity propagation~\cite{Frey2007} (AP), on synthetic similarity graphs generated using Gaussian mixtures (see main text) which is full connected. 
    Parameters are $q=4$ groups, In-group and out-group distribuions are Gaussian distribution with unity variance and mean $\rho_{in}=5-\delta$ and $\rho_{in}=5-\delta$ respectively, where $\delta$ controls the hardness of the clustering problem and varyed as the X-axis. The accuracy of clustering in Y-axis are evalutated using the relative normalized mutual information between the obtained partition and the ground-truth~\cite{rnmi}.
    Each point in the figure is averaged over $10$ instances and the implementation of affinity propagation was downloaded from the official website.
    \label{fig:ap}
    }
\end{figure}

For similarity problems there is a very popular and classic method, name Affinity Propagation (AP)~\cite{Frey2007} which is also based on message passing, so we would like to compare the performance of our algorithm against affinity propagation. Since affinity propagation is designed for fully-connected similarity graphs, to make a fair comparison we use the Gaussian mixtures with full similarity matrix. 
The results are shown in Fig.~\ref{fig:ap} where we can see that even on fully-connected similarity graphs, our algorithm significantly outperforms the affinity propagation. We note that over the past several years there have been many variants of the AP algorithm, while what we have compared is the original version.

\subsection{Community detection in weighted networks and directed networks.}
As we described in the Sec.~\ref{sec:intro}, the data clustering problem in the sparse regime is closely related to the community detection in weighted networks, when we are interested in the assortative structures where weights on edges can be understood as similarities between nodes. 
In the field of community detection, many algorithms have been proposed, and there exist benchmark networks having ground-true communities for evaluating the performance of algorithms. In this work, we consider the widely used LFR benchmark networks~\cite{lancichinetti2008benchmark}.

The LFR model generates networks having a power-law degree distribution with exponent $\gamma$. 
A topological mixing parameter $\mu_{t}$ is introduced to tune the ratio between internal degree (number of edges connecting to the nodes in the same group) $k_i^{\text{in}}$ and total degree $k_i$ of a node $i$
$k_{i}^{in}=(1-\mu_t)k_i$.
The weights of node $i$ is assigned by
$s_i=\sum_j\omega_{ij}=k_i^\alpha$,
with $\alpha$ being a parameter controlling the strength of weights.
Then a third parameter $\mu_{\omega}$ is used to assign the internal strength $s_{i}^{in}=(1-\mu_{\omega})s_{i}$. 
In our numerical experiments, we fix $\gamma=2$ , $\alpha=1$, $\mu_t=\mu_\omega$ and vary $\mu_{t}$, number of nodes $n$ and average degree $c$, to show the performances of our algorithm in different situations and compare them with existing algorithms. With other $\gamma$, $\alpha$, and $\mu_t$ values we see a qualitatively similar behavior of performance comparison.

We conduct experiments on weighted networks generated by the LFR model with $n=10^4$ nodes, average degree $c=10$ and $q^*=2$ groups, and compare the performance of our algorithm with three other algorithms, Infomap~\cite{Rosvall2008}, Oslom~\cite{lancichinetti2011finding}, and Louvain~\cite{Blondel2008}. 
In all experiments, the group number $q^*=2$ is not provided to $4$ algorithms, thus every algorithm determines the number of groups by itself.
Since usually different algorithms eventually find different $q$ values, which are quite different from $q^*$, it is not possible to use overlap (Eq.~\eqref{eq:ovl}) to characterize the accuracy of community detection. So we adopt the relative normalized mutual information (rNMI) \cite{rnmi} between the ground-true partition and the detected partition to characterize the accuracy.
The reason that we use rNMI rather than the normalized mutual information (NMI) \cite{danon2005comparing} is that: Infomap usually gives a large number of groups, and NMI has a systematic bias to a large group number~\cite{rnmi}. The rNMI is actually the NMI with the bias extracted, hence is more suitable for measurement for this task.

\begin{figure}[h]
\centering
\subfigure[]{
\includegraphics[width=\tsize\textwidth]{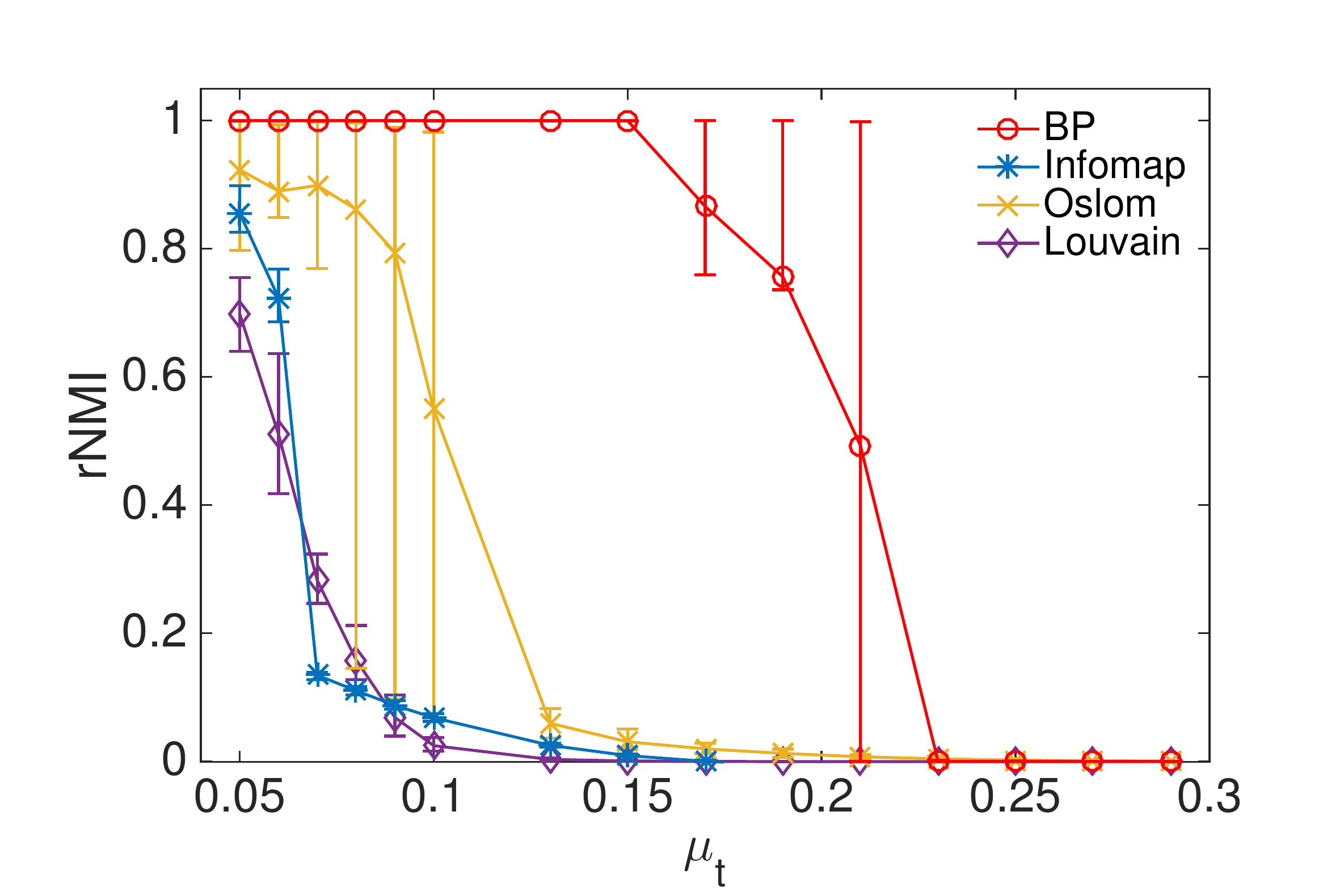}
}
\subfigure[]{
\includegraphics[width=\tsize\textwidth]{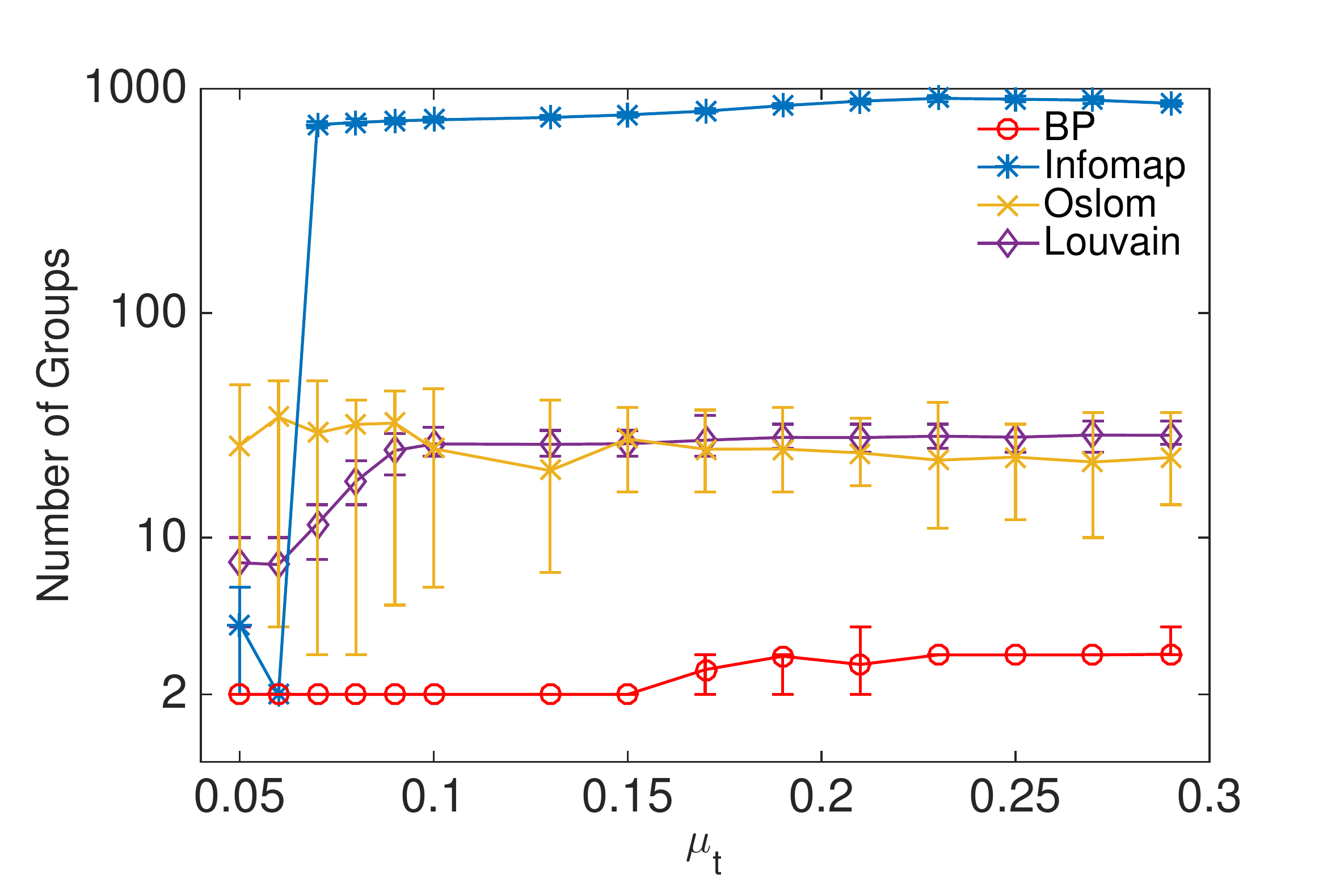}
}
\caption{Performance comparison of different algorithms in community detection in weighted networks. Networks are generated by LFR benchmarks (see text for details) with $n=10^4$ nodes, maximum degree $30$, average degree $c=10$, true number of groups $q^*=2$ and  $\beta=1$. Panel (a) shows the relative normalized mutual information (rNMI~\cite{rnmi}, the larger the better) between the true community and that given by algorithms. Panel (b) illustrates the numbers of groups detected by four algorithms. Every point in the figures is averaged $50$ instances.\label{fig:lfr}}
\end{figure}

\begin{figure}[h]
\centering
\subfigure[]{
\includegraphics[width=\tsize\textwidth]{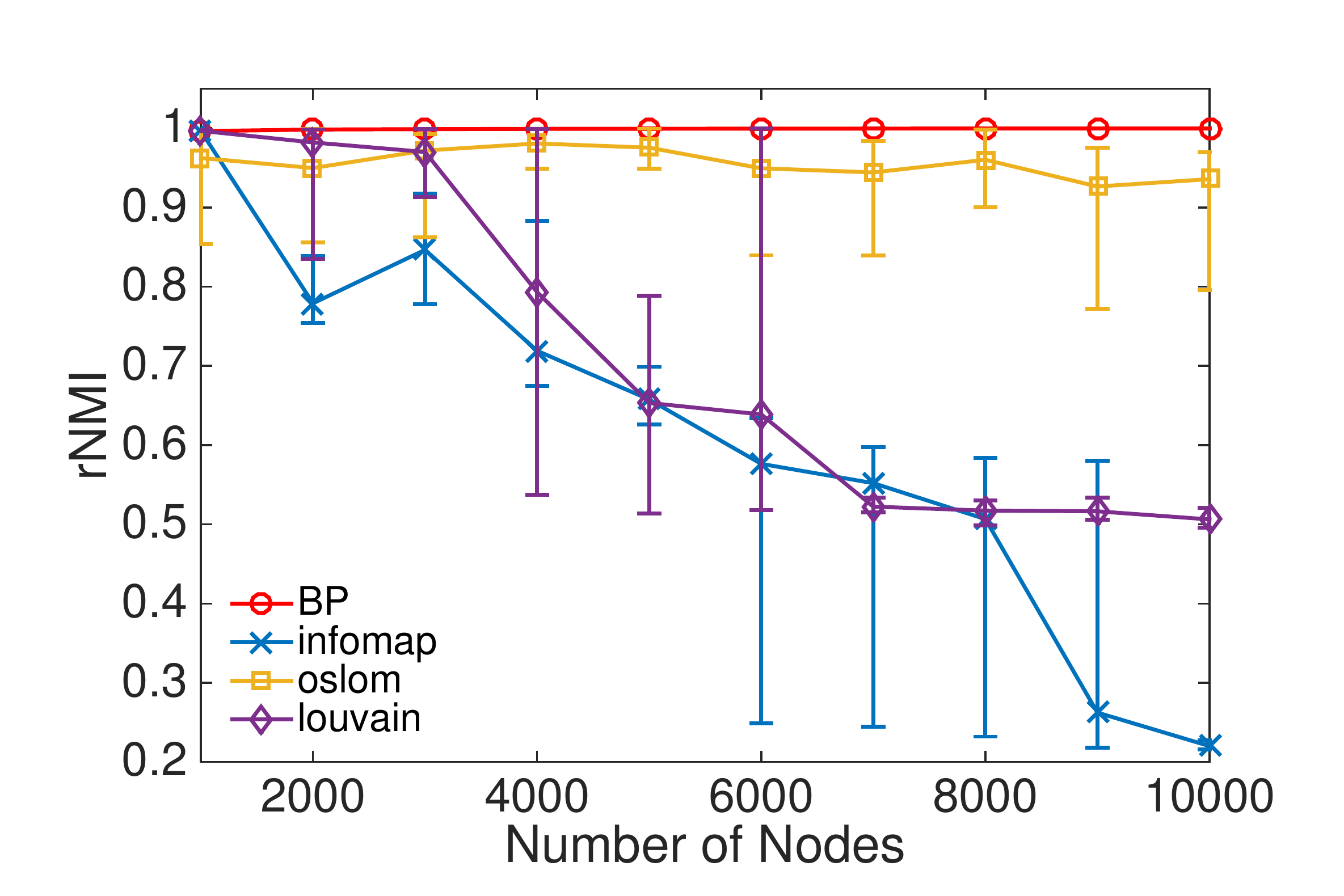}
}
\subfigure[]{
\includegraphics[width=\tsize\textwidth]{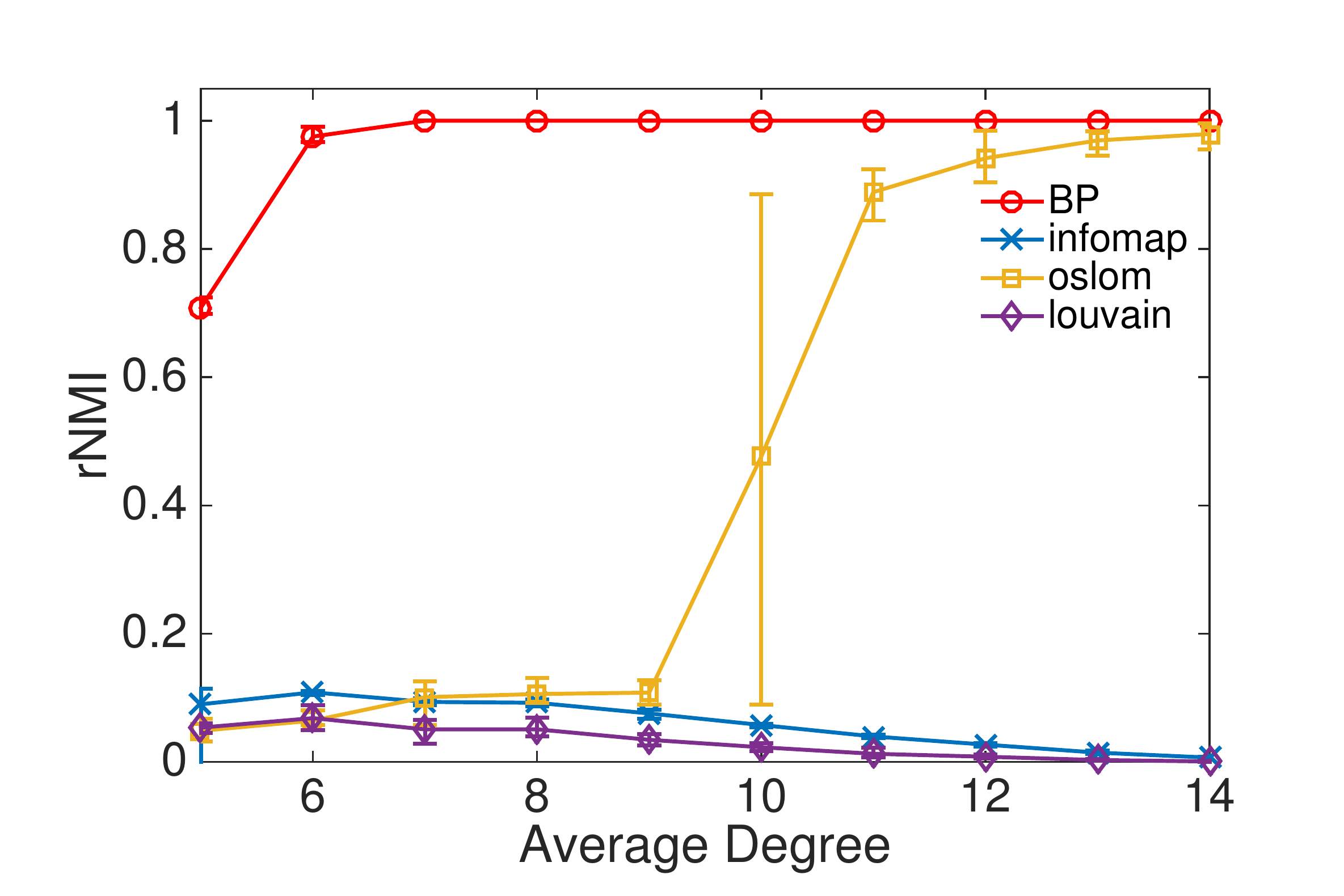}
}
\caption{
Performance comparison of different algorithms on community detection in weighted networks generated by the LFR model, with varying size (a) and varying sparsity (b).
In panel (a), the true number of groups is $q^*=4$, average degree is fixed to $c=10$, $\mu_t=\mu_w=0.1$, number of nodes is varying. In panel (b), number of nodes is $n=10^4$, true number of groups is $q^*=2$, $\mu_t=\mu_w=0.1$, and average degree of networks are varying. Every point in the figures is averaged $50$ instances.\label{fig:lfr.2}}
\end{figure}

The results are shown in Fig.~\ref{fig:lfr}. 
From the panel (a) we can see that
our algorithm always gives a larger rNMI value than Infomap, Oslom and Louvain. In the parameter regime with $\mu_t>0.15$, Infomap, Oslom and Louvain do not work at all, finding partitions that have almost $0$ rNMI, while our algorithm gives a large rNMI until $\mu_t>0.21$.
In panel (b) of Fig.~\ref{fig:lfr} we plot the number of groups that are determined by different algorithms on networks with $q^*=2$ groups and varying $\mu_t$. We can see that the number of groups found by our algorithm is very close to the ground-true value $q^*=2$, while other algorithms report significantly much larger values. Particularly, the Infomap algorithm gives a huge number of groups which means it actually tends to break the large network into many small parts.
Figure~\ref{fig:lfr.2} again gives a different view of the results, with the left panel comparing the performance of algorithms on LFR networks with a varying number of nodes $n$ and right panel comparing that with varying average degree $c$. From the panel (a) we can see that our algorithm is not sensitive to system size, giving accuracy close to $1$ in all cases, while the accuracy of Infomap and Louvain fall quickly when system size is increased (while average degree is fixed). The panel (b) of Fig.~\ref{fig:lfr.2} shows that 
Infomap and Louvain do not work in the whole regime, and Oslom requires average degree larger than $10$, while our algorithm performs well in all regime, giving almost perfect detection when the average degree is greater than $6$.

\begin{figure}[h]
    \centering
    \includegraphics[width=0.45\textwidth]{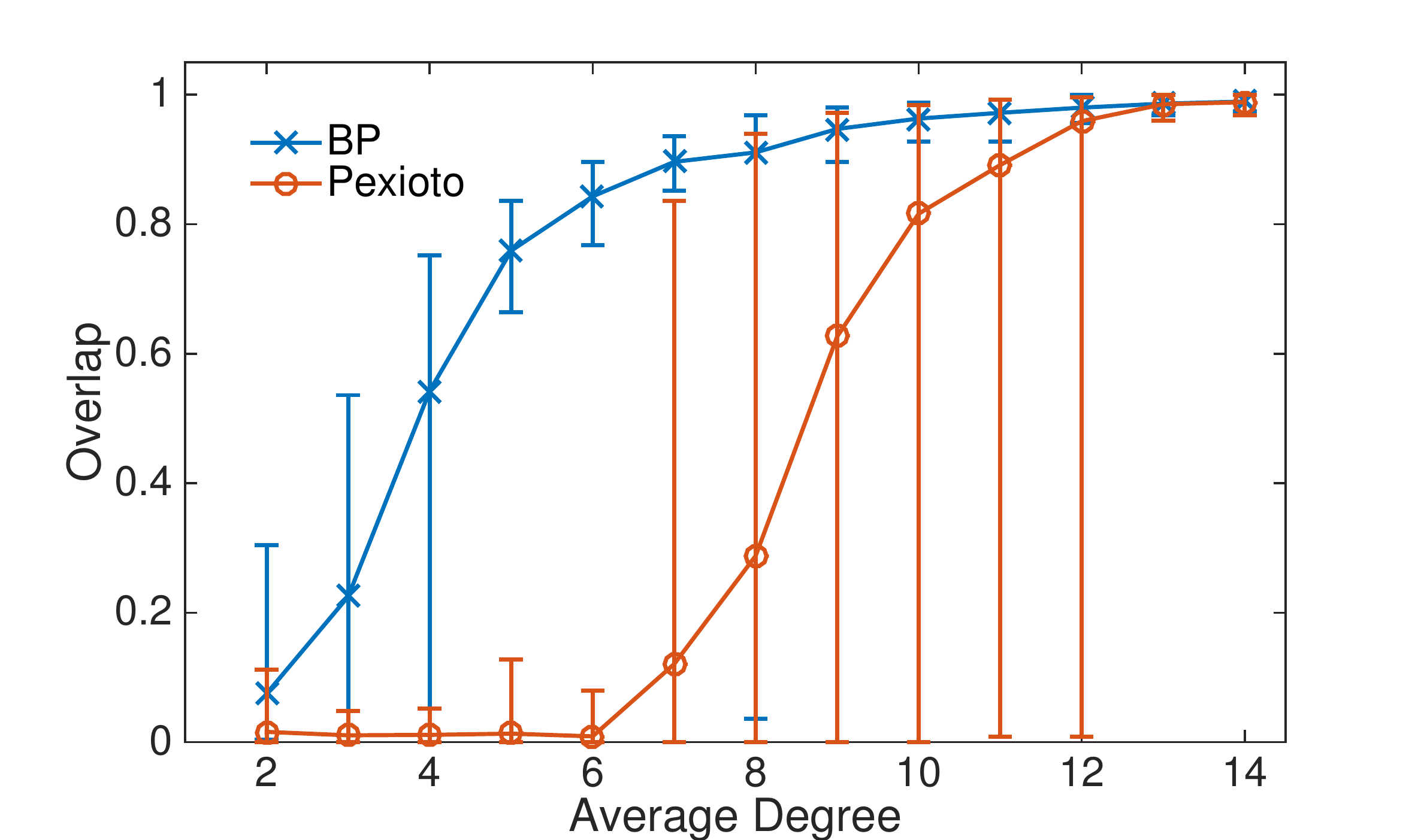}
    \includegraphics[width=0.45\textwidth]{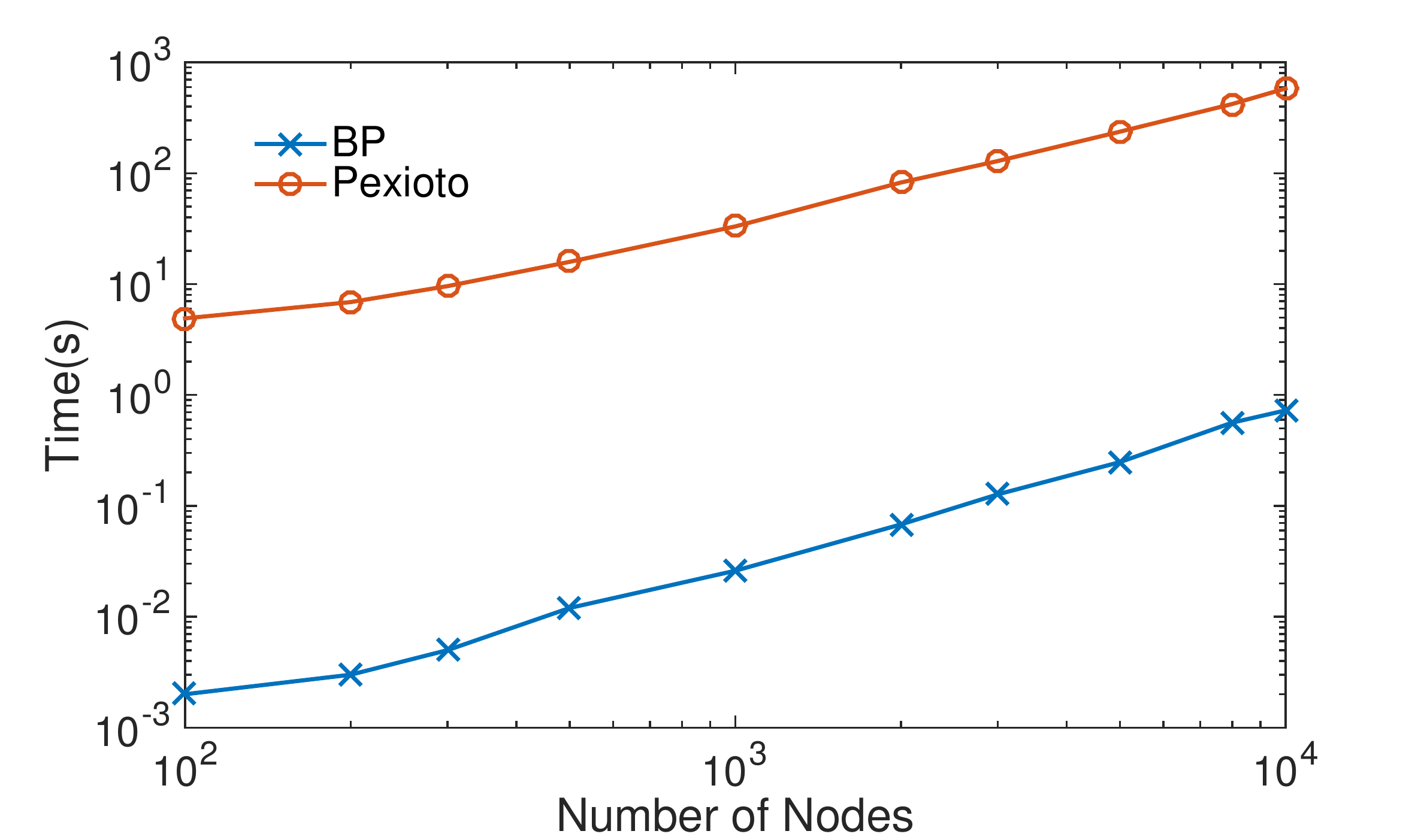}
    \caption{Performance comparison between our belief propagation algorithm (BP) and hierarchical Bayesian method of Peixoto~\cite{peixoto2017nonparametric} on sparse Gaussian mixtures, In-group and out-group Gaussian distribuions have unity variance and mean $\rho_{in}=5.75$ and $\rho_{out}=4.25$ respectively. In the left panel we plot the accuracy of clustering using overlap by restricting two methods to use $q=2$ groups. The graph has $n=500$ nodes, and varying average degree. 
    In the right panel, the computational time for two algorithms are plotted for networks with average degree fixed to $c=15$ and varying system size.
    Each point in the figure is averaged over $10$ instances, the implementation of ~\cite{peixoto2017nonparametric} was downloaded from Tiago Peixoto's website.
    \label{fig:tiago}
    }
\end{figure}

We also compared our algorithm with the recently developed method of Peixoto~\cite{peixoto2017nonparametric}. 
Peixoto's method is based on series of his work~\cite{PhysRevE.89.012804,peixoto2013parsimonious,Peixoto2014} of hierarhical non-parametric Baysian methods in community detection, which become quite popular in recent years. 
We have performed extensive comparison between our algorithm and method in ~\cite{peixoto2017nonparametric}, the typical result is that on dense networks two algorithm perform similarly, while in sparse networks our algorithm performs considerably better. For an example the comparison results on sparse Gaussian mixtures are shown in Fig.~\ref{fig:tiago}, where we can see from the left panel that with average degree low, our algorithm works much better. We notice that in the figure one can observe that Peixoto's method has a large variance. We think this is a quite general phenomenon when an algorithm, which does not target particularly the sparsity, is applied to sparse graphs. Actually notice that in Fig.~\ref{fig:direct} where the Leight-Newman algorithm is applied to sparse graphs, and in Fig.~\ref{fig:bp:tap} where the Thouless-Anderson-Palmer equation (which are essentially a dense approximation to our BP method) is applied to sparse graphs, the similar phenomenons are observed as well. We have also seen the similar phenomenon in clustering sparse graphs using spectral algorithm based on the adjacency matrix, see for example in Fig.4 of reference~\cite{Krzakala2013}.

Moreover, the right panel of Fig.~\ref{fig:tiago} illustrates that our method is much (almost $1000$ times) faster than method in~\cite{peixoto2017nonparametric}, especially in large networks, due to the difference in efficiency between our message passing technique and the MCMC method used in ~\cite{peixoto2017nonparametric}.

\begin{figure}[h]
\subfigure[]{
\includegraphics[width=0.38\textwidth]{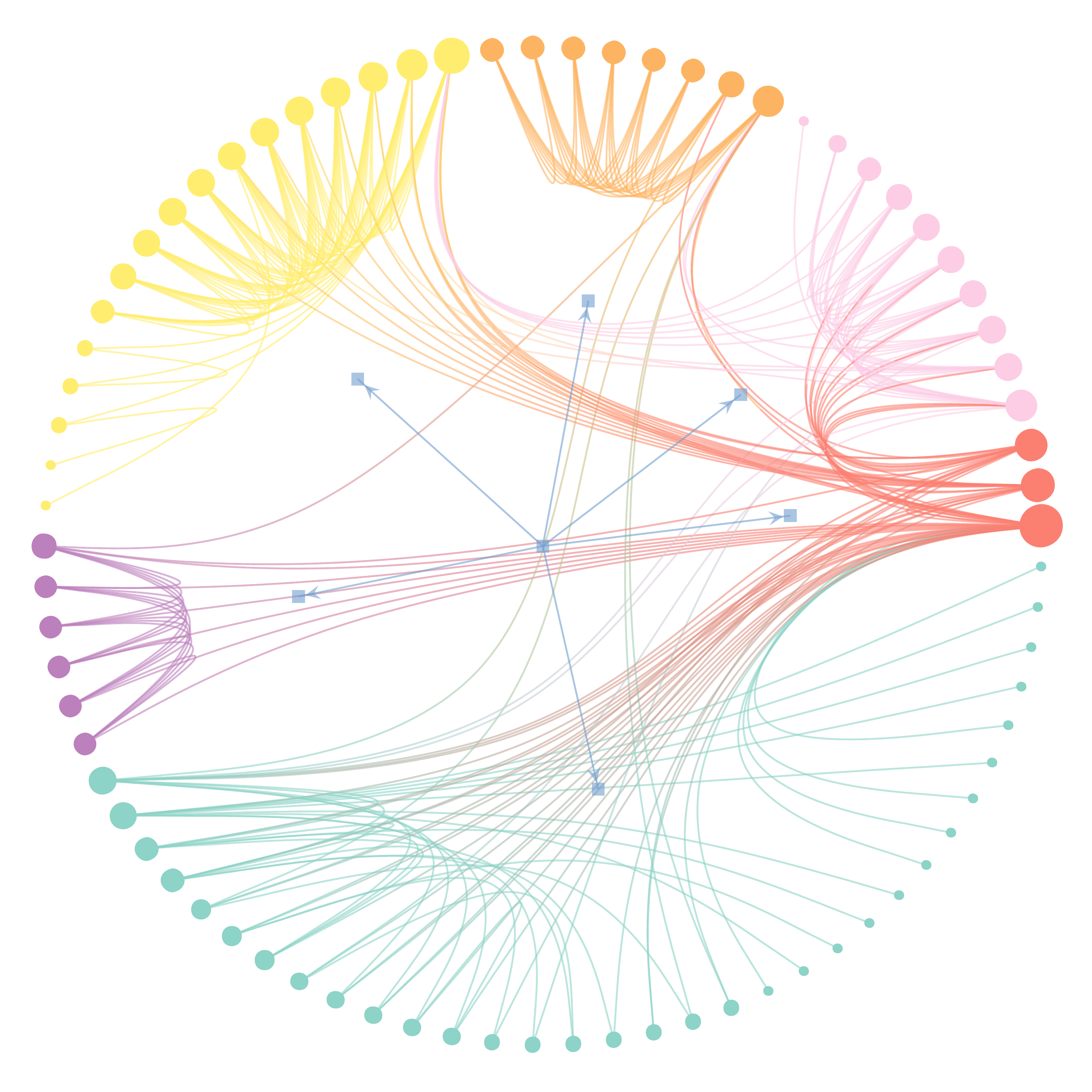}
}
\subfigure[]{
\includegraphics[width=0.38\textwidth]{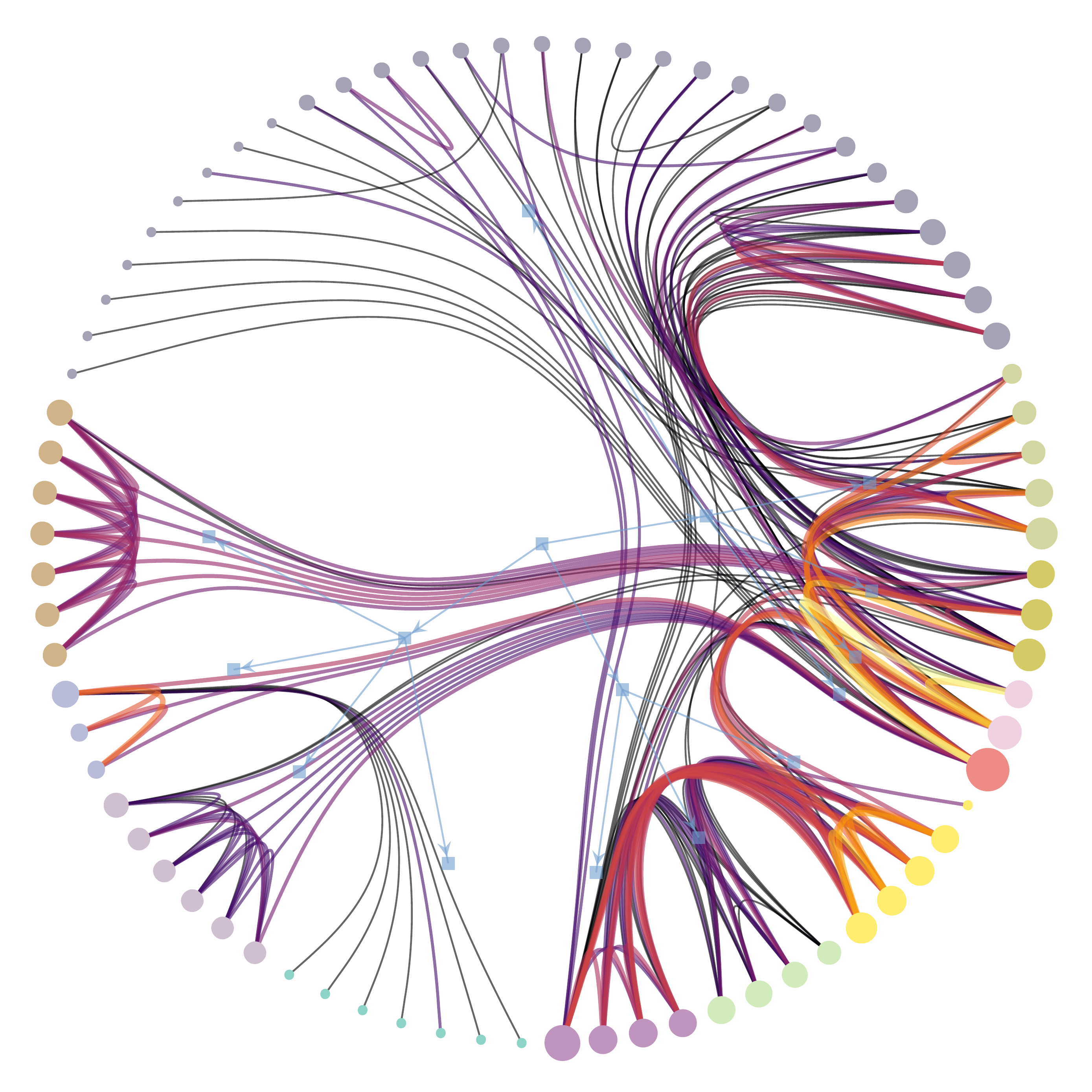}
}
\caption{Clustering results given by our algorithm on unweighted version (a) and weighted version (b) of network composed of co-occurances of major characters in Victor Hugo's novel 'Les Mis\'erables'. A node represents a character and an edge between two nodes shows that these two characters appeared in the same chapter of the book. The weight of each link indicates how often such a co-appearance occured. The figures are drawn using Tiago Peixoto's software graph-tool (graph-tool.skewed.de).
\label{fig:real}
}
\end{figure}

To test our algorithm on real-world network, we ran our algorithm on the network composed of co-occurances of major characters in Victor Hugo's novel 'Les Mis\'erables'. In the network a node represents a character and an edge between two nodes shows that these two characters appeared in the same chapter of the book. 
The weight of each link indicates how often such a co-appearance occured. 
We have tested our algorithm on both weighted version and unweighted version (which sets all the weights to $1$) of the network, and found that the results (which are plotted in Fig.~\ref{fig:real}) are quite different: 
In results obtained by our algorithm on the unweighted version, the Bishop Myriel, his sister, their housemaid together with seven other characters, including Myriel's father and the general of Myriel's father in war, are in one big community; while in the results on the weighted version, Myriel, his sister, and their housemaid are in one group, while the other seven characters in another group. 
We noticed that the three characters, Myriel, his sister, and their housemaid frequently interact in the book, which corresponds to heavy weights on edges between them, while the other seven characters only communicate with them separately once or twice, which corresponds to very light weights. 
In the weighted version, the frequency of interaction between characters are taken into account, therefore the three semi-important characters,  Myriel, his sister, and their housemaid are grouped together, while the other seven characters are in another group; however in the unweighted version where all links are considered to be equal, the information of the strength of interactions are lost. Therefore the ten characters are considered to belong to one group inappropriately.
We then conclude that on the network of 'Les Mis\'erables' our algorithm on weighted network successfully captures the community information given by weights on the edges and give more reasonable results then simply using the unweighted network.

The last application we examine is the community detection in directed networks. It is well known that one way to detect communities in directed networks is to treat it as a weighted network~\cite{malliaros2013clustering}, giving different weights to direct (having one direction) and indirect edges (having two directions). Here we simply give an edge having both direction weight $2$, and give an edge having only one direction weight $1$. It is a relatively simple strategy but we find it already works well with our algorithm for the converted weighted networks that we have tested. We evaluate our method by comparing its performance to the famous Leicht-Newman algorithm~\cite{leicht2008community} which is a spectral algorithm using eigenvectors of the directed Modularity matrix.
The benchmark networks we use are directed LFR networks~\cite{lancichinetti2008benchmark}, the results are plotted in Fig.~\ref{fig:direct}. From the figure, we can find that our algorithm performs better in all regimes than Leicht-Newman algorithm with a different average degree, and with different noise parameters $\mu_t$. 

\begin{figure}[h]
\centering
\subfigure[]{
\label{Fig.sub.1}
\includegraphics[width=\tsize\textwidth]{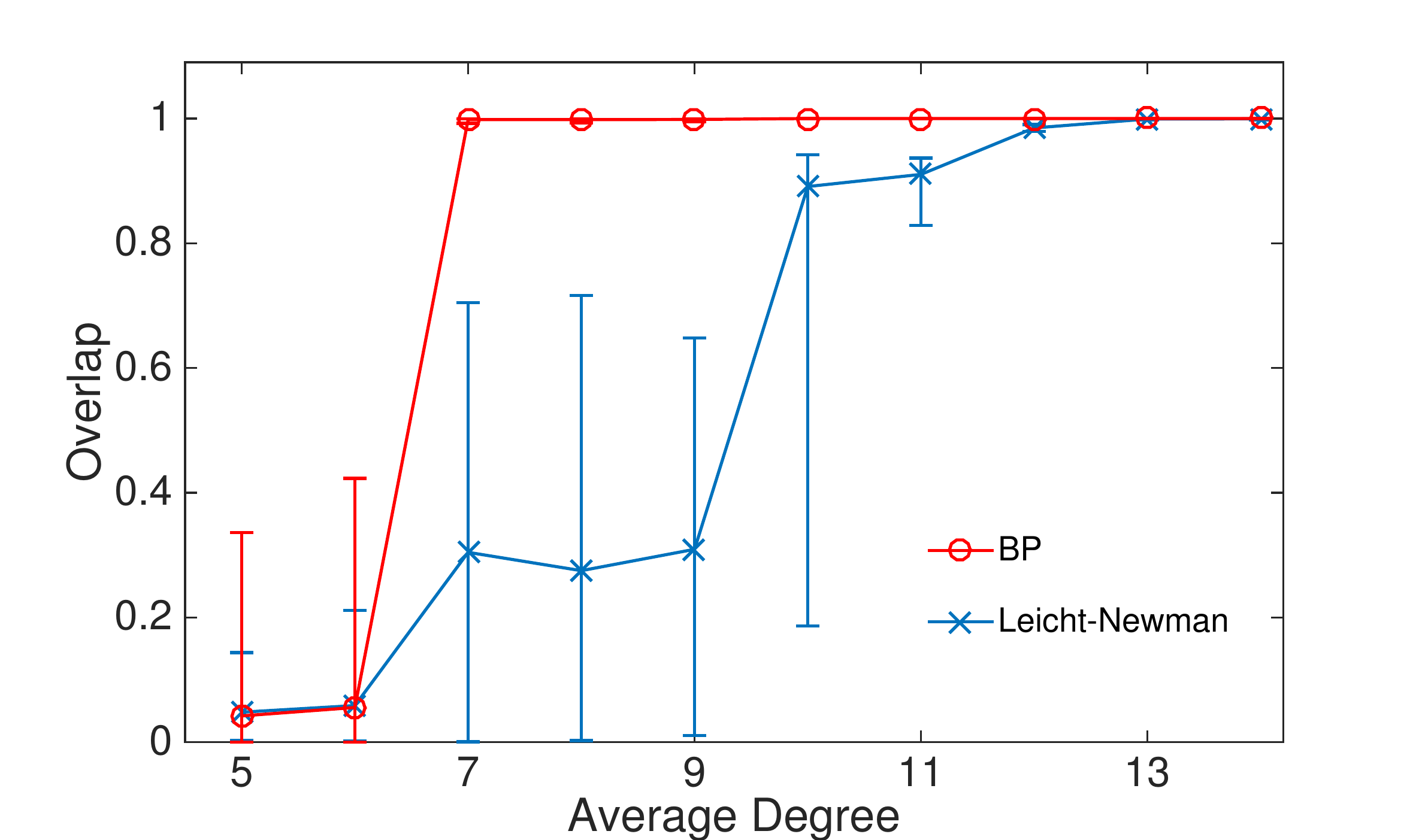}
}
\subfigure[]{
\label{Fig.sub.2}
\includegraphics[width=\tsize\textwidth]{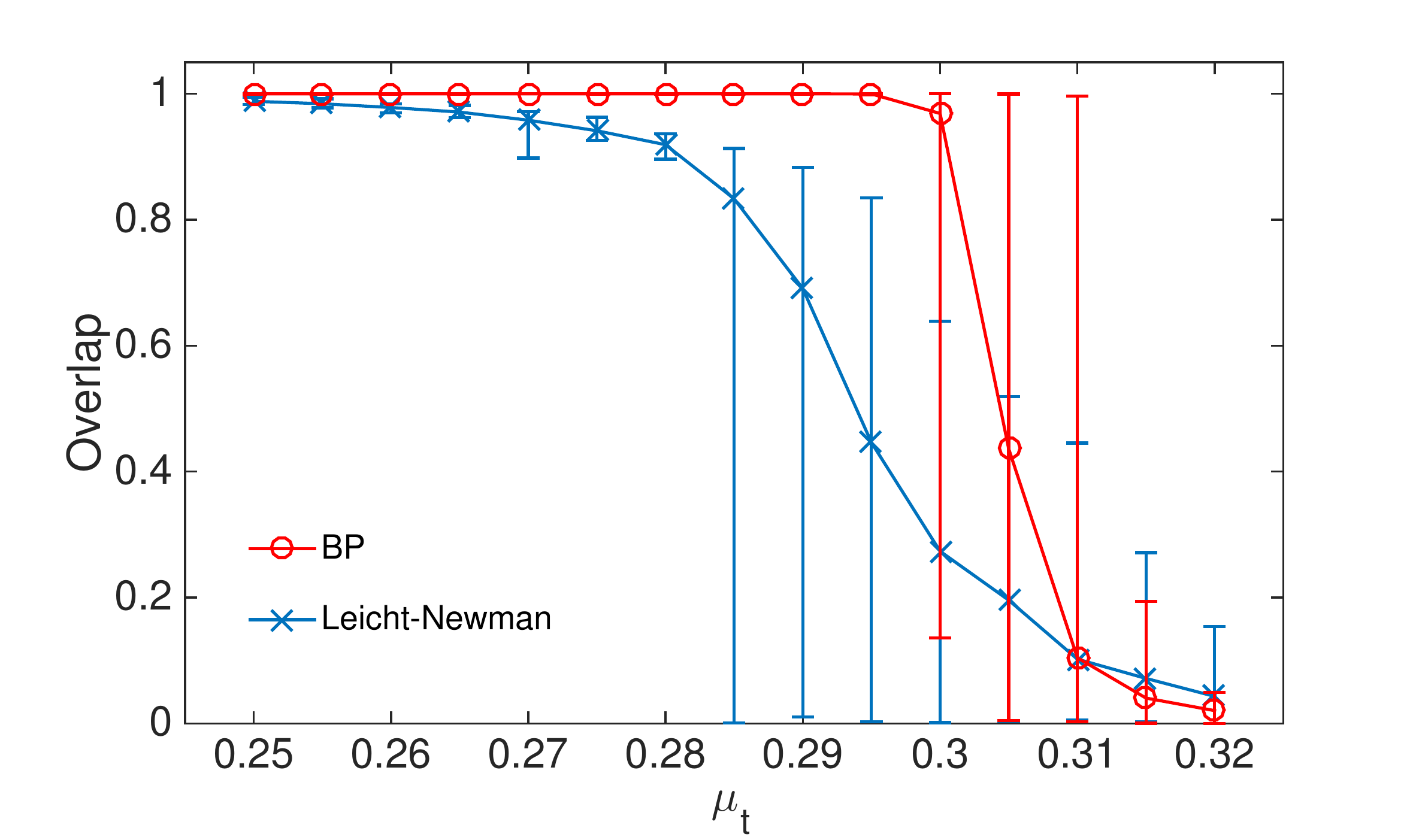}
}
\caption{Comparison of performance (in overlap Eq.~\eqref{eq:ovl}) between our algorithm (BP) and Leicht-Newman algorithm~\cite{lancichinetti2008benchmark} in benchmark networks generated by the directed LFR model~\cite{lancichinetti2008benchmark} with $\mu_{t}$ denoting the mixing parameter. The networks have $n=10000$ nodes, $q^*=2$ groups with same size and maximum degree $50$. In the panel (a), $\mu_{t}$ is fixed to $0.28$. In panel (b) the average degree is fixed to $c=10$. Each result is averaged $50$ instances.\label{fig:direct}}
\end{figure}

\section{Clustering on geometric datasets and semi-supervised clustering}\label{sec:semi}
\begin{figure*}
\centering
\subfigure[]{
\includegraphics[width=0.31\textwidth]{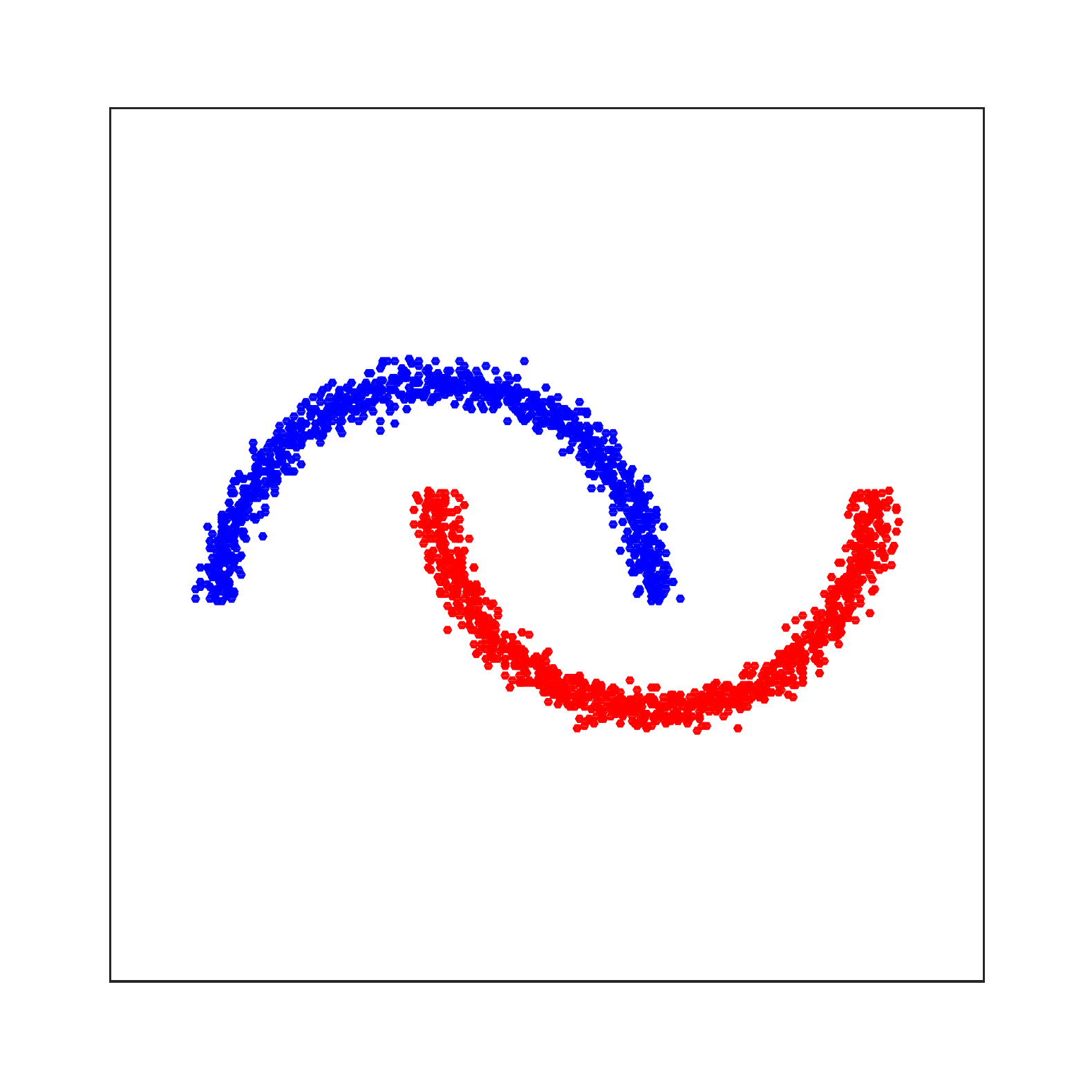}
}
\subfigure[]{
\includegraphics[width=0.31\textwidth]{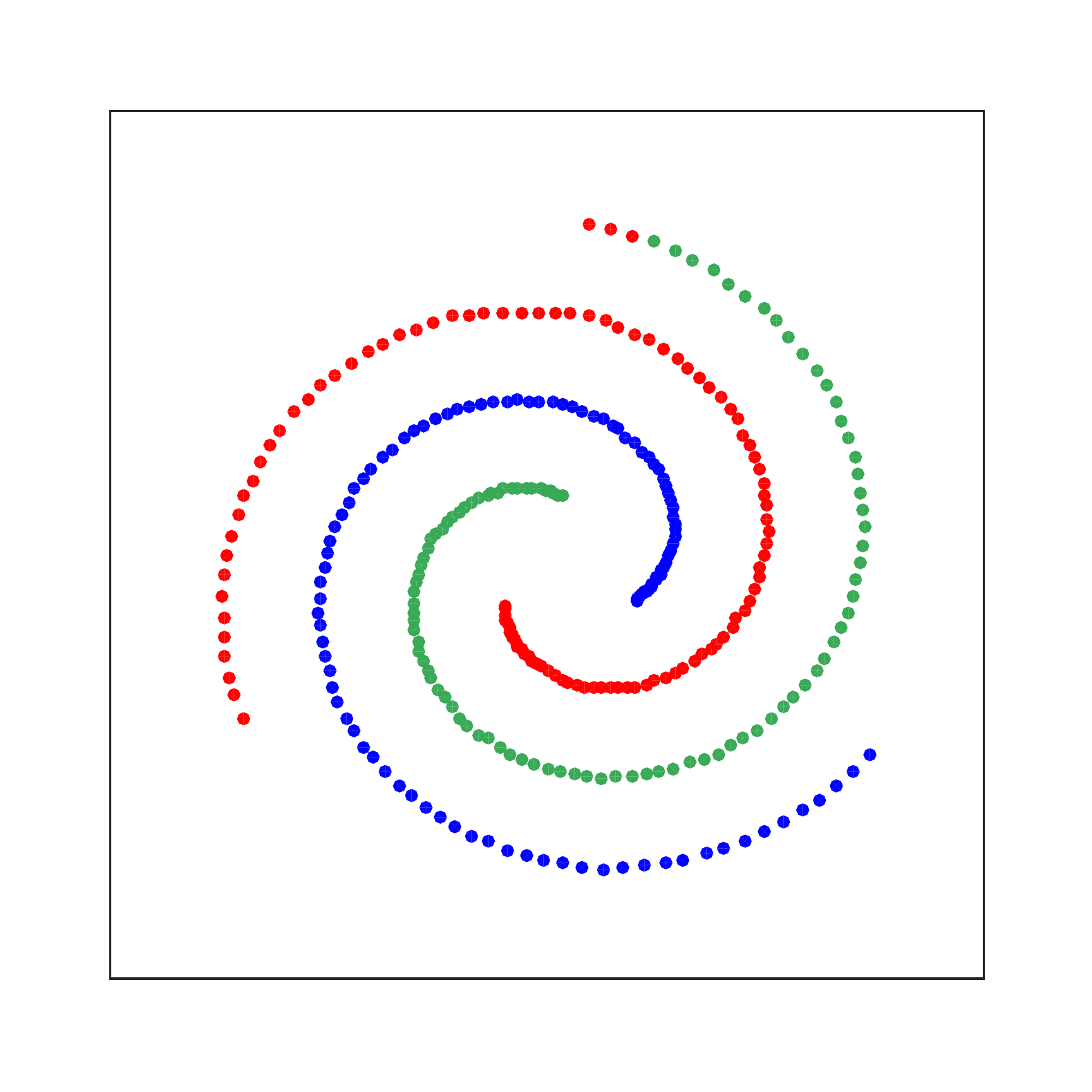}
}
\subfigure[]{
\includegraphics[width=0.31\textwidth]{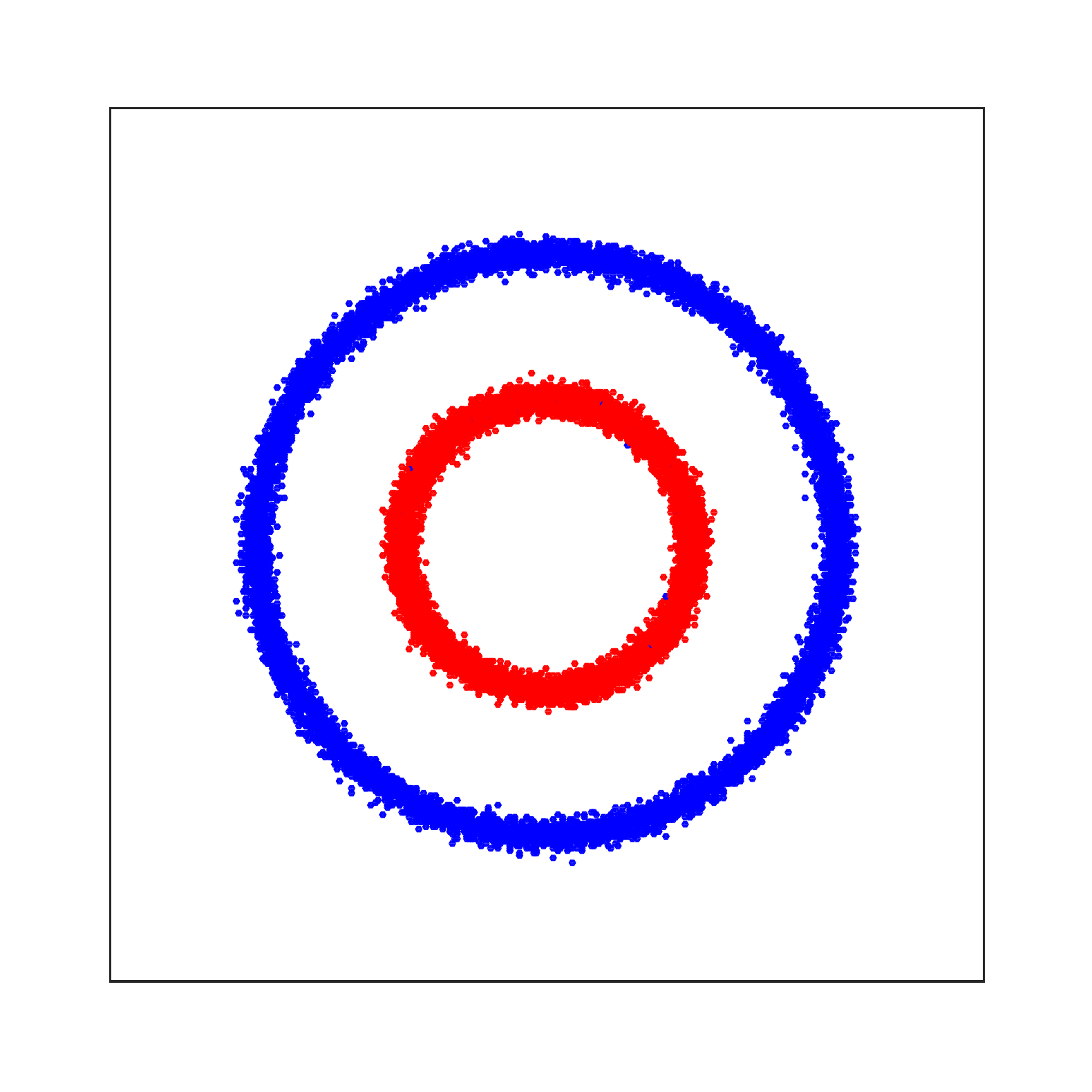}
}
\caption{Clustering results given by belief propagation using inverse of geodesic distance ~\cite{tenenbaum2000global} (on the $5$ nearest-neighbor graph) as similarity. Similarity graph in all three figures are sparse with average degree $10$. Different colors represent different groups found by BP. (a) Moon clustering, $n=2000$ points, similarity graph is sparse with degree $10$. (b)  Spiral with $n=312$ points. (3) Two cycles, $n=20000$ points, $2$ true labels are used in semi-supervision.\label{fig:semi}
}
\end{figure*}

In some real-world clustering problems, a good objective function or similarity measures are hard to choose. For example in the classic datasets of moons, spirals and two rings, as shown in Fig.~\ref{fig:semi},
although the clusters are very clear to human beings, those commonly used objective functions such as distance between items in the same group do not work well, due to the non-spherical properties of clustering structures.

In this case, data does not come with well-defined similarities, we need to choose a $W_{ij}$ using distance between a pair of nodes.
In our experiments, we have tried several definitions of similarities, and we found that the inverse geodesic distance 
~\cite{tenenbaum2000global} works best than other choices such as Gaussian kernels.

In some hard instances such as Spirals, as shown in Fig.~\ref{fig:semi} BP still does not give good-enough results, this means the similarity function or kernel function is not chosen optimally. In these cases, we need to use a small fraction of labels (true group memberships) to guide the algorithm. This approach is called semi-supervised clustering~\cite{basu2004probabilistic}. In our method, the carefully designed objective function can be integrated smoothly into our belief propagation equations, as we only need to redefine the similarities. For semi-supervised clustering using true label, $t_i$ is also easy in our method, because in BP (as adopted in ~\cite{Zhang2014phase}) we can fix the marginals of item $i$ and cavity marginals that being sent from item $i$ to its neighbors $j$ in such a way that,
\begin{align}
\psi^i_{t_i^*}&=\psi_{t_i^*}^{i\to j}=1\nonumber\\
\psi^i_{t\neq t_i^*}&=\psi_{t\neq t_i^*}^{i\to j}=0.
\end{align}

Fig.~\ref{fig:semi} shows results of semi-supervised clustering using our method on classic toy datasets of moons, spirals, and two circles. Different colors represent different groups given by our method using the inverse of geodesic distance ~\cite{tenenbaum2000global} as similarity measure. 
From figures, we can see that BP successfully detect the human-intuitive clusterings in panel (a) and (b).
In panel (c) of the figure using geodesic distance does not give a good result, so we further use $2$ true-labels, each in a circle, in the semi-supervised procedure. This leads to a perfect clustering into two circles.

\section{Clustering in the dense networks using the Thouless-Anderson-Palmer equations}
In above sections, we have shown the performance of our method in several applications in the sparse regime, where the number of measurements, i.e. the connectivity of the weighted network is small. However, when the weighted network is dense, belief propagation algorithm which scales linearly in a number of edges would be too slow. In this section we target at resolving the non-scalable problem of BP in dense graphs by deriving the Thouless-Anderson-Palmer (TAP) equations~\cite{thouless1977solution} which gives a good approximation to BP in dense graphs while is much faster.

TAP equations have been widely used for studying macroscopic properties of mean-field systems in retrieval phase and in spin glass phases. There are at least two ways to derive the TAP equations, first way is via the Plefka expansions where the Legendre transform of the free energy at high temperature is expanded and truncated to second-order term; and the second way is adopted here, where belief propagation equation is first approximated at high-temperature to relaxed belief propagation, then eliminate all the cavity messages using the Onsager reaction terms. The reason that we derive TAP is that it is much faster than BP while only a little less in-accurate than BP in the dense graphs, as illustrated in Fig.9.

Notice that although the update equations are derived using high-temperature expansions, it is a function of $\beta$, which is not necessarily small. In other words, at a high temperature (low beta), TAP equations is equivalent to belief propagation equations. While at a low temperature, TAP equations is only an approximation to belief propagation, as it is derived by expanding belief propagation at high temperature.

Actually the approximatio is very good in the dense graphs, as the paramagnetic-spin glass transition $\beta^*$ is a function of inverse average connectivity $\frac{1}{c}$, hence is very small for dense graphs, vanishing at thermodynamic limit. This means a very small $\beta$ value could still be enough for system to stay in the retrieval phase.

Observe that when average degree of the network is large, marginal probability $\psi_{t_i}^i$ in~Eq.\eqref{eq:bp:marginal} is close to cavity 
probability in~Eq.\eqref{eq:bp}. Thus when one can capture the difference approximately hence eliminate cavity probabilities for reducing the computational complexity. First we expand cavity messages around $0$ to second order, and obtained the so-called \textit{Relaxed Belief Propagation} (RBP)
\begin{align}\label{eq:rbp}
\psi_{t_i}^{i\rightarrow k}
&\approx \frac{e^{h(t_i)}}{Z_{i\rightarrow k}}\prod_{j\in \partial i \backslash k} e^{\psi_{t_i}^{j \rightarrow i}\beta \omega_{ij}+\frac{1}{2}[\psi_{t_i}^{j \rightarrow i}-(\psi_{t_i}^{j \rightarrow i})^2] \beta^2 \omega_{ij}^2}\\
\psi_{t_i}^i&=\frac{ e^{h(t_i)}}{Z_{i}}\prod_{j\in \partial i } e^{\psi_{t_i}^{j \rightarrow i}\beta \omega_{ij}+\frac{1}{2}[\psi_{t_i}^{j}-(\psi_{t_i}^{j})^2] \beta^2 \omega_{ij}^2}.
\end{align}

Then by assuming that $\beta$ is small (which is true in the retrieval phase with average degree large), and by approximating $\psi_{t_i}^{i\to k}$ using marginal probabilities we arrive at TAP equations:
\begin{equation}\label{eq:TAP}
\psi_{t_i}^i = \frac{ e^{h(t_i)}}{Z_{i}}\prod_{j\in \partial i }e^{\psi_{t_i}^{j}+\beta^2 \omega_{ij}^2 (\psi_{t_i}^{j}  (\sum_{s} \psi_{s}^j \psi_{s}^{i}- \psi_{t_i}^{i}) +\frac{1}{2}(\psi_{t_i}^{j}-(\psi_{t_i}^{j})^2)) }.
\end{equation}
The detailed derivations from BP to TAP can be found at Appendix.~\ref{sec:tap}.

\begin{figure}[h]
\centering
\subfigure[]{
\includegraphics[width=0.31\textwidth]{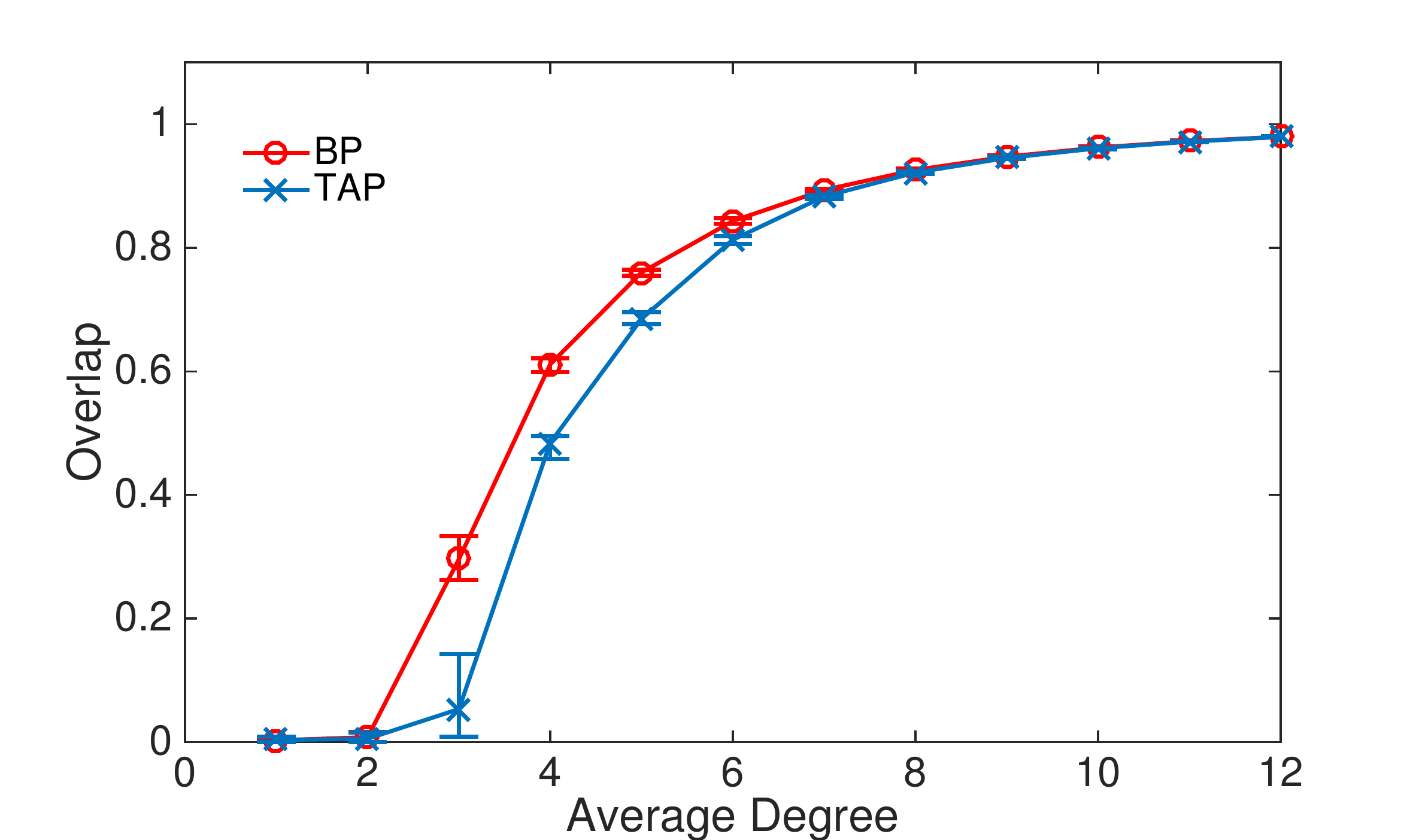}
}
\subfigure[]{
\includegraphics[width=0.31\textwidth]{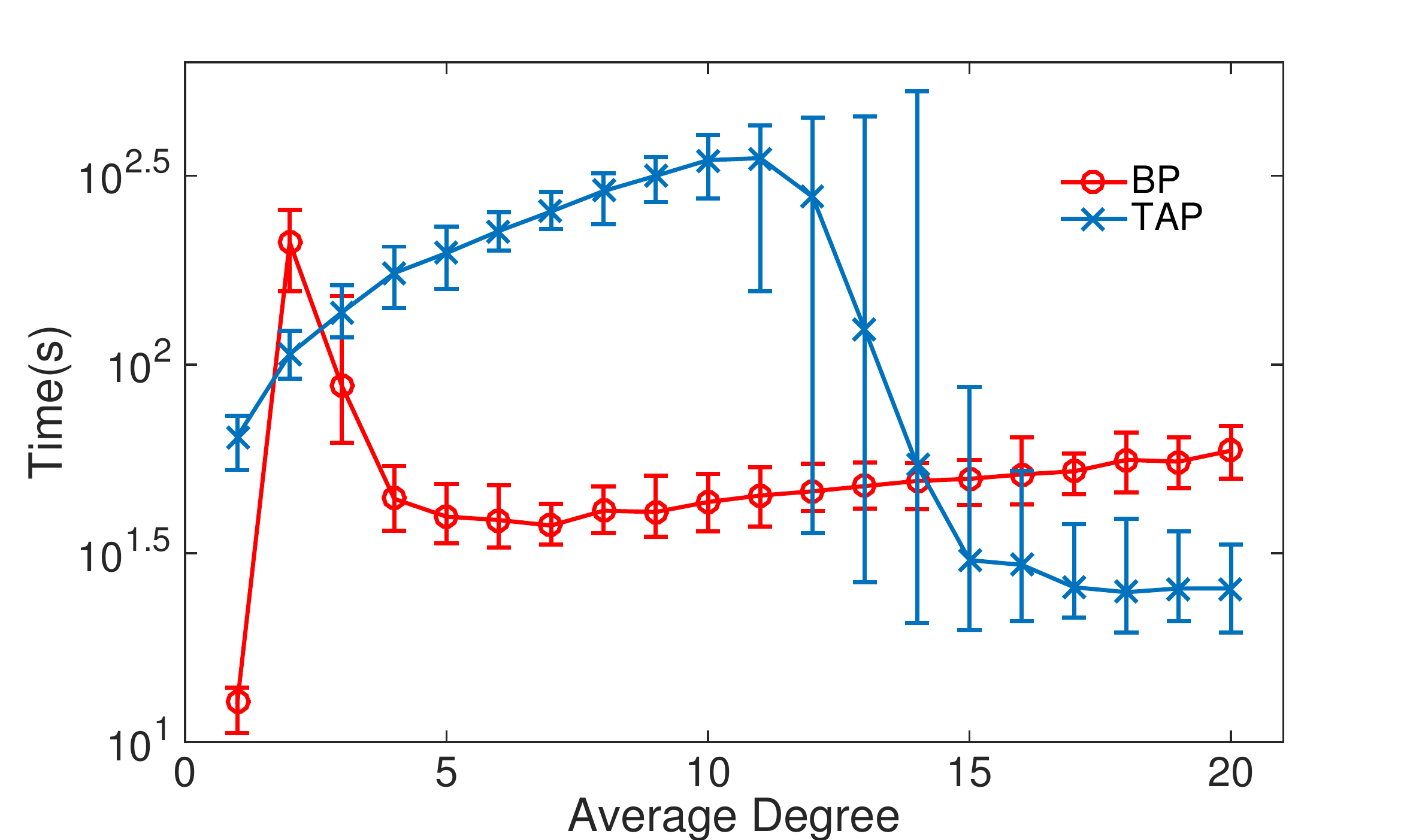}
}
\subfigure[]{
\includegraphics[width=0.31\textwidth]{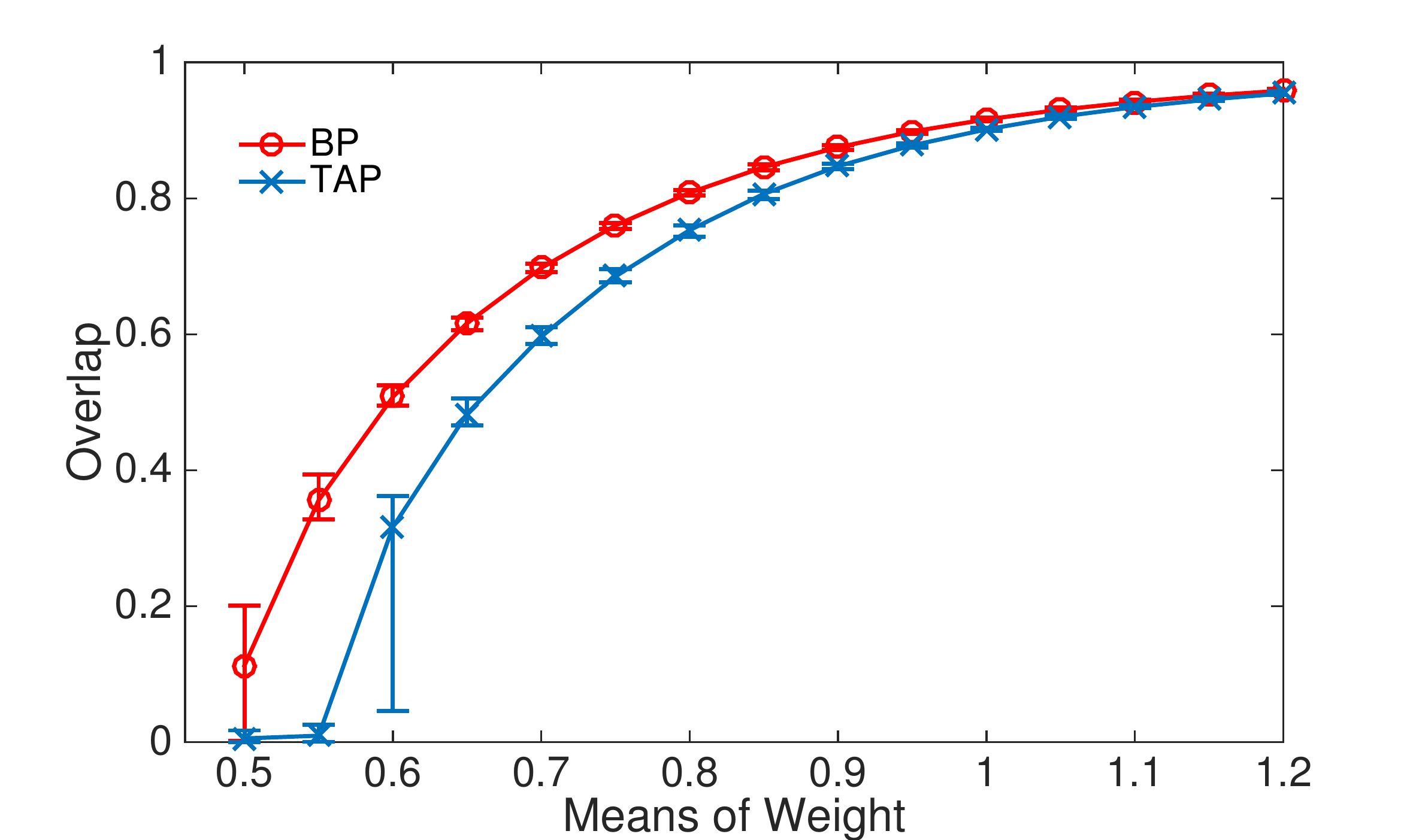}
}

\caption{Comparison of accuracy and time used between BP and TAP. Each point is averaged over 20 realizations of size n = 100,000 (a) in the weighted SBM network with Gaussian weight distributions $\omega_{in}$,$\omega_{out}$ with means $0.75$ and $-0.75$ respectively and unit variance. (b) Computational time used by BP and TAP with conditions in (a) . All points are at the spin-glass transition temperature  $\beta^*$ and the maximal iteration time is 1000. (c) Average degree is fixed to $5$, the x axis show the means of weight $\langle  \omega_{in} \rangle=-\langle \omega_{out}\rangle$. \label{fig:bp:tap}}
\end{figure}

We can see from Eq.~\eqref{eq:TAP} that cavity probabilities are completely eliminated, and there are no messages passing along edges of the graph. Thus the number of messages is reduced from $m$ in BP to $n$ in TAP, as a consequence we can imagine that TAP are much faster than BP in dense graphs. 
Then the question remaining is how TAP works as compared to BP. In Fig.~\ref{fig:bp:tap} we compare accuracy and computational time of BP and TAP in the Gaussian mixture problem. In Fig.~\ref{fig:bp:tap} (a) the mean degree of the similarity graph is varying, we can see that surprisingly with degree larger than $6$ there are already no differences between BP and TAP in accuracy. Even for very sparse similarity graphs with average degree ranging from $2$ to $6$, the difference in accuracy is very small. 
In Fig.~\ref{fig:bp:tap} (b) computation time of BP and TAP are compared with varying average degrees, and we can see that with small average degree, TAP is slower than BP. This is because in sparse graphs TAP is a worse approximation than BP, hence requires more iterations to converge. The figure also illustrates that with average degree larger than $15$, TAP becomes much faster than BP. Most importantly the computational time of TAP stays almost constant with average degree increasing, while the computational time of BP is linear increases with average degree.

In Fig~\ref{fig:bp:tap} (c) we compare the performance of BP and TAP in accuracy with average degree fixed to $5$ and varying mean weights.
We can see that even in such a sparse network, TAP works closer and closer to BP when the clustering problem becomes easier with a larger mean weights.

We notice that the original derivation of TAP equations (e.g. in the Sherrington-Kirkpatrick model) was obtained by minimizing the TAP free energy, which is obtained by taking Legendre transform of the free energy at a high temperature then perform truncating at the second term. This would give identical equations to ours in our model. Notice if one performs truncating of free energy at the first term, the Na\"ive mean-field equations (NMF) can be derived, which is the TAP equations without the Onsager reaction term. In the Appendix we derived also the NMF equations and did an overall comparison of BP, RBP, NMF, and TAP, to complete our picture of using approximations related to Belief Propagation.

\section{Conclusions and discussions}\label{sec:con}
We have presented a message-passing algorithm and its related spectral algorithm for the data clustering problem. Our method is based on Potts spin-glass model in statistical physics, treating a selected objective function as Hamiltonian and apply cavity method. Instead of tempting to directly maximize the objective function, our method use belief propagation to look for a retrieval phase in the whole temperature regime where the statistically significant clustering emerges. This is amount to find the consensus of many partitions with a good objective function, rather than looking for a single partition that maximizes the function. Finding consensus gives us a robust method to overcome the over-fitting problem that the optimization procedure is usually prone to. We further showed that this can be done efficiently by checking the convergence of belief propagation algorithm at the spin glass transition point and gives a mathematically principled way for determining the number of groups.

We applied our method to several clustering problems including a mixture of Gaussian distributions in the sparse regime, semi-supervised clustering, and community detection in weighted and directed networks. We validated the efficiency of our method in the sparse Gaussian mixtures by showing that it works all the way down to the theoretical limits, and works as well as the Bayes optimal inference. In community detection problem in weighted and directed networks, we show that our method significantly outperforms existing algorithms.

The belief propagation algorithm we use has computational complexity linear to the number of edges of the similarity graph. This could be too heavy if the similarity graph is dense and large. In fact, when the similarity graph is dense, weak similarities (or weights) are usually enough to guarantee that the clusterings are detectable. In this case, we can approximate the belief propagation equations using the Thouless-Anderson-Palmer equations~\cite{thouless1977solution} (which is also known as approximated message passing when variables are continuous) whose computational complexity is linear with the number of nodes, rather than a number of edges. We leave this for future work.

A C++ implementation of the proposed algorithms can be found at \\
http://lib.itp.ac.cn/html/panzhang/.

\begin{acknowledgements}
C.S. would like to acknowledge Tao Zhou for support.
\end{acknowledgements}
\bibliography{ref}

\begin{thebibliography}{43}%
\makeatletter
\providecommand \@ifxundefined [1]{%
 \@ifx{#1\undefined}
}%
\providecommand \@ifnum [1]{%
 \ifnum #1\expandafter \@firstoftwo
 \else \expandafter \@secondoftwo
 \fi
}%
\providecommand \@ifx [1]{%
 \ifx #1\expandafter \@firstoftwo
 \else \expandafter \@secondoftwo
 \fi
}%
\providecommand \natexlab [1]{#1}%
\providecommand \enquote  [1]{``#1''}%
\providecommand \bibnamefont  [1]{#1}%
\providecommand \bibfnamefont [1]{#1}%
\providecommand \citenamefont [1]{#1}%
\providecommand \href@noop [0]{\@secondoftwo}%
\providecommand \href [0]{\begingroup \@sanitize@url \@href}%
\providecommand \@href[1]{\@@startlink{#1}\@@href}%
\providecommand \@@href[1]{\endgroup#1\@@endlink}%
\providecommand \@sanitize@url [0]{\catcode `\\12\catcode `\$12\catcode
  `\&12\catcode `\#12\catcode `\^12\catcode `\_12\catcode `\%12\relax}%
\providecommand \@@startlink[1]{}%
\providecommand \@@endlink[0]{}%
\providecommand \url  [0]{\begingroup\@sanitize@url \@url }%
\providecommand \@url [1]{\endgroup\@href {#1}{\urlprefix }}%
\providecommand \urlprefix  [0]{URL }%
\providecommand \Eprint [0]{\href }%
\providecommand \doibase [0]{http://dx.doi.org/}%
\providecommand \selectlanguage [0]{\@gobble}%
\providecommand \bibinfo  [0]{\@secondoftwo}%
\providecommand \bibfield  [0]{\@secondoftwo}%
\providecommand \translation [1]{[#1]}%
\providecommand \BibitemOpen [0]{}%
\providecommand \bibitemStop [0]{}%
\providecommand \bibitemNoStop [0]{.\EOS\space}%
\providecommand \EOS [0]{\spacefactor3000\relax}%
\providecommand \BibitemShut  [1]{\csname bibitem#1\endcsname}%
\let\auto@bib@innerbib\@empty
\bibitem [{\citenamefont {Hartigan}\ and\ \citenamefont
  {Hartigan}(1975)}]{hartigan1975clustering}%
  \BibitemOpen
  \bibfield  {author} {\bibinfo {author} {\bibfnamefont {J.~A.}\ \bibnamefont
  {Hartigan}}\ and\ \bibinfo {author} {\bibfnamefont {J.}~\bibnamefont
  {Hartigan}},\ }\href@noop {} {\emph {\bibinfo {title} {Clustering
  algorithms}}},\ Vol.\ \bibinfo {volume} {209}\ (\bibinfo  {publisher} {Wiley
  New York},\ \bibinfo {year} {1975})\BibitemShut {NoStop}%
\bibitem [{\citenamefont {Jain}\ \emph {et~al.}(1999)\citenamefont {Jain},
  \citenamefont {Murty},\ and\ \citenamefont {Flynn}}]{jain1999data}%
  \BibitemOpen
  \bibfield  {author} {\bibinfo {author} {\bibfnamefont {A.~K.}\ \bibnamefont
  {Jain}}, \bibinfo {author} {\bibfnamefont {M.~N.}\ \bibnamefont {Murty}}, \
  and\ \bibinfo {author} {\bibfnamefont {P.~J.}\ \bibnamefont {Flynn}},\
  }\href@noop {} {\bibfield  {journal} {\bibinfo  {journal} {ACM computing
  surveys (CSUR)}\ }\textbf {\bibinfo {volume} {31}},\ \bibinfo {pages} {264}
  (\bibinfo {year} {1999})}\BibitemShut {NoStop}%
\bibitem [{\citenamefont {Frey}\ and\ \citenamefont {Dueck}(2007)}]{Frey2007}%
  \BibitemOpen
  \bibfield  {author} {\bibinfo {author} {\bibfnamefont {B.~J.}\ \bibnamefont
  {Frey}}\ and\ \bibinfo {author} {\bibfnamefont {D.}~\bibnamefont {Dueck}},\
  }\href {\doibase 10.1126/science.1136800} {\ \textbf {\bibinfo {volume}
  {315}},\ \bibinfo {pages} {972} (\bibinfo {year} {2007})}\BibitemShut
  {NoStop}%
\bibitem [{\citenamefont {Rodriguez}\ and\ \citenamefont
  {Laio}(2014)}]{rodriguez2014clustering}%
  \BibitemOpen
  \bibfield  {author} {\bibinfo {author} {\bibfnamefont {A.}~\bibnamefont
  {Rodriguez}}\ and\ \bibinfo {author} {\bibfnamefont {A.}~\bibnamefont
  {Laio}},\ }\href@noop {} {\bibfield  {journal} {\bibinfo  {journal}
  {Science}\ }\textbf {\bibinfo {volume} {344}},\ \bibinfo {pages} {1492}
  (\bibinfo {year} {2014})}\BibitemShut {NoStop}%
\bibitem [{\citenamefont {Jain}(2010)}]{jain2010data}%
  \BibitemOpen
  \bibfield  {author} {\bibinfo {author} {\bibfnamefont {A.~K.}\ \bibnamefont
  {Jain}},\ }\href@noop {} {\bibfield  {journal} {\bibinfo  {journal} {Pattern
  recognition letters}\ }\textbf {\bibinfo {volume} {31}},\ \bibinfo {pages}
  {651} (\bibinfo {year} {2010})}\BibitemShut {NoStop}%
\bibitem [{\citenamefont {Holland}\ \emph {et~al.}(1983)\citenamefont
  {Holland}, \citenamefont {Laskey},\ and\ \citenamefont
  {Leinhardt}}]{holland1983stochastic}%
  \BibitemOpen
  \bibfield  {author} {\bibinfo {author} {\bibfnamefont {P.~W.}\ \bibnamefont
  {Holland}}, \bibinfo {author} {\bibfnamefont {K.~B.}\ \bibnamefont {Laskey}},
  \ and\ \bibinfo {author} {\bibfnamefont {S.}~\bibnamefont {Leinhardt}},\
  }\href@noop {} {\bibfield  {journal} {\bibinfo  {journal} {Social networks}\
  }\textbf {\bibinfo {volume} {5}},\ \bibinfo {pages} {109} (\bibinfo {year}
  {1983})}\BibitemShut {NoStop}%
\bibitem [{\citenamefont {Luxburg}\ \emph {et~al.}(2007)\citenamefont
  {Luxburg}, \citenamefont {Belkin}, \citenamefont {Bousquet},\ and\
  \citenamefont {Pertinence}}]{Luxburg2007}%
  \BibitemOpen
  \bibfield  {author} {\bibinfo {author} {\bibfnamefont {U.~V.}\ \bibnamefont
  {Luxburg}}, \bibinfo {author} {\bibfnamefont {M.}~\bibnamefont {Belkin}},
  \bibinfo {author} {\bibfnamefont {O.}~\bibnamefont {Bousquet}}, \ and\
  \bibinfo {author} {\bibnamefont {Pertinence}},\ }\href@noop {} {\bibfield
  {journal} {\bibinfo  {journal} {Stat. Comput}\ } (\bibinfo {year}
  {2007})}\BibitemShut {NoStop}%
\bibitem [{\citenamefont {Shi}\ and\ \citenamefont {Malik}(1997)}]{Shi1997}%
  \BibitemOpen
  \bibfield  {author} {\bibinfo {author} {\bibfnamefont {J.}~\bibnamefont
  {Shi}}\ and\ \bibinfo {author} {\bibfnamefont {J.}~\bibnamefont {Malik}},\
  }\href@noop {} {\bibfield  {journal} {\bibinfo  {journal} {IEEE Transactions
  on Pattern Analysis and Machine Intelligence}\ }\textbf {\bibinfo {volume}
  {22}},\ \bibinfo {pages} {888} (\bibinfo {year} {1997})}\BibitemShut
  {NoStop}%
\bibitem [{\citenamefont {Ng}\ \emph {et~al.}(2001)\citenamefont {Ng},
  \citenamefont {Jordan},\ and\ \citenamefont {Weiss}}]{Ng2001}%
  \BibitemOpen
  \bibfield  {author} {\bibinfo {author} {\bibfnamefont {A.~Y.}\ \bibnamefont
  {Ng}}, \bibinfo {author} {\bibfnamefont {M.~I.}\ \bibnamefont {Jordan}}, \
  and\ \bibinfo {author} {\bibfnamefont {Y.}~\bibnamefont {Weiss}},\
  }\href@noop {} {\bibfield  {journal} {\bibinfo  {journal} {Proceedings of
  Advances in Neural Information Processing Systems. Cambridge, MA: MIT Press}\
  }\textbf {\bibinfo {volume} {14}},\ \bibinfo {pages} {849} (\bibinfo {year}
  {2001})}\BibitemShut {NoStop}%
\bibitem [{\citenamefont {Zhang}(2016)}]{zhang2016robust}%
  \BibitemOpen
  \bibfield  {author} {\bibinfo {author} {\bibfnamefont {P.}~\bibnamefont
  {Zhang}},\ }in\ \href
  {http://papers.nips.cc/paper/6491-robust-spectral-detection-of-global-structures-in-the-data-by-learning-a-regularization.pdf}
  {\emph {\bibinfo {booktitle} {Advances In Neural Information Processing
  Systems 29}}},\ \bibinfo {editor} {edited by\ \bibinfo {editor}
  {\bibfnamefont {D.~D.}\ \bibnamefont {Lee}}, \bibinfo {editor} {\bibfnamefont
  {U.~V.}\ \bibnamefont {Luxburg}}, \bibinfo {editor} {\bibfnamefont
  {I.}~\bibnamefont {Guyon}}, \ and\ \bibinfo {editor} {\bibfnamefont
  {R.}~\bibnamefont {Garnett}}}\ (\bibinfo  {publisher} {Curran Associates,
  Inc.},\ \bibinfo {year} {2016})\ pp.\ \bibinfo {pages} {541--549}\BibitemShut
  {NoStop}%
\bibitem [{\citenamefont {Zhang}\ and\ \citenamefont
  {Moore}(2014)}]{Zhang2014pnas}%
  \BibitemOpen
  \bibfield  {author} {\bibinfo {author} {\bibfnamefont {P.}~\bibnamefont
  {Zhang}}\ and\ \bibinfo {author} {\bibfnamefont {C.}~\bibnamefont {Moore}},\
  }\href {\doibase 10.1073/pnas.1409770111} {\bibfield  {journal} {\bibinfo
  {journal} {Proceedings of the National Academy of Sciences}\ }\textbf
  {\bibinfo {volume} {111}},\ \bibinfo {pages} {18144} (\bibinfo {year}
  {2014})},\ \Eprint
  {http://arxiv.org/abs/http://www.pnas.org/content/111/51/18144.full.pdf+html}
  {http://www.pnas.org/content/111/51/18144.full.pdf+html} \BibitemShut
  {NoStop}%
\bibitem [{\citenamefont {Newman}(2016)}]{newman2016community}%
  \BibitemOpen
  \bibfield  {author} {\bibinfo {author} {\bibfnamefont {M.}~\bibnamefont
  {Newman}},\ }\href@noop {} {\bibfield  {journal} {\bibinfo  {journal} {arXiv
  preprint arXiv:1606.02319}\ } (\bibinfo {year} {2016})}\BibitemShut {NoStop}%
\bibitem [{\citenamefont {Achlioptas}\ and\ \citenamefont
  {McSherry}(2007)}]{achlioptas2007fast}%
  \BibitemOpen
  \bibfield  {author} {\bibinfo {author} {\bibfnamefont {D.}~\bibnamefont
  {Achlioptas}}\ and\ \bibinfo {author} {\bibfnamefont {F.}~\bibnamefont
  {McSherry}},\ }\href@noop {} {\bibfield  {journal} {\bibinfo  {journal}
  {Journal of the ACM (JACM)}\ }\textbf {\bibinfo {volume} {54}},\ \bibinfo
  {pages} {9} (\bibinfo {year} {2007})}\BibitemShut {NoStop}%
\bibitem [{\citenamefont {Weigt}\ \emph {et~al.}(2009)\citenamefont {Weigt},
  \citenamefont {White}, \citenamefont {Szurmant}, \citenamefont {Hoch},\ and\
  \citenamefont {Hwa}}]{Weigt2009}%
  \BibitemOpen
  \bibfield  {author} {\bibinfo {author} {\bibfnamefont {M.}~\bibnamefont
  {Weigt}}, \bibinfo {author} {\bibfnamefont {R.~A.}\ \bibnamefont {White}},
  \bibinfo {author} {\bibfnamefont {H.}~\bibnamefont {Szurmant}}, \bibinfo
  {author} {\bibfnamefont {J.~A.}\ \bibnamefont {Hoch}}, \ and\ \bibinfo
  {author} {\bibfnamefont {T.}~\bibnamefont {Hwa}},\ }\href@noop {} {\bibfield
  {journal} {\bibinfo  {journal} {PNAS}\ }\textbf {\bibinfo {volume} {106}},\
  \bibinfo {pages} {67} (\bibinfo {year} {2009})}\BibitemShut {NoStop}%
\bibitem [{\citenamefont {Blatt}\ \emph
  {et~al.}(1996{\natexlab{a}})\citenamefont {Blatt}, \citenamefont {Wiseman},\
  and\ \citenamefont {Domany}}]{blatt1996superparamagnetic}%
  \BibitemOpen
  \bibfield  {author} {\bibinfo {author} {\bibfnamefont {M.}~\bibnamefont
  {Blatt}}, \bibinfo {author} {\bibfnamefont {S.}~\bibnamefont {Wiseman}}, \
  and\ \bibinfo {author} {\bibfnamefont {E.}~\bibnamefont {Domany}},\
  }\href@noop {} {\bibfield  {journal} {\bibinfo  {journal} {Physical review
  letters}\ }\textbf {\bibinfo {volume} {76}},\ \bibinfo {pages} {3251}
  (\bibinfo {year} {1996}{\natexlab{a}})}\BibitemShut {NoStop}%
\bibitem [{\citenamefont {Blatt}\ \emph
  {et~al.}(1996{\natexlab{b}})\citenamefont {Blatt}, \citenamefont {Wiseman},\
  and\ \citenamefont {Domany}}]{blatt1996clustering}%
  \BibitemOpen
  \bibfield  {author} {\bibinfo {author} {\bibfnamefont {M.}~\bibnamefont
  {Blatt}}, \bibinfo {author} {\bibfnamefont {S.}~\bibnamefont {Wiseman}}, \
  and\ \bibinfo {author} {\bibfnamefont {E.}~\bibnamefont {Domany}},\ }in\
  \href@noop {} {\emph {\bibinfo {booktitle} {Advances in Neural Information
  Processing Systems}}}\ (\bibinfo {year} {1996})\ pp.\ \bibinfo {pages}
  {416--422}\BibitemShut {NoStop}%
\bibitem [{\citenamefont {Wiseman}\ \emph {et~al.}(1998)\citenamefont
  {Wiseman}, \citenamefont {Blatt},\ and\ \citenamefont
  {Domany}}]{wiseman1998superparamagnetic}%
  \BibitemOpen
  \bibfield  {author} {\bibinfo {author} {\bibfnamefont {S.}~\bibnamefont
  {Wiseman}}, \bibinfo {author} {\bibfnamefont {M.}~\bibnamefont {Blatt}}, \
  and\ \bibinfo {author} {\bibfnamefont {E.}~\bibnamefont {Domany}},\
  }\href@noop {} {\bibfield  {journal} {\bibinfo  {journal} {Physical Review
  E}\ }\textbf {\bibinfo {volume} {57}},\ \bibinfo {pages} {3767} (\bibinfo
  {year} {1998})}\BibitemShut {NoStop}%
\bibitem [{\citenamefont {Krzakala}\ \emph {et~al.}(2013)\citenamefont
  {Krzakala}, \citenamefont {Moore}, \citenamefont {Mossel}, \citenamefont
  {Neeman}, \citenamefont {Sly}, \citenamefont {Zdeborov\'a},\ and\
  \citenamefont {Zhang}}]{Krzakala2013}%
  \BibitemOpen
  \bibfield  {author} {\bibinfo {author} {\bibfnamefont {F.}~\bibnamefont
  {Krzakala}}, \bibinfo {author} {\bibfnamefont {C.}~\bibnamefont {Moore}},
  \bibinfo {author} {\bibfnamefont {E.}~\bibnamefont {Mossel}}, \bibinfo
  {author} {\bibfnamefont {J.}~\bibnamefont {Neeman}}, \bibinfo {author}
  {\bibfnamefont {A.}~\bibnamefont {Sly}}, \bibinfo {author} {\bibfnamefont
  {L.}~\bibnamefont {Zdeborov\'a}}, \ and\ \bibinfo {author} {\bibfnamefont
  {P.}~\bibnamefont {Zhang}},\ }\href {\doibase 10.1073/pnas.1312486110}
  {\bibfield  {journal} {\bibinfo  {journal} {Proc. Natl. Acad. Sci. USA}\
  }\textbf {\bibinfo {volume} {110}},\ \bibinfo {pages} {20935} (\bibinfo
  {year} {2013})},\ \Eprint
  {http://arxiv.org/abs/http://www.pnas.org/content/110/52/20935.full.pdf+html}
  {http://www.pnas.org/content/110/52/20935.full.pdf+html} \BibitemShut
  {NoStop}%
\bibitem [{\citenamefont {Girvan}\ and\ \citenamefont
  {Newman}(2002)}]{Girvan2002}%
  \BibitemOpen
  \bibfield  {author} {\bibinfo {author} {\bibfnamefont {M.}~\bibnamefont
  {Girvan}}\ and\ \bibinfo {author} {\bibfnamefont {M.~E.~J.}\ \bibnamefont
  {Newman}},\ }\href@noop {} {\bibfield  {journal} {\bibinfo  {journal}
  {Proceedings of the National Academy of Sciences}\ }\textbf {\bibinfo
  {volume} {99}},\ \bibinfo {pages} {7821} (\bibinfo {year}
  {2002})}\BibitemShut {NoStop}%
\bibitem [{\citenamefont {Mezard}\ and\ \citenamefont
  {Montanari}(2009)}]{Mezard2009}%
  \BibitemOpen
  \bibfield  {author} {\bibinfo {author} {\bibfnamefont {M.}~\bibnamefont
  {Mezard}}\ and\ \bibinfo {author} {\bibfnamefont {A.}~\bibnamefont
  {Montanari}},\ }\href@noop {} {\emph {\bibinfo {title} {Information, Physics
  and Computation}}}\ (\bibinfo  {publisher} {Oxford University press},\
  \bibinfo {year} {2009})\BibitemShut {NoStop}%
\bibitem [{\citenamefont {Yedidia}\ \emph {et~al.}(2001)\citenamefont
  {Yedidia}, \citenamefont {Freeman},\ and\ \citenamefont
  {Weiss}}]{Yedidia2001}%
  \BibitemOpen
  \bibfield  {author} {\bibinfo {author} {\bibfnamefont {J.}~\bibnamefont
  {Yedidia}}, \bibinfo {author} {\bibfnamefont {W.}~\bibnamefont {Freeman}}, \
  and\ \bibinfo {author} {\bibfnamefont {Y.}~\bibnamefont {Weiss}},\ }in\
  \href@noop {} {\emph {\bibinfo {booktitle} {International Joint Conference on
  Artificial Intelligence (IJCAI)}}}\ (\bibinfo {year} {2001})\BibitemShut
  {NoStop}%
\bibitem [{\citenamefont {Zdeborov\'a}(2009)}]{Zdeborova2009}%
  \BibitemOpen
  \bibfield  {author} {\bibinfo {author} {\bibfnamefont {L.}~\bibnamefont
  {Zdeborov\'a}},\ }\href@noop {} {\bibfield  {journal} {\bibinfo  {journal}
  {Acta Phys. Slov.}\ }\textbf {\bibinfo {volume} {59}},\ \bibinfo {pages}
  {169} (\bibinfo {year} {2009})}\BibitemShut {NoStop}%
\bibitem [{\citenamefont {Mossel}\ \emph {et~al.}(2015)\citenamefont {Mossel},
  \citenamefont {Neeman},\ and\ \citenamefont
  {Sly}}]{mossel2015reconstruction}%
  \BibitemOpen
  \bibfield  {author} {\bibinfo {author} {\bibfnamefont {E.}~\bibnamefont
  {Mossel}}, \bibinfo {author} {\bibfnamefont {J.}~\bibnamefont {Neeman}}, \
  and\ \bibinfo {author} {\bibfnamefont {A.}~\bibnamefont {Sly}},\ }\href@noop
  {} {\bibfield  {journal} {\bibinfo  {journal} {Probability Theory and Related
  Fields}\ }\textbf {\bibinfo {volume} {162}},\ \bibinfo {pages} {431}
  (\bibinfo {year} {2015})}\BibitemShut {NoStop}%
\bibitem [{\citenamefont {Zhang}(2015{\natexlab{a}})}]{dogmat}%
  \BibitemOpen
  \bibfield  {author} {\bibinfo {author} {\bibfnamefont {P.}~\bibnamefont
  {Zhang}},\ }\href {\doibase 10.1103/PhysRevE.91.042120} {\bibfield  {journal}
  {\bibinfo  {journal} {Phys. Rev. E}\ }\textbf {\bibinfo {volume} {91}},\
  \bibinfo {pages} {042120} (\bibinfo {year} {2015}{\natexlab{a}})}\BibitemShut
  {NoStop}%
\bibitem [{\citenamefont {Peixoto}(2014{\natexlab{a}})}]{Peixoto2014}%
  \BibitemOpen
  \bibfield  {author} {\bibinfo {author} {\bibfnamefont {T.~P.}\ \bibnamefont
  {Peixoto}},\ }\href {\doibase 10.1103/PhysRevX.4.011047} {\bibfield
  {journal} {\bibinfo  {journal} {Phys. Rev. X}\ }\textbf {\bibinfo {volume}
  {4}},\ \bibinfo {pages} {011047} (\bibinfo {year}
  {2014}{\natexlab{a}})}\BibitemShut {NoStop}%
\bibitem [{\citenamefont {Decelle}\ \emph {et~al.}(2011)\citenamefont
  {Decelle}, \citenamefont {Krzakala}, \citenamefont {Moore},\ and\
  \citenamefont {Zdeborov\'a}}]{Decelle2011}%
  \BibitemOpen
  \bibfield  {author} {\bibinfo {author} {\bibfnamefont {A.}~\bibnamefont
  {Decelle}}, \bibinfo {author} {\bibfnamefont {F.}~\bibnamefont {Krzakala}},
  \bibinfo {author} {\bibfnamefont {C.}~\bibnamefont {Moore}}, \ and\ \bibinfo
  {author} {\bibfnamefont {L.}~\bibnamefont {Zdeborov\'a}},\ }\href {\doibase
  10.1103/PhysRevE.84.066106} {\bibfield  {journal} {\bibinfo  {journal} {Phys.
  Rev. E}\ }\textbf {\bibinfo {volume} {84}},\ \bibinfo {pages} {066106}
  (\bibinfo {year} {2011})}\BibitemShut {NoStop}%
\bibitem [{\citenamefont {Saade}\ \emph {et~al.}(2016)\citenamefont {Saade},
  \citenamefont {Lelarge}, \citenamefont {Krzakala},\ and\ \citenamefont
  {Zdeborov{\'a}}}]{saade2016clustering}%
  \BibitemOpen
  \bibfield  {author} {\bibinfo {author} {\bibfnamefont {A.}~\bibnamefont
  {Saade}}, \bibinfo {author} {\bibfnamefont {M.}~\bibnamefont {Lelarge}},
  \bibinfo {author} {\bibfnamefont {F.}~\bibnamefont {Krzakala}}, \ and\
  \bibinfo {author} {\bibfnamefont {L.}~\bibnamefont {Zdeborov{\'a}}},\ }in\
  \href@noop {} {\emph {\bibinfo {booktitle} {Information Theory (ISIT), 2016
  IEEE International Symposium on}}}\ (\bibinfo {organization} {IEEE},\
  \bibinfo {year} {2016})\ pp.\ \bibinfo {pages} {780--784}\BibitemShut
  {NoStop}%
\bibitem [{\citenamefont {Heimlicher}\ \emph {et~al.}(2012)\citenamefont
  {Heimlicher}, \citenamefont {Lelarge},\ and\ \citenamefont
  {Massouli{\'e}}}]{heimlicher2012community}%
  \BibitemOpen
  \bibfield  {author} {\bibinfo {author} {\bibfnamefont {S.}~\bibnamefont
  {Heimlicher}}, \bibinfo {author} {\bibfnamefont {M.}~\bibnamefont {Lelarge}},
  \ and\ \bibinfo {author} {\bibfnamefont {L.}~\bibnamefont {Massouli{\'e}}},\
  }\href@noop {} {\bibfield  {journal} {\bibinfo  {journal} {arXiv preprint
  arXiv:1209.2910}\ } (\bibinfo {year} {2012})}\BibitemShut {NoStop}%
\bibitem [{\citenamefont {Zhang}(2015{\natexlab{b}})}]{rnmi}%
  \BibitemOpen
  \bibfield  {author} {\bibinfo {author} {\bibfnamefont {P.}~\bibnamefont
  {Zhang}},\ }\href {http://stacks.iop.org/1742-5468/2015/i=11/a=P11006}
  {\bibfield  {journal} {\bibinfo  {journal} {Journal of Statistical Mechanics:
  Theory and Experiment}\ }\textbf {\bibinfo {volume} {2015}},\ \bibinfo
  {pages} {P11006} (\bibinfo {year} {2015}{\natexlab{b}})}\BibitemShut
  {NoStop}%
\bibitem [{\citenamefont {Lancichinetti}\ \emph {et~al.}(2008)\citenamefont
  {Lancichinetti}, \citenamefont {Fortunato},\ and\ \citenamefont
  {Radicchi}}]{lancichinetti2008benchmark}%
  \BibitemOpen
  \bibfield  {author} {\bibinfo {author} {\bibfnamefont {A.}~\bibnamefont
  {Lancichinetti}}, \bibinfo {author} {\bibfnamefont {S.}~\bibnamefont
  {Fortunato}}, \ and\ \bibinfo {author} {\bibfnamefont {F.}~\bibnamefont
  {Radicchi}},\ }\href@noop {} {\bibfield  {journal} {\bibinfo  {journal}
  {Physical Review E}\ }\textbf {\bibinfo {volume} {78}},\ \bibinfo {pages}
  {046110} (\bibinfo {year} {2008})}\BibitemShut {NoStop}%
\bibitem [{\citenamefont {Rosvall}\ and\ \citenamefont
  {Bergstrom}(2008)}]{Rosvall2008}%
  \BibitemOpen
  \bibfield  {author} {\bibinfo {author} {\bibfnamefont {M.}~\bibnamefont
  {Rosvall}}\ and\ \bibinfo {author} {\bibfnamefont {C.~T.}\ \bibnamefont
  {Bergstrom}},\ }\href@noop {} {\bibfield  {journal} {\bibinfo  {journal}
  {Proceedings of the National Academy of Sciences}\ }\textbf {\bibinfo
  {volume} {105}},\ \bibinfo {pages} {1118} (\bibinfo {year}
  {2008})}\BibitemShut {NoStop}%
\bibitem [{\citenamefont {Lancichinetti}\ \emph {et~al.}(2011)\citenamefont
  {Lancichinetti}, \citenamefont {Radicchi}, \citenamefont {Ramasco},\ and\
  \citenamefont {Fortunato}}]{lancichinetti2011finding}%
  \BibitemOpen
  \bibfield  {author} {\bibinfo {author} {\bibfnamefont {A.}~\bibnamefont
  {Lancichinetti}}, \bibinfo {author} {\bibfnamefont {F.}~\bibnamefont
  {Radicchi}}, \bibinfo {author} {\bibfnamefont {J.~J.}\ \bibnamefont
  {Ramasco}}, \ and\ \bibinfo {author} {\bibfnamefont {S.}~\bibnamefont
  {Fortunato}},\ }\href@noop {} {\bibfield  {journal} {\bibinfo  {journal}
  {PloS one}\ }\textbf {\bibinfo {volume} {6}},\ \bibinfo {pages} {e18961}
  (\bibinfo {year} {2011})}\BibitemShut {NoStop}%
\bibitem [{\citenamefont {Blondel}\ \emph {et~al.}(2008)\citenamefont
  {Blondel}, \citenamefont {Guillaume}, \citenamefont {Lambiotte},\ and\
  \citenamefont {Lefebvre}}]{Blondel2008}%
  \BibitemOpen
  \bibfield  {author} {\bibinfo {author} {\bibfnamefont {V.~D.}\ \bibnamefont
  {Blondel}}, \bibinfo {author} {\bibfnamefont {J.-L.}\ \bibnamefont
  {Guillaume}}, \bibinfo {author} {\bibfnamefont {R.}~\bibnamefont
  {Lambiotte}}, \ and\ \bibinfo {author} {\bibfnamefont {E.}~\bibnamefont
  {Lefebvre}},\ }\href@noop {} {\bibfield  {journal} {\bibinfo  {journal} {J.
  Stat. Mech.}\ }\textbf {\bibinfo {volume} {2008}},\ \bibinfo {pages} {P10008}
  (\bibinfo {year} {2008})}\BibitemShut {NoStop}%
\bibitem [{\citenamefont {Danon}\ \emph {et~al.}(2005)\citenamefont {Danon},
  \citenamefont {Diaz-Guilera}, \citenamefont {Duch},\ and\ \citenamefont
  {Arenas}}]{danon2005comparing}%
  \BibitemOpen
  \bibfield  {author} {\bibinfo {author} {\bibfnamefont {L.}~\bibnamefont
  {Danon}}, \bibinfo {author} {\bibfnamefont {A.}~\bibnamefont {Diaz-Guilera}},
  \bibinfo {author} {\bibfnamefont {J.}~\bibnamefont {Duch}}, \ and\ \bibinfo
  {author} {\bibfnamefont {A.}~\bibnamefont {Arenas}},\ }\href@noop {}
  {\bibfield  {journal} {\bibinfo  {journal} {Journal of Statistical Mechanics:
  Theory and Experiment}\ }\textbf {\bibinfo {volume} {2005}},\ \bibinfo
  {pages} {P09008} (\bibinfo {year} {2005})}\BibitemShut {NoStop}%
\bibitem [{\citenamefont {Peixoto}(2017)}]{peixoto2017nonparametric}%
  \BibitemOpen
  \bibfield  {author} {\bibinfo {author} {\bibfnamefont {T.~P.}\ \bibnamefont
  {Peixoto}},\ }\href@noop {} {\bibfield  {journal} {\bibinfo  {journal} {arXiv
  preprint arXiv:1708.01432}\ } (\bibinfo {year} {2017})}\BibitemShut {NoStop}%
\bibitem [{\citenamefont {Peixoto}(2014{\natexlab{b}})}]{PhysRevE.89.012804}%
  \BibitemOpen
  \bibfield  {author} {\bibinfo {author} {\bibfnamefont {T.~P.}\ \bibnamefont
  {Peixoto}},\ }\href {\doibase 10.1103/PhysRevE.89.012804} {\bibfield
  {journal} {\bibinfo  {journal} {Phys. Rev. E}\ }\textbf {\bibinfo {volume}
  {89}},\ \bibinfo {pages} {012804} (\bibinfo {year}
  {2014}{\natexlab{b}})}\BibitemShut {NoStop}%
\bibitem [{\citenamefont {Peixoto}(2013)}]{peixoto2013parsimonious}%
  \BibitemOpen
  \bibfield  {author} {\bibinfo {author} {\bibfnamefont {T.~P.}\ \bibnamefont
  {Peixoto}},\ }\href@noop {} {\bibfield  {journal} {\bibinfo  {journal}
  {Physical review letters}\ }\textbf {\bibinfo {volume} {110}},\ \bibinfo
  {pages} {148701} (\bibinfo {year} {2013})}\BibitemShut {NoStop}%
\bibitem [{\citenamefont {Malliaros}\ and\ \citenamefont
  {Vazirgiannis}(2013)}]{malliaros2013clustering}%
  \BibitemOpen
  \bibfield  {author} {\bibinfo {author} {\bibfnamefont {F.~D.}\ \bibnamefont
  {Malliaros}}\ and\ \bibinfo {author} {\bibfnamefont {M.}~\bibnamefont
  {Vazirgiannis}},\ }\href@noop {} {\bibfield  {journal} {\bibinfo  {journal}
  {Physics Reports}\ }\textbf {\bibinfo {volume} {533}},\ \bibinfo {pages} {95}
  (\bibinfo {year} {2013})}\BibitemShut {NoStop}%
\bibitem [{\citenamefont {Leicht}\ and\ \citenamefont
  {Newman}(2008)}]{leicht2008community}%
  \BibitemOpen
  \bibfield  {author} {\bibinfo {author} {\bibfnamefont {E.~A.}\ \bibnamefont
  {Leicht}}\ and\ \bibinfo {author} {\bibfnamefont {M.~E.~J.}\ \bibnamefont
  {Newman}},\ }\href@noop {} {\bibfield  {journal} {\bibinfo  {journal}
  {Physical review letters}\ }\textbf {\bibinfo {volume} {100}},\ \bibinfo
  {pages} {118703} (\bibinfo {year} {2008})}\BibitemShut {NoStop}%
\bibitem [{\citenamefont {Tenenbaum}\ \emph {et~al.}(2000)\citenamefont
  {Tenenbaum}, \citenamefont {De~Silva},\ and\ \citenamefont
  {Langford}}]{tenenbaum2000global}%
  \BibitemOpen
  \bibfield  {author} {\bibinfo {author} {\bibfnamefont {J.~B.}\ \bibnamefont
  {Tenenbaum}}, \bibinfo {author} {\bibfnamefont {V.}~\bibnamefont {De~Silva}},
  \ and\ \bibinfo {author} {\bibfnamefont {J.~C.}\ \bibnamefont {Langford}},\
  }\href@noop {} {\bibfield  {journal} {\bibinfo  {journal} {science}\ }\textbf
  {\bibinfo {volume} {290}},\ \bibinfo {pages} {2319} (\bibinfo {year}
  {2000})}\BibitemShut {NoStop}%
\bibitem [{\citenamefont {Basu}\ \emph {et~al.}(2004)\citenamefont {Basu},
  \citenamefont {Bilenko},\ and\ \citenamefont
  {Mooney}}]{basu2004probabilistic}%
  \BibitemOpen
  \bibfield  {author} {\bibinfo {author} {\bibfnamefont {S.}~\bibnamefont
  {Basu}}, \bibinfo {author} {\bibfnamefont {M.}~\bibnamefont {Bilenko}}, \
  and\ \bibinfo {author} {\bibfnamefont {R.~J.}\ \bibnamefont {Mooney}},\ }in\
  \href@noop {} {\emph {\bibinfo {booktitle} {Proceedings of the tenth ACM
  SIGKDD international conference on Knowledge discovery and data mining}}}\
  (\bibinfo {organization} {ACM},\ \bibinfo {year} {2004})\ pp.\ \bibinfo
  {pages} {59--68}\BibitemShut {NoStop}%
\bibitem [{\citenamefont {Zhang}\ \emph {et~al.}(2014)\citenamefont {Zhang},
  \citenamefont {Moore},\ and\ \citenamefont {Zdeborov\'a}}]{Zhang2014phase}%
  \BibitemOpen
  \bibfield  {author} {\bibinfo {author} {\bibfnamefont {P.}~\bibnamefont
  {Zhang}}, \bibinfo {author} {\bibfnamefont {C.}~\bibnamefont {Moore}}, \ and\
  \bibinfo {author} {\bibfnamefont {L.}~\bibnamefont {Zdeborov\'a}},\ }\href
  {\doibase 10.1103/PhysRevE.90.052802} {\bibfield  {journal} {\bibinfo
  {journal} {Phys. Rev. E}\ }\textbf {\bibinfo {volume} {90}},\ \bibinfo
  {pages} {052802} (\bibinfo {year} {2014})}\BibitemShut {NoStop}%
\bibitem [{\citenamefont {Thouless}\ \emph {et~al.}(1977)\citenamefont
  {Thouless}, \citenamefont {Anderson},\ and\ \citenamefont
  {Palmer}}]{thouless1977solution}%
  \BibitemOpen
  \bibfield  {author} {\bibinfo {author} {\bibfnamefont {D.~J.}\ \bibnamefont
  {Thouless}}, \bibinfo {author} {\bibfnamefont {P.~W.}\ \bibnamefont
  {Anderson}}, \ and\ \bibinfo {author} {\bibfnamefont {R.~G.}\ \bibnamefont
  {Palmer}},\ }\href@noop {} {\bibfield  {journal} {\bibinfo  {journal}
  {Philosophical Magazine}\ }\textbf {\bibinfo {volume} {35}},\ \bibinfo
  {pages} {593} (\bibinfo {year} {1977})}\BibitemShut {NoStop}%
\end{thebibliography}%

\appendix
\section{Deriving Eq.~\eqref{eq:bp} from Eq.~\eqref{eq:bp:iter}}\label{sec:derive}
First notice that on sparse graphs $\overline{\omega}=2\sum_{\langle ij\rangle\in\mathcal{E}}\omega_{ij}/n^2$ is weak, because number of edges $|\mathcal{E}|$ is much smaller than $n$. Thus the Eq.~\eqref{eq:bp:iter} can be well-approximated as
\begin{align}
\psi_{t_i}^{i\rightarrow k}
&=\frac{1}{Z_{i\rightarrow k}}\prod_{j\in \partial i \backslash k}\sum _{t_j=1}^{q} e^{\beta \omega_{ij} \delta_{t_i t_j}}\psi_{t_j}^{j \rightarrow i} 
\prod_{l\neq i}\sum _{t_l=1}^{q} e^{-\beta\overline{\omega} \delta_{t_i t_l} }\psi_{t_l}^{l \rightarrow i}\nonumber\\
&=\frac{1}{Z_{i\rightarrow k}}\prod_{j\in \partial i \backslash k}\left( e^{\beta \omega_{ij} }\psi_{t_i}^{j \rightarrow i} +\sum _{t_j\neq t_i} e^{0}\psi_{t_j}^{j \rightarrow i} \right)
\prod_{l\neq i}\sum _{t_l=1}^{q} e^{-\beta\overline{\omega} \delta_{t_i t_l} }\psi_{t_l}^{l \rightarrow i}\nonumber\\
&\approx\frac{1}{Z_{i\rightarrow k}}\prod_{j\in \partial i \backslash k}\left(  e^{\beta \omega_{ij} }\psi_{t_i}^{j \rightarrow i} + 1-\psi_{t_i}^{j \rightarrow i}\right)
\prod_{l\neq i}\sum _{t_l=1}^{q} (1-\beta\overline{\omega} \delta_{t_i t_l} )\psi_{t_l}^{l \rightarrow i}\nonumber\\
&=\frac{1}{Z_{i\rightarrow k}}\prod_{j\in \partial i \backslash k}\left( 1+ \psi_{t_i}^{j \rightarrow i}(e^{\beta \omega_{ij} }-1)\right)
\prod_{l\neq i}(1-\beta\overline{\omega} \sum_{t_l=1}^q\delta_{t_i t_l} \psi_{t_l}^{l \rightarrow i})\nonumber\\
&=\frac{1}{Z_{i\rightarrow k}}\prod_{j\in \partial i \backslash k}\left( 1+ \psi_{t_i}^{j \rightarrow i}(e^{\beta \omega_{ij} }-1)\right)
\prod_{l\neq i}(1-\beta\overline{\omega}  \psi_{t_i}^{l \rightarrow i}).
\end{align}
By using the fact that $\overline\omega$ is of small order, and putting the last term in the right hand side of the last equation to  we get
\begin{align}
\psi_{t_i}^{i\rightarrow k}
&\approx\frac{1}{Z_{i\rightarrow k}}\prod_{j\in \partial i \backslash k}\left( 1+ \psi_{t_i}^{j \rightarrow i}(e^{\beta \omega_{ij} }-1)\right)
\prod_{l\neq i}e^{-\beta\overline{\omega}  \psi_{t_i}^{l \rightarrow i}},
\end{align}
which is equivalent to Eq.~\eqref{eq:bp}.

\section{Deriving the Thouless-Anderson-Palmer equations}\label{sec:tap}
The belief propagation equations are written as
\begin{align}
\psi_{t_i}^{i\rightarrow k}&=\frac{e^{h(t_i)}}{Z_{i\rightarrow k}}\prod_{j\in \partial i \backslash k} (1+\psi_{t_i}^{j \rightarrow i}(e^{\beta\omega_{ij}}-1)), \nonumber\\ 
\psi_{t_i}^i&=\frac{ e^{h(t_i)}}{Z_{i}}\prod_{j\in \partial i } (1+\psi_{t_i}^{j \rightarrow i}(e^{\beta\omega_{ij}}-1)).\nonumber
\end{align}
We see from that when $\beta$ is small, we can expand the $\psi_{t_i}^i$ to second order and approximate the cavity probabilities as:
\begin{align}
\psi_{t_i}^i & \approx \frac{ e^{h(t_i)}}{Z_{i}}\prod_{j\in \partial i } (e^{\psi_{t_i}^{j \rightarrow i}\beta \omega_{ij}+\frac{1}{2}[\psi_{t_i}^{j \rightarrow i}-(\psi_{t_i}^{j \rightarrow i})^2] \beta^2 \omega_{ij}^2})\nonumber\\
& \approx \frac{ e^{h(t_i)}}{Z_{i}}\prod_{j\in \partial i } (e^{\psi_{t_i}^{j \rightarrow i}\beta \omega_{ij}+\frac{1}{2}[\psi_{t_i}^{j}-(\psi_{t_i}^{j})^2] \beta^2 \omega_{ij}^2})\nonumber
\end{align}

In dense graphs when $\psi_{t_i}^{i}$ is close to $\psi_{t_i}^{i\rightarrow k}$, the $\psi_{t_i}^{i\rightarrow k}$ can be expressed by
\begin{align}
\psi_{t_i}^{i\rightarrow k}&=\psi_{t_i}^{i}-\frac{e^{H_{t_i}^{i}}}{Z_{i}}+\frac{e^{H_{t_i}^{i}-u_{t_i}^{k\rightarrow i}}}{\sum_{s}{e^{H_{s}^{i}-u_{s}^{k\rightarrow i}}}}\nonumber\\
& \approx \psi_{t_i}^{i}-\frac{e^{H_{t_i}^{i}}}{Z_{i}}+\frac{e^{H_{t_i}^{i}-e^{H_{t_i}^{i}} u_{t_i}^{k\rightarrow i}}}{Z_i-\sum_{s}{(e^{H_{s}^{i}} u_{s}^{k\rightarrow i})}}\nonumber\\
& \approx \psi_{t_i}^{i} + \frac{e^{H_{t_i}^{i}} \sum_{s} {e^{H_{s}^{i}}u_{s}^{k\rightarrow i}} - e^{H_{t_i}^{i}}u_{t_i}^{k\rightarrow i} Z_i}{Z_i^2}\nonumber\\
&=\psi_{t_i}^{i}+\psi_{t_i}^{i} (\sum_{s} \psi_{s}^i u_{s}^{k\rightarrow i} - u_{t_i}^{k\rightarrow i})\nonumber\\
&=\psi_{t_i}^{i}+\psi_{t_i}^{i} (\sum_{s} \psi_{s}^i \ln(1+\psi_{s}^{k \rightarrow i}(e^{\beta\omega_{ik}}-1)) - \ln(1+\psi_{t_i}^{k \rightarrow i}(e^{\beta\omega_{ik}}-1)))\nonumber\\
& \approx \psi_{t_i}^{i}+\psi_{t_i}^{i} (\sum_{s} \psi_{s}^i \ln(1+\psi_{s}^{k}(e^{\beta\omega_{ik}}-1)) - \ln(1+\psi_{t_i}^{k}(e^{\beta\omega_{ik}}-1)))\nonumber\\
& \approx \psi_{t_i}^{i}+\psi_{t_i}^{i} (\sum_{s} \psi_{s}^i \psi_{s}^{k} - \psi_{t_i}^{k})\beta\omega_{ik}
\end{align}

Where $$H_{t_i}^{i\rightarrow k}=e^{h(t_i)}+\sum_{j\in \partial i \backslash k} u_{t_i}^{j\rightarrow i},$$
 $$u_{t_i}^{j\rightarrow i}=\ln(1+\psi_{t_i}^{j \rightarrow i}(e^{\beta\omega_{ij}}-1)),$$ and  
$$H_{t_i}^{i}=e^{h(t_i)}+\sum_{j\in \partial i} u_{t_i}^{j\rightarrow i}.$$ 

Then by substituting $\psi_{t_i}^{i}$ into the BP equations, We finally arrive at the TAP equation:
\begin{equation}
\psi_{t_i}^i = \frac{ e^{h(t_i)}}{Z_{i}}\prod_{j\in \partial i }e^{\psi_{t_i}^{j}+\beta^2 \omega_{ij}^2 (\psi_{t_i}^{j}  (\sum_{s} \psi_{s}^j \psi_{s}^{i}- \psi_{t_i}^{i}) +\frac{1}{2}(\psi_{t_i}^{j}-(\psi_{t_i}^{j})^2)) }
\end{equation}

\begin{figure}
\subfigure[]{
\label{Fig.sub.1}
\includegraphics[width=\tsize\textwidth]{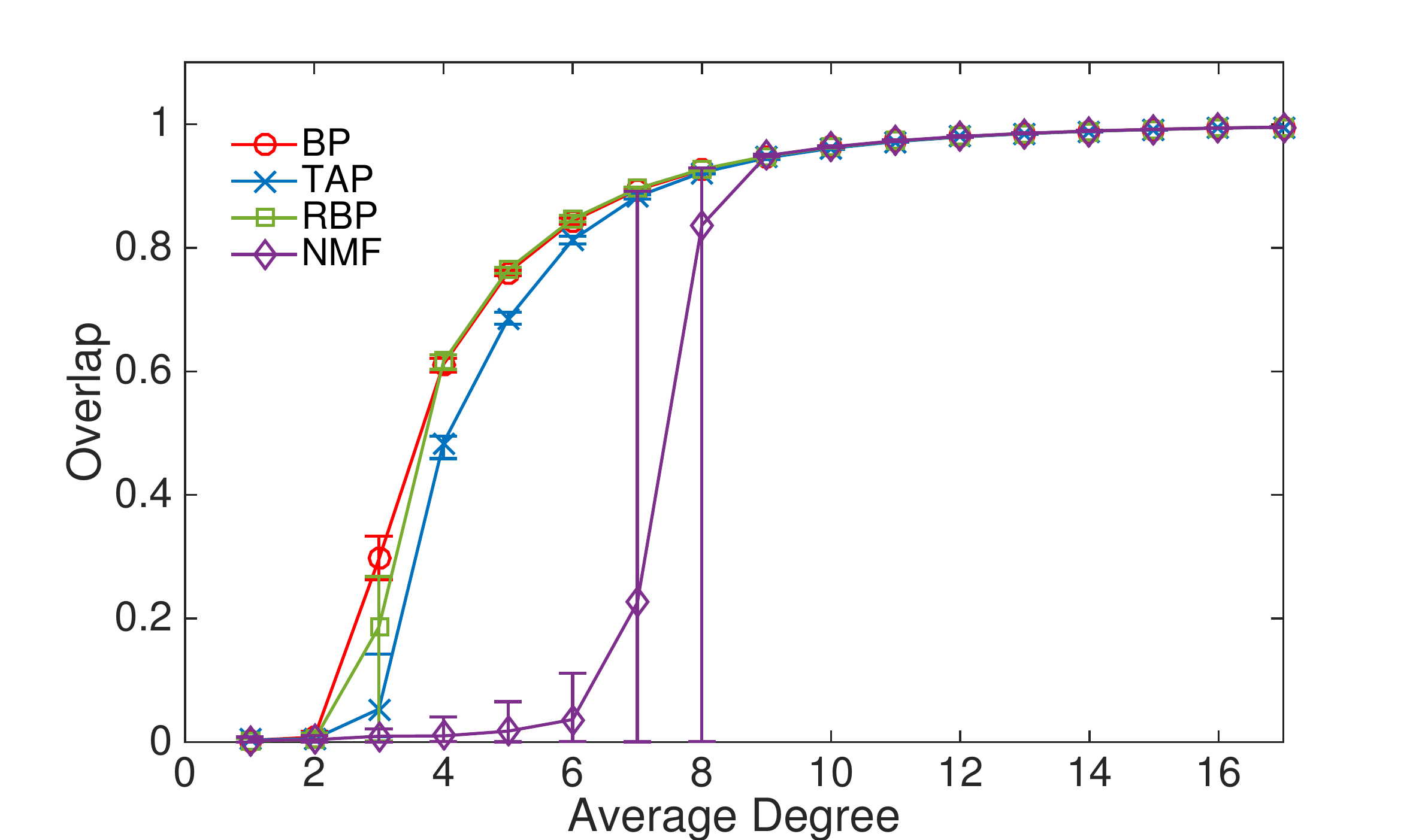}
}
\subfigure[]{
\label{Fig.sub.2}
\includegraphics[width=\tsize\textwidth]{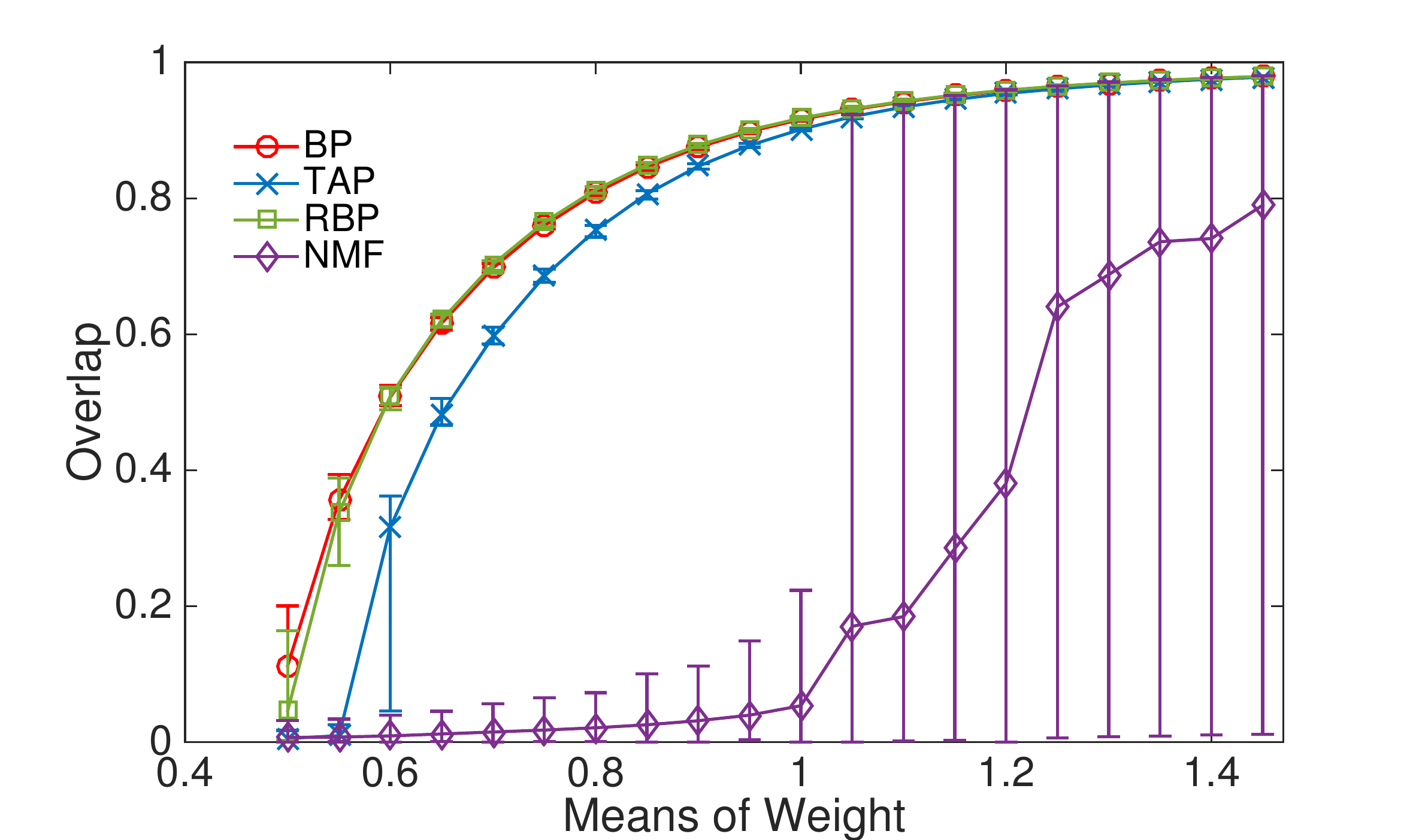}
}
\caption{Comparison of accuracy and time used by Belief Propagation (BP), Relaxed Belief Propagation (RBP), Na\"ive Mean Field (NMF) and Thouless-Anderson-Palmer (TAP). Each point is averaged over 20 realizations of networks with size $n = 100,000$ nodes. (a) In the Gaussian mixture model with Gaussian weight distributions $\omega_{in}$,$\omega_{out}$, which have unit variance and mean $0.75$ and $-0.75$ respectively. (b) The same with (a), but with average degree fixed to $5$, the x-axis shows difference between the means of two distributions, i.e. $\langle  \omega_{in} \rangle-\langle \omega_{out}\rangle$. The maximal iteration time is $10000$ for NMF and $1000$ for others.\label{fig:approx}}
\end{figure}

As noted in the main text, the Na\"ive mean-field equations for this problem can be derived as
\begin{equation}
\psi_{t_i}^i=\frac{ e^{h(t_i)}}{Z_{i}}\prod_{j\in \partial i } e^{\psi_{t_i}^{j}\beta \omega_{ij}}.
\end{equation}
In Fig.~\ref{fig:approx} we did an overall comparison of BP, RBP, NMF and TAP in the Gaussian mixture problem. In panel (a) we vary average degree and in panel (b) we fix average degree and vary mean weight. We can see from the figure that BP almost works best, RBP is very close to BP. although TAP performs not far from BP, NMF which has the same computational complexity work much worse than TAP.

\end{document}